\documentclass[aps,prx,twocolumn,groupaddresses,superscriptaddress,longbibliography]{revtex4-1}
\usepackage{amsmath}
\usepackage{graphicx}
\usepackage{amsfonts}
\usepackage{amssymb}
\usepackage{amsthm}
\usepackage{hyperref}
\newtheorem{theorem}{Theorem}
\newtheorem*{theorem*}{Theorem}
\newtheorem{lemma}[theorem]{Lemma}
\newtheorem{proposition}[theorem]{Proposition}

\DeclareMathOperator*{\argmin}{argmin}
\DeclareMathOperator*{\argmax}{argmax}
\DeclareMathOperator*{\Tr}{Tr}
\newcommand{\ket}[1]{|#1 \rangle}
\newcommand{\bra}[1]{\langle #1|}

\usepackage{epstopdf}
\usepackage{xcolor}
\usepackage[normalem]{ulem}
\definecolor{JP}{RGB}{0,200,51}

\makeatletter
\def\p@subsection{}
\makeatother
\begin{document}

\title{Convex optimization of programmable quantum computers}
\author{Leonardo Banchi}
\email{leonardo.banchi@unifi.it}
\affiliation{ Department of Physics and Astronomy, University of Florence, via G. Sansone 1, I-50019 Sesto Fiorentino (FI), Italy}
\affiliation{ INFN Sezione di Firenze, via G.Sansone 1, I-50019 Sesto Fiorentino (FI), Italy }
\author{Jason Pereira}
\affiliation{Department of Computer Science, University of York, York YO10 5GH, UK}
\author{Seth Lloyd}
\affiliation{Department of Mechanical Engineering, Massachusetts Institute of Technology (MIT), Cambridge MA 02139, USA}
\affiliation{Research Laboratory of Electronics, Massachusetts Institute of Technology (MIT), Cambridge MA 02139, USA}
\author{Stefano Pirandola}
\affiliation{Department of Computer Science, University of York, York YO10 5GH, UK} \affiliation{Research Laboratory of Electronics, Massachusetts Institute of Technology (MIT), Cambridge MA 02139, USA}

\begin{abstract}
A fundamental model of quantum computation is the programmable quantum gate
array. This is a quantum processor which is fed by a \textquotedblleft
program\textquotedblright\ state that induces a corresponding quantum
operation on input states. While being programmable, any finite-dimensional
design of this model is known to be non-universal, meaning that the processor
cannot perfectly simulate an arbitrary quantum channel over the input.
Characterizing how close the simulation is and finding the optimal program
state have been open questions for the last 20 years. Here we answer these
questions by showing that the search for the optimal program state is a convex
optimization problem that can be solved via semidefinite programming and
gradient-based methods commonly employed for machine learning. 
We apply this general result to
different types of processors, from a shallow design based on quantum
teleportation, to deeper schemes relying on port-based teleportation and
parametric quantum circuits.
\end{abstract}
\maketitle

\section*{Introduction}

Back in 1997 a seminal work by Nielsen and
Chuang~\cite{nielsen1997programmable} proposed a quantum version of the
programmable gate array that has become a fundamental model for quantum
computation~\cite{nielsen2000quantum}. This is a quantum processor where a
fixed quantum operation is applied to an input state together with a program
state. The aim of the program state is to induce the processor to apply some
target quantum gate or channel~\cite{watrous2018theory} to the input state.
Such a desired feature of quantum programmability comes with a cost: The model
cannot be universal, unless the program state is allowed to have an infinite
dimension, i.e., infinite
qubits~\cite{nielsen1997programmable,knill2001scheme}. Even though this
limitation has been known for many years, there is still no exact
characterization on how well a finite-dimensional programmable quantum
processor can generate or simulate an arbitrary quantum channel. Also there is
no literature on how to find the corresponding optimal program state or even
to show that this state can indeed be found by some optimization procedure.
Here we show the solutions to these long-standing open problems.

Here we show that the optimization of programmable quantum computers
is a convex problem for which the solution can always be found by means of
classical semidefinite programming (SDP) and classical gradient-based 
methods that are commonly employed for machine learning applications. 
Machine learning (ML) methods have found wide applicability across many disciplines
~\cite{MLbook}, and we are currently witnessing 
the development of new hybrid areas of investigation where ML methods 
are interconnected with quantum information theory, such as
quantum-enhanced machine
learning~\cite{QML1,QML2,dunjko2018machine,schuld2015introduction,ciliberto2018quantum}
(e.g., quantum neural
networks, quantum annealing etc.), protocols of quantum-inspired machine
learning (e.g., for recommendation systems~\cite{QIML1} or component analysis
and supervised clustering~\cite{QIML2}) and classical learning methods applied
to quantum computers, as explored here in this manuscript.

In our work, we quantify the error between an arbitrary target channel and its
programmable simulation in terms of the diamond 
distance~\cite{kitaev2002classical,watrous2018theory}  and other suitable
cost functions, including the trace distance and the quantum fidelity. For all
the considered cost functions, we are able to show that the minimization of
the simulation error is a convex optimization problem in the space of the
program states. This already solves an outstanding problem which affects
various models of quantum computers (e.g., variational quantum circuits) where
the optimization over classical parameters is non-convex and therefore not
guaranteed to converge to a global optimum. By contrast, because our problem
is proven to be convex, we can use SDP to minimize the diamond distance and
always find the optimal program state for the simulation of a target channel,
therefore optimizing the programmable quantum processor. Similarly, we may
find suboptimal solutions by minimizing the trace distance or the quantum
fidelity by means of gradient-based techniques adapted 
from the ML literature, such as the projected
subgradient method~\cite{boyd2003subgradient} and the conjugate gradient
method~\cite{jaggi2011convex,jaggi2013revisiting}.
We note indeed that the minimization of the $\ell_1$-norm, mathematically related
to the quantum trace distance, is widely employed
in many ML tasks \cite{duchi2008efficient,liu2013tensor}, so many of those techniques
can be adapted for learning program states.

With these general results in our hands, we first discuss the optimal learning
of arbitrary unitaries with a generic programmable quantum processor. Then, we
consider specific designs of the processor, from a shallow scheme based on the
teleportation protocol, to higher-depth designs based on
port-based teleportation (PBT)~\cite{ishizaka2008asymptotic,ishizaka2009quantum,ishizaka2015some} and
parametric quantum circuits (PQCs)~\cite{lloyd1996universal}, introducing a
suitable convex reformulation of the latter. In the various cases, we
benchmark the processors for the simulation of basic unitary gates (qubit
rotations) and various basic channels, including the amplitude damping channel
which is known to be the most difficult to
simulate~\cite{pirandola2017fundamental,commREVIEW}. For the deeper designs,
we find that the optimal program states do not correspond to the Choi matrices
of the target channels, which is rather counter-intuitive and unexpected.



\section*{RESULTS}

We first present our main theoretical results on how to train the program state of programmable 
quantum processors, either via convex optimization or first-order 
gradient based algorithms. We then apply our general methods to study the learning of 
arbitrary unitaries, and the simulation of different channels via processors
built either from quantum teleportation and its generalization, or from parametric quantum 
circuits.

\subsection*{Programmable quantum computing}\label{s:program}

Let us consider an arbitrary mapping from $d$-dimensional input states into 
$d^{\prime}$-dimensional output states, where $d^{\prime}\neq d$ in the
general case. This is described by a quantum channel $\mathcal E$ that 
may represent the overall action of a quantum
computation and does not need to be a unitary transformation. 
Any channel $\mathcal{E}$ can be simulated by means of a programmable quantum processor,
which is modeled in general by a fixed completely positive trace-preserving
(CPTP) map $Q$ which is applied to both the input state and a variable
program state $\pi$. In this way, the processor transforms the input state by
means of an approximate channel $\mathcal{E}_{\pi}$ as
\begin{equation}
\mathcal{E}_{\pi}(\rho)=\Tr_{2}\left[  Q(\rho\otimes\pi)\right]
,\label{channelDEF}%
\end{equation}
where $\Tr_{2}$ is the partial trace over the program state.
A fundamental result \cite{nielsen1997programmable}
is that there is no {\it fixed} quantum ``processor'' $Q$ that is able 
to {\it exactly} simulate any quantum channel $\mathcal E$. In other terms, 
given $\mathcal E$, we cannot find the corresponding program $\pi$ such that 
$\mathcal E \equiv \mathcal E_\pi$. Yet  simulation can be achieved in an 
approximate sense, where the quality of the simulation may increase for larger 
program dimension.
In general, the open problem is to determine the optimal program state $\tilde{\pi}$
that minimizes the simulation error, that can be quantified by the cost function
\begin{equation}
C_{\diamond}(\pi):=\left\Vert \mathcal{E}-\mathcal{E}_{\pi}\right\Vert
_{\diamond},\label{Cdiamond}%
\end{equation}
namely the diamond distance~\cite{kitaev2002classical,watrous2018theory}
between the target channel $\mathcal{E}$ and its simulation $\mathcal{E}_{\pi
}$. 
In other words,
\begin{equation}
\text{{Find} }\tilde{\pi}\text{ {such that} }C_{\diamond}(\tilde{\pi}%
)=\min_{\pi}C_{\diamond}(\pi). \label{sol1}%
\end{equation}
From theory~\cite{nielsen1997programmable,knill2001scheme} we know that we
cannot achieve $C_{\diamond}=0$ for arbitrary $\mathcal{E}$ unless $\pi$ and
$Q$ have infinite dimensions. As a result, for any finite-dimensional
realistic design of the quantum processor, finding the optimal program state
$\tilde{\pi}$ is an open problem.
Recall that the diamond distance is defined by $\left\Vert \mathcal{E}%
-\mathcal{E}_{\pi}\right\Vert _{\diamond}:=\max_{\varphi}\left\Vert
\mathcal{I}\otimes\mathcal{E}(\varphi)-\mathcal{I}\otimes\mathcal{E}_{\pi
}(\varphi)\right\Vert _{1}$, where $\mathcal{I}$ is the identity map and
$\left\Vert O\right\Vert _{1}:=\mathrm{Tr}\sqrt{O^{\dagger}O}$ is the trace
norm~\cite{nielsen2000quantum}.

\begin{figure}[t]
\vspace{-0.0cm}
\par
\begin{center}
\includegraphics[width=0.40\textwidth] {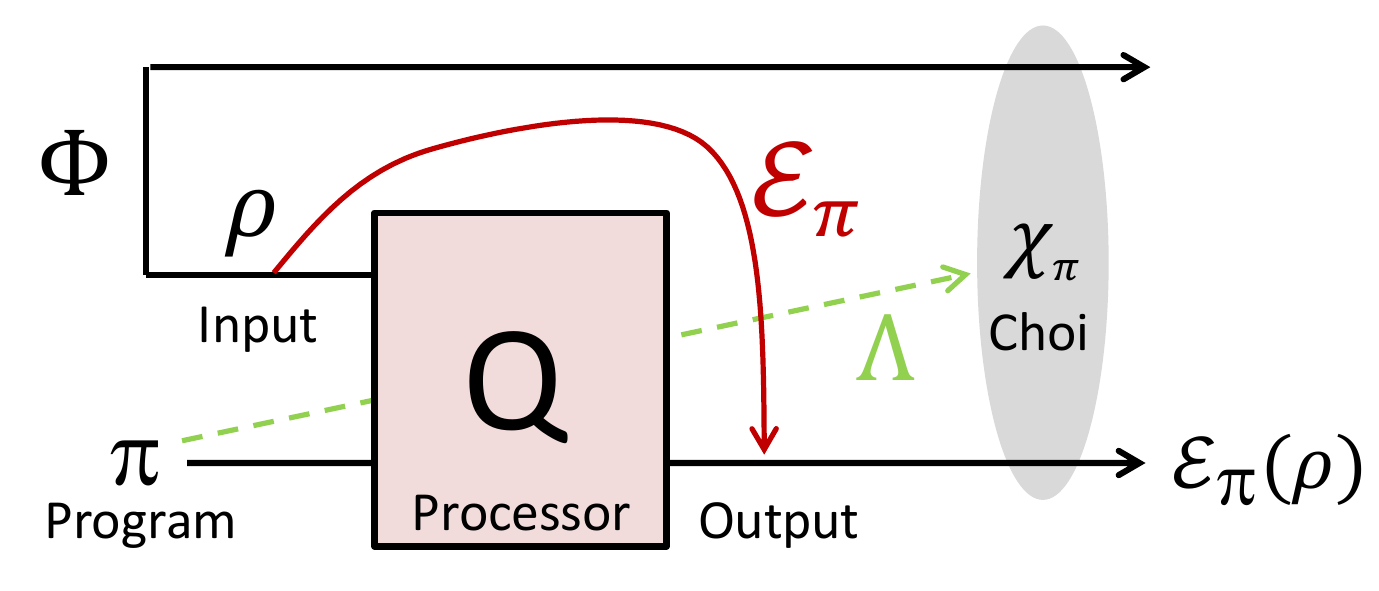}
\end{center}
\par
\vspace{-0.5cm}\caption{Quantum processor $Q$ with program state $\pi$\ which
simulates a quantum channel $\mathcal{E}_{\pi}$\ from input to output. We also
show the CPTP map $\Lambda$ of the processor, from the program state $\pi$ to
the output Choi matrix $\chi_{\pi}$ (generated by partial transmission of the
maximally-entangled state $\Phi$).}%
\label{progP}%
\end{figure}

It is important to note that this problem can be reduced to a simpler one by
introducing the channel's Choi matrix 
\begin{align}
\chi_{\mathcal{E}_{\pi}}  &  =\mathcal{I}\otimes\mathcal{E}_{\pi}%
(\Phi)\nonumber\\
&  =d^{-1}%
{\textstyle\sum\nolimits_{ij}}
\left\vert i\right\rangle \!\left\langle j\right\vert \otimes\mathrm{Tr}%
_{2}\left[  Q(\left\vert i\right\rangle \!\left\langle j\right\vert \otimes
\pi)\right]  ,
\end{align}
where $\Phi:=\left\vert \Phi\right\rangle\!
\left\langle \Phi\right\vert $ is a $d$-dimensional maximally-entangled state.
From this expression, it is clear that the Choi matrix $\chi_{\mathcal{E}%
_{\pi}}$ is linear in the program state $\pi$. More precisely, the Choi matrix
$\chi_{\mathcal{E}_{\pi}}$ at the output of the processor $Q$ can be directly
written as a CPTP linear map $\Lambda$ acting on the space of the program
states $\pi$, i.e.,%
\begin{equation}
\chi_{\pi}:=\chi_{\mathcal{E}_{\pi}}=\Lambda(\pi). \label{lambdadef}%
\end{equation}%
This map is also depicted in Fig.~\ref{progP}
and fully describes the action of the processor $Q$.
Then, using results from
Refs.~\cite{watrous2018theory,Karol,fuchs1999cryptographic}, we may write
\begin{equation}
C_{\diamond}(\pi)\leq d~C_{1}(\pi)\leq2d\sqrt{C_{F}(\pi)},
\end{equation}
where 
\begin{equation}
C_{1}(\pi):=\left\Vert \chi_{\mathcal{E}}-\chi_{\pi}\right\Vert _{1}%
,\label{traceD}%
\end{equation}
is the trace distance~\cite{nielsen2000quantum} between target and simulated
Choi matrices, and 
\begin{equation}
C_{F}(\pi)=1-F(\pi)^{2}, \label{Cf}%
\end{equation}
where $F(\pi)$ is Bures' fidelity between the two Choi matrices $\chi
_{\mathcal{E}}$ and $\chi_{\pi}$, i.e.,%
\begin{equation}
F(\pi):=\left\Vert \sqrt{\chi_{\mathcal{E}}}\sqrt{\chi_{\pi}}\right\Vert
_{1}=\mathrm{Tr}\sqrt{\sqrt{\chi_{\mathcal{E}}}\chi_{\pi}\sqrt{\chi
_{\mathcal{E}}}}. \label{fidCC}%
\end{equation}
Another possible upper bound can be written using the quantum Pinsker's
inequality~\cite{pinsker1964information,carlen2012bounds}. In fact, we may
write $C_{1}(\pi)\leq(2\ln\sqrt{2})\sqrt{C_{R}(\pi)}$, where%
\begin{equation}
C_{R}(\pi):=\min\left\{  S(\chi_{\mathcal{E}}||\chi_{\pi}),S(\chi_{\pi}%
||\chi_{\mathcal{E}})\right\}  , \label{REcost}%
\end{equation}
and $S(\rho||\sigma):=\mathrm{Tr}[\rho(\log_{2}\rho-\log_{2}\sigma)]$ is the
quantum relative entropy between $\rho$ and $\sigma$.
In Supplementary Note~\ref{a:cost} we also introduce a cost function $C_p(\pi)$ based on 
the Schatten p-norm.

\subsection*{Convex optimization}
One of the main problems in the optimization
of reconfigurable quantum chips is that the relevant cost functions are not
convex in the set of classical parameters. This problem is completely solved
here thanks to the fact that the optimization of a programmable quantum
processor is done\ with respect to a quantum state. In fact,  
in the methods section we prove the following


\begin{theorem}
\label{t:convexC1CF} Consider the simulation of a target quantum channel
$\mathcal{E}$ by means of a programmable quantum processor $Q$. The
optimization of the cost functions $C_{\diamond}$, $C_{1}$, $C_{F}$, $C_R$ or $C_p$ is a
convex problem in the space of program states $\pi$. In particular, the global
minimum $\tilde{\pi}$ for $C_{\diamond}$ can always be found as a local
minimum.
\end{theorem}

This convexity result is generally valid for any cost function which is convex
in $\pi$. This is the case for any desired norm, not only the trace norm, but
also the Frobenius norm, or any Schatten p-norm. It also applies to the
relative entropy. Furthermore, the result can also be extended to any convex
parametrization of the program states. 

When dealing with convex optimization with respect to positive operators, 
the standard approach is to map the problem to a form that is solvable via
semi-definite programming (SDP) \cite{watrous2009semidefinite,watrous2013simpler}.
Since the optimal program is the one minimizing the cost function, it 
is important to write the computation of the cost function itself
as a minimization. For the case of the diamond distance, this 
can be achieved by using the dual formulation \cite{watrous2009semidefinite}.
More precisely, 
consider the
linear map $\Omega_{\pi}:=\mathcal{E}-\mathcal{E}_{\pi}$ with Choi matrix
$\chi_{\Omega_{\pi}}=\chi_{\mathcal{E}}-\chi_{\pi}=\chi_{\mathcal{E}}%
-\Lambda({\pi})$, and the spectral norm $\left\Vert O\right\Vert
_{\infty }:=\max\{\left\Vert Ou\right\Vert
:u\in\mathbb{C}^{d},\left\Vert u\right\Vert \leq1\}$, which is the
maximum eigenvalue of $\sqrt{O^{\dagger}O}$. Then, by the strong
duality of the diamond norm, $C_{\diamond}(\pi)=\left\Vert \Omega_{\pi}\right\Vert _{\diamond}$
is given by the SDP~\cite{Watrous}
\begin{gather}
\mathrm{Minimize~}2\left\Vert \mathrm{Tr}_{2}Z\right\Vert _{\infty
},\nonumber\\
\text{\textrm{Subject to}~}Z\geq0~\text{\textrm{and} }Z\geq d
(\chi_{\mathcal E}-\Lambda(\pi)). \label{SDP}%
\end{gather}
The importance of the above dual formulation is
that the diamond distance is a \textit{minimization}, rather than a
maximization over a set of matrices. In order to find the optimal program
$\tilde{\pi}$ we apply the unique minimization of Eq.~\eqref{SDP} where $\pi$
is variable and satisfies the additional constraints $\pi\geq0$ and
$\mathrm{Tr}(\pi)=1$.

In the methods section we show that other cost functions such as 
$C_1$ and $C_F$ can also be written as SDPs. Correspondingly, the optimal 
programs $\tilde\pi$ can be obtained by numerical SDP solvers.
Most numerical packages implement second-order algorithms
such as the interior point method \cite{vandenberghe1996semidefinite}.
However, second order methods tend to be computationally heavy for 
large problem sizes \cite{chao2013first,monteiro2003first}, namely when $\tilde\pi$ 
contains many qudits. 
In the following section we introduce first order methods, that are better suited 
for larger program states. 
It is important to remark that there also exist 
zeroth-order (derivative-free) methods, 
such as the simultaneous perturbation stochastic
approximation method \cite{spall2000adaptive}, which was utilized for a quantum problem 
in \cite{zhuang2019physical}. However, it is known that zeroth-order methods 
normally have slower convergence times \cite{harrow2019low} compared to first-order 
methods. 

\subsection*{Gradient based optimization } \label{s:firstord}

In machine learning applications, where a large amount of data is commonly available, 
there have been several works that study the minimization of suitable matrix norms 
for different purposes 
\cite{liu2013tensor,duchi2008efficient,cai2010singular,recht2010guaranteed}. 
First-order methods, are preferred for large dimensional problems, as they 
are less computationally intensive and require less memory. 
Here we show how to apply first-order  (gradient-based) algorithms, 
which are widely employed in machine learning applications, 
to find the optimal quantum program.

For this purpose, we need to introduce the subgradient of the cost function $C$ 
at any point $\pi\in\mathcal{S}$, which is the
set%
\begin{equation}
\partial C(\pi)=\{Z:C(\sigma)-C(\pi)\geq\text{\textrm{Tr}}[Z(\sigma
-\pi)],~\forall\sigma\in\mathcal{S}\},
\label{subgradient}
\end{equation}
where $Z$ is Hermitian~\cite{nesterov2013introductory,coutts2018certifying}.
If $C$ is differentiable, then $\partial C(\pi)$ contains a single
element: its gradient $\nabla C(\pi)$. We explicitly compute this
gradient for an arbitrary programmable quantum processor \eqref{channelDEF} whose 
Choi matrix $\chi_{\mathcal E_\pi} \equiv \chi_{\pi}= \Lambda(\pi)$, can be written as a quantum 
channel $\Lambda$ that maps a generic program state to the processor's Choi matrix. 
This map can be defined by its Kraus
decomposition $\Lambda(\pi)=\sum_{k}A_{k}\pi A_{k}^{\dagger}$ for some
operators $A_{k}$. In fact, let us call $\Lambda^{\ast}(\rho)=\sum_{k}%
A_{k}^{\dagger}\rho A_{k}$ the dual map, then in 
the methods section we prove the following

\begin{theorem}
\label{t:gradients}Consider an arbitrary quantum channel $\mathcal{E}$\ with Choi
matrix $\chi_{\mathcal{E}}$ which is simulated by a quantum processor $Q$ with
map $\Lambda(\pi)=\chi_{\pi}$ (and dual map $\Lambda^{\ast}$). Then, we may
write the following gradients for the trace distance cost $C_{1}(\pi)$ and the
infidelity cost $C_{F}(\pi)$%
\begin{align}
\nabla C_{1}(\pi) &  =\sum_{k}\mathrm{sign}(\lambda_{k})\Lambda^{\ast}%
(P_{k}),\\
\nabla C_{F}(\pi) &  =-2\sqrt{1-C_{F}(\pi)}\nabla F(\pi),\\
\nabla F(\pi) &  =\frac{1}{2}\Lambda^{\ast}\left[  \sqrt{\chi_{\mathcal{E}}%
}\left(  \sqrt{\chi_{\mathcal{E}}}\,\Lambda(\pi)\,\sqrt{\chi_{\mathcal{E}}%
}\right)  ^{-\frac{1}{2}}\sqrt{\chi_{\mathcal{E}}}\right]  ,\label{fidGRAD}%
\end{align}
where $\lambda_{k}$ ($P_{k}$) are the eigenvalues (eigenprojectors) of the
Hermitian operator $\chi_{\pi}-\chi_{\mathcal{E}}$. When $C_{1}(\pi)$ or
$C_{F}(\pi)$ are not differentiable in $\pi$, then the above expressions
provide an element of the subgradient $\partial C(\pi)$.
\end{theorem}

Once we have the (sub)gradient of the cost function $C$, we can solve the
optimization $\min_{\pi\in\mathcal{S}}C(\pi)$ using the projected subgradient
method~\cite{nesterov2013introductory,boyd2003subgradient}. Let $\mathcal{P}%
_{\mathcal{S}}$ be the projection onto the set of program states $\mathcal{S}%
$, namely $\mathcal{P}_{\mathcal{S}}(X)=\mathrm{argmin}_{\pi\in
S}\Vert X-\pi\Vert_{2}$, that we show to be computable from the
spectral decomposition of any Hermitian $X$~(see 
Theorem~\ref{t:proj} in the methods section).
 Then, we iteratively apply the steps
\begin{equation}%
\begin{array}
[c]{l}%
1)~\mathrm{Select~an~operator~}g_{i}{~\mathrm{from~}}\partial C(\pi_{i}),\\
2)~\pi_{i+1}=\mathcal{P}_{\mathcal{S}}\left(  \pi_{i}-\alpha_{i}g_{i}\right)
,
\end{array}
\label{projsubgrad}%
\end{equation}
where $i$ is the iteration index, $\alpha_{i}$ is what is called 
``learning rate'', and 
Theorem~\ref{t:gradients} can be employed to find $g_{i}$ at each step.
It is simple to show that $\pi_i$ 
converges to the optimal program state $\tilde{\pi}$ in $\mathcal{O}%
(\epsilon^{-2})$ steps, for any desired precision $\epsilon$ such that 
$|C(\pi)-C(\tilde{\pi})|\leq\epsilon$. Another
approach is the conjugate gradient
method~\cite{jaggi2011convex,nesterov2013introductory}, sometimes called
Frank-Wolfe algorithm. Here, we apply
\begin{equation}%
\begin{array}
[c]{l}%
1)~\text{\textrm{Find the smallest eigenvalue}}~\left\vert \sigma
_{i}\right\rangle ~\text{\textrm{of}}~\nabla C(\pi_{i}),\\
2)~\pi_{i+1}=\frac{i}{i+2}\pi_{i}+\frac{2}{i+2}\left\vert \sigma
_{i}\right\rangle \left\langle \sigma_{i}\right\vert .
\end{array}
\label{FrankWolfe}%
\end{equation}
When the gradient of $f$ is Lipschitz continuous with constant $L$, the method
converges after $\mathcal{O}(L/\epsilon)$
steps~\cite{jaggi2013revisiting,nesterov2005smooth}. To justify the
applicability of this method a suitable smoothening of the cost function must
be employed~\footnote{The downside of the conjugate gradient method is that it
necessarily requires a differentiable cost function $C$, with gradient $\nabla
C$. Specifically, this may create problems for the trace distance cost $C_{1}$
which is generally non-smooth. A solution to this problem is to define the cost
function in terms of the smooth trace distance $C_{\mu}(\pi)=\mathrm{Tr}%
\left[  h_{\mu}\left(  \chi_{\pi}-\chi_{\mathcal{E}}\right)
\right]  $ where $h_{\mu}$ is the so-called Huber penalty function
$h_{\mu}(x):=x^{2}/(2\mu) \mathrm{~if~}|x|<\mu$ and $|x|-\mu/2
\mathrm{~if~}|x|\geq\mu.$
This quantity satisfies $C_{\mu}(\pi)\leq C_{1}(\pi)\leq
C_{\mu}(\pi)+\mu d/2 $ and is a convex function over program
states, with gradient $\nabla
C_{\mu}(\pi)=\Lambda^{\ast}[h_{\mu}^{\prime}(\chi_{\pi}-\chi
_{\mathcal{E}})] $.}.

\subsection*{Learning of arbitrary unitaries}
One specific application is the
simulation of quantum gates or, more generally, unitary transformations
\cite{lloyd1996universal,khaneja2005optimal,banchi2016quantum,innocenti2018supervised,mitarai2018quantum}.
Here, the infidelity provides the most convenient cost function, as the optimal
program can be found analytically.  In fact,
suppose we use a quantum processor with map $\Lambda$\ to simulate a target
unitary $U$. Because the Choi matrix of $U$ is pure $|\chi_{U}\rangle
\langle\chi_{U}|$, we first note that $F(\pi)^2=\langle\chi_{U}|\Lambda
(\pi)|\chi_{U}\rangle$ and then we see that Eq.~(\ref{fidGRAD}) drastically
simplifies to $\nabla F(\pi)=\Lambda^{\ast}\left(  |\chi_{U}\rangle\langle
\chi_{U}|\right)  /\sqrt{4F(\pi)^2}$. As a result, we find
\begin{equation}
\nabla C_{F}(\pi)=-\Lambda^{\ast}\left[  |\chi_{U}\rangle\langle\chi
_{U}|\right]  ,
\end{equation}
where there is no dependence on $\pi$. Therefore, using the conjugate gradient
method in Eq.~\eqref{FrankWolfe}, we see that the optimal program state
$\tilde{\pi}$ for the infidelity cost function $C_{F}$\ is a fixed point of
the iteration and is equal to the maximum eigenvector of $\Lambda^{\ast
}\left[  |\chi_{U}\rangle\langle\chi_{U}|\right]  $.

\subsection*{Teleportation processor}

Once we have shown how to optimize a
generic programmable quantum processor, we discuss some specific designs, over
which we will test the optimization procedure. One possible (shallow) design
for the quantum processor $Q$ is a generalized teleportation
protocol~\cite{bennett1993teleporting} over an arbitrary program state $\pi$.
In dimension $d$, the protocol involves a basis of $d^{2}$ maximally-entangled
states $|\Phi_{i}\rangle$ and a basis $\{U_{i}\}$ of teleportation unitaries
such that $\mathrm{Tr}(U_{i}^{\dagger}U_{j})=d\delta_{ij}$~\cite{teleREVIEW}.
An input $d$-dimensional state $\rho$ and the $A$ part of the program
$\pi_{AB}$ are subject to the projector $|\Phi_{i}\rangle\left\langle \Phi
_{i}\right\vert $. The classical outcome $i$ is communicated to the $B$ part
of $\pi_{AB}$\ where the correction $U_{i}^{-1}$ is applied.

The above procedure defines the teleportation channel $\mathcal{E}_{\pi}$ over $\rho$
\begin{equation}
	\mathcal{E}^{\rm tele}_{\pi}(\rho)=\sum_{i}U_{i}^{B}\langle\Phi_{i}^{SA}|\rho^{S}%
\otimes\pi^{AB}|\Phi_{i}^{SA}\rangle U_{i}^{B\dagger}.
\end{equation}
Its Choi matrix can be written as $\chi_{\pi}=\Lambda_{\text{tele}}(\pi)$,
where the map of the teleportation processor is equal to%
\begin{equation}
\Lambda_{\text{tele}}(\pi)=d^{-2}\sum_{i}\left(  U_{i}^{\ast}\otimes
U_{i}\right)  \pi\left(  U_{i}^{\ast}\otimes U_{i}\right)  ^{\dagger},
\label{eqAB}%
\end{equation}
which is clearly self-dual $\Lambda^{\ast}=\Lambda$. Given a target quantum
channel $\mathcal{E}$ which is
teleportation-covariant~\cite{pirandola2017fundamental,commREVIEW}, 
namely when $[\pi,U_{i}^{\ast}\otimes
U_{i}]=0$, then we know that
that its simulation is perfect and the optimal program $\tilde{\pi}$ is the
channel's Choi matrix, i.e., one of the fixed points of the map $\Lambda
_{\text{tele}}$. For a general channel, the optimal program $\tilde{\pi}$ can
be approximated by using the cost functions in our Theorem~\ref{t:gradients} with
$\Lambda$ being given in Eq.~(\ref{eqAB}), or directly found by optimizing
$C_{\diamond}(\pi) $.

\subsection*{Port-based teleportation}

\begin{figure}[t]
\centering
\includegraphics[width=0.95\linewidth]{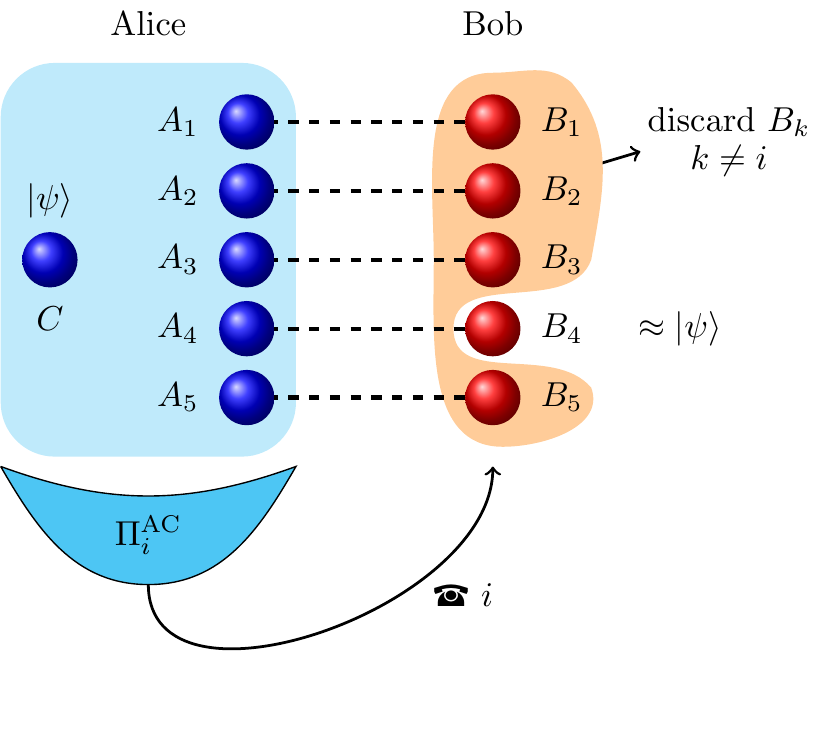} \caption{PBT
scheme. Two distant parties, Alice and Bob, share $N$ maximally entangled
pairs $\{A_{k},B_{k}\}_{k=1}^{N}$. Alice also has another system $C$ in the
state $|\psi\rangle$. To teleport $C$, Alice performs the POVM $\{\Pi
_{i}^{\mathbf{A}C}\}$ on all her local systems $\mathbf{A}=\{A_{k}\}_{k=1}^{N}$
and $C$. She then communicates the outcome $i$ to Bob. Bob discards all his
systems $\mathbf{B}=\{B_{k}\}_{k=1}^{N}$ with the exception of $B_{i}$. After
these steps, the state $|\psi\rangle$ is approximately teleported to $B_{i}$.
Similarly, an arbitrary channel $\mathcal{E}$ is simulated with $N$ copies of
the Choi matrix $\chi_{\mathcal{E}}^{A_{k}B_{k}}$. The figure shows an example
with $N=5$, where $i=4$ is selected. }%
\label{fig:pbt}%
\end{figure}

A deeper design is provided by a PBT\ processor, whose overall protocol is illustrated 
in Fig.~\ref{fig:pbt}. Here we consider a more
general formulation of the original
PBT\ protocol~\cite{ishizaka2008asymptotic,ishizaka2009quantum} where the
resource entangled pairs are replaced by an arbitrary program state $\pi$. In
a PBT\ processor, each party has $N$ systems (or `ports'), $\mathbf{A}%
=\{A_{1},\dots,A_{N}\}$ for Alice and $\mathbf{B}=\{B_{1},\dots,B_{N}\}$ for
Bob. These are prepared in a program state $\pi_{\mathbf{AB}}$. To teleport an
input state $\rho_{C}$, Alice performs a joint positive operator-value measurement (POVM)
$\{\Pi_{i}\}$
\cite{ishizaka2008asymptotic} on system $C$ and the $\mathbf{A}$-ports. She
then communicates the outcome $i$ to Bob, who discards all ports except
$B_{i}$ which is the output $B_{\text{out}}$. 
 The resulting PBT channel $\mathcal{P}_{\pi
}:\mathcal{H}_{C}\mapsto\mathcal{H}_{B_{\mathrm{out}}}$ is then
\begin{align}
\mathcal{P}_{\pi}(\rho) 
&  =\sum_{i=1}^{N}\Tr_{\mathbf{A\bar{B}}_{i}C}\left[  \sqrt{\Pi_{i}}%
(\pi_{\mathbf{AB}}\otimes\rho_{C})\sqrt{\Pi_{i}}\right]  _{B_{i}\rightarrow
B_{\mathrm{out}}}\nonumber
\\
&  = \sum_{i=1}^{N}\Tr_{\mathbf{A\bar{B}}_{i}C}\left[
\Pi_{i}(\pi_{\mathbf{AB}}\otimes\rho_{C})\right]  _{B_{i}\rightarrow
B_{\mathrm{out}}},\label{PBTchan}
\end{align}
where $\mathbf{\bar{B}}_{i}=\mathbf{B}\backslash B_{i}=\{B_{k}:k\neq i\}$. 

In the standard PBT protocol~\cite{ishizaka2008asymptotic,ishizaka2009quantum}, 
the program state is fixed as 
$\pi_{\mathbf{AB}}=\bigotimes_{k=1}^{N}\Phi_{A_{k}B_{k}}$,
where $|\Phi_{A_{k}B_{k}}\rangle$ are Bell states, and 
the following POVM is used%
\begin{equation}
\Pi_{i}=\tilde{\Pi}_{i}+\frac{1}{N}\left(  \openone-\sum_{k}\tilde{\Pi}%
_{k}\right)  ,\label{POVMchoice}%
\end{equation}
where
\begin{align}
\tilde{\Pi}_{i}  & =\sigma_{\mathbf{A}C}^{-1/2}\Phi_{A_{i}C}{\sigma
}_{\mathbf{A}C}^{-1/2},\\
\sigma_{\mathbf{A}C}  & :=\sum_{i=1}^{N}\Phi_{A_{i}C},
\end{align}
and $\sigma^{-1/2}$ is an operator defined only on the support of $\sigma$.
The PBT\ protocol is formulated for $N\geq2$ ports. However, we also include
here the trivial case for $N=1$, corresponding to the process where Alice's
input is traced out and the output is the reduced state of Bob's port, i.e., a
maximally mixed state.
In the limit $N\rightarrow\infty$, the standard PBT protocol  
approximates an identity channel
$\mathcal{P}_{\pi}(\rho)\approx\rho$, with 
fidelity~\cite{ishizaka2008asymptotic,ishizaka2015some}
$ F_{\pi}=1-\mathcal{O}\left(  \frac{1}{N}\right) $,
so perfect simulation is possible only in the limit $N\rightarrow\infty$. 
Since the standard PBT-protocol provides an approximation to the identity channel,
we call it $\mathcal{I}_N$.

From the PBT-simulation of the identity channel  it is possible to 
approximate any general channel $\mathcal E$ by noting that 
$\mathcal{E}$ can be written as a composition
$\mathcal{E}\circ\mathcal{I}$, where $\mathcal{I}$ is the identity channel.
This is done by replacing the identity
channel $\mathcal{I}$ with its PBT simulation $\mathcal{I}_{N}$, and then
applying $\mathcal{E}$ to $B_{i}$. However, since Bob does not perform any
post-processing on his systems $\mathbf{B}$, aside from discarding all ports
$B_{k}$ with $k\neq i$, he can also apply \textit{first} the channel
$\mathcal{E}^{\otimes N}$ to all his ports and \textit{then} discard all the
ports $B_{k}$ with $k\neq i$.
In doing so, he changes the program state to
\begin{equation}
\pi_{\mathbf{AB}}=\openone_{A}\otimes\mathcal{E}_{B}^{\otimes N}\left[
\bigotimes_{k=1}^{N}\Phi_{A_{k}B_{k}}\right]  =\bigotimes_{k=1}^{N}%
\chi_{\mathcal{E}}^{A_{k}B_{k}}.\label{e:pbtchoiprogram}%
\end{equation}
In other terms, any channel $\mathcal{E}$ can be PBT-approximated by $N$
copies of its Choi matrix $\chi_{\mathcal{E}}$ as program state. 
Since PBT-simulation can be decomposed as 
$\mathcal{E}_{\pi}=\mathcal{E}\circ\mathcal{I}_{N}$,
the error $C^N_\diamond = \|\mathcal E-\mathcal E_\pi\|_\diamond$ 
in simulating the channel $\mathcal E\equiv\mathcal E\circ\mathcal I$
satisfies
\begin{equation}
	C^N_\diamond = \|\mathcal E\circ\mathcal I-\mathcal E\circ\mathcal I_N\| \le 
\|\mathcal{I}-\mathcal{I}_{N}\|_{\diamond}\leq2d(d-1)N^{-1}~.
\label{simerr}
\end{equation}
where we used the data processing inequality and an upper bound from 
~\cite{pirandola2018fundamental}.
While the channel's Choi matrix assures that $C^{N}_{\diamond} \rightarrow 0$ for large $N$, for any finite $N$ it does not represent the optimal program state.
In general, for any finite $N$, finding
the optimal program state $\pi_{\mathbf{AB}}$ simulating a channel
$\mathcal{E}$ with PBT\ is an open problem, and no explicit solutions or
procedures are known.

We employ our convex optimization procedures to find the optimal program
state. This can be done either exactly by minimizing the diamond distance cost
function $C_{\diamond}$ via SDP, or approximately, by determining the optimal
program state via the minimization of the trace distance cost function $C_{1}$
via either SDP or the gradient-based techniques discussed above. For this second
approach, we need to derive the map $\Lambda$ of the PBT processor, between
the program state $\pi$ to output Choi matrix as in Eq.~\eqref{lambdadef}.
%
To compute 
the Choi matrix and CP-map $\Lambda$, we consider an input
maximally-entangled state $|\Phi_{DC}\rangle$ and a basis $|e_{j}^{i}\rangle$
of $\mathbf{A}\{\mathbf{B}\backslash B_{i}\}C$. Then, by using Eq.~\ref{PBTchan}
and the definition $\Lambda(\pi)  =\chi_{\mathcal{P}_{\pi}}=\openone_{D}\otimes\mathcal{P}_{\pi
}[\Phi_{DC}]
$ we find 
the map $\Lambda _{\mathbf{AB}\rightarrow DB_{\text{out}}}$ of a PBT\ processor 
\begin{equation}
\Lambda(\pi)=%
{\textstyle\sum_{ij}}
K_{ij}\pi K_{ij}^{\dagger},~K_{ij}:=\langle{e_{j}^{i}}|\sqrt{\Pi_{i}}%
\otimes\openone_{BD}|\Phi_{DC}\rangle. \label{PBTmap2}%
\end{equation}

Note that a general program state for PBT consists of $2N$ qudits, and hence
the parameter space has exponential size $d^{4N}$. However, because the PBT
protocol is symmetric under permutation of port labels, we show in 
Supplementary Note~\ref{compressSEC} that one can exploit this
symmetry and reduce the number of free parameters to the binomial coeffficient
$\binom{N+d^{4}-1}{d^{4}-1}$, which is polynomial in the number of ports
$N$. Despite this exponential reduction, the scaling in the
number of parameters still represents a practical limiting factor, even for
qubits for which $\mathcal{O}\left(  N^{15}\right)  $. A sub-optimal strategy
consists in reducing the space of program states to a convex set that we call
the \textquotedblleft Choi space\textquotedblright\ $\mathcal{C} $. Consider
an arbitrary probability distribution $\{p_{k}\}$\ and then define
\begin{equation}
\mathcal{C}=\{\pi:\pi=%
{\textstyle\sum_{k}}
p_{k}\rho_{AB}^{k\otimes N},~\mathrm{Tr}_{B}(\rho_{AB}^{k})=d^{-1}\openone\}.
\label{Choispace}%
\end{equation}
One can show (see Supplementary Note~\ref{compressSEC}) that a global minimum in $\mathcal{C}$ is a global
minimum in the extremal (non-convex) subspace for $p_{k}=\delta_{k,1}$
consisting of tensor-products of Choi matrices $\rho_{AB}^{\otimes N}$. Among
these states, there is the $N$-copy Choi matrix of the target channel
$\chi_{\mathcal{E}}^{\otimes N}=[\mathcal{I}\otimes\mathcal{E}(|\Phi
\rangle\left\langle \Phi\right\vert )]^{\otimes N}$ which is not necessarily
the optimal program, as we show below. 


\subsection*{Parametric quantum circuits}

Another deep design of quantum processor is based on
PQCs~\cite{lloyd1995almost,lloyd1996universal}. A PQC is a sequence of unitary
matrices $U(t)=U_{N}(t_{N})\dots U_{2}(t_{2})U_{1}(t_{1})$, where $U_{j}%
(t_{j})=\exp(it_{j}H_{j})$ for some Hamiltonian $H_{j}$ and time interval
$t_{j}$.  The problem with PQCs is that the cost functions in the classical
parameters~\cite{khaneja2005optimal} are not convex, so that numerical
algorithms are not guaranteed to converge to the global optimum. Here we fix
this issue by introducing a convex formulation of PQCs where classical
parameters are replaced by a quantum program. This results in a programmable
PQC processor which is optimizable by our methods.

The universality of PQCs can be employed for universal channel simulation.
Indeed, thanks to Stinespring's dilation theorem, any channel can be written
as a unitary evolution on a bigger space, 
$
\mathcal{E}(\rho_{A})=\mathrm{Tr}_{R_{0}}[U(\rho_{A}\otimes\theta
_{0})U^{\dagger}],
$
where the system is paired to an extra register $R_{0}$ and 
$\theta_{0}$ belongs to $R_{0}$. In the Stinespring representation
$U$ acts on system $A$ and register $R_{0}$. 
In Ref.~\cite{lloyd1995almost} it has been shown that 
sequences of two unitaries, $U_{0}$ and $U_{1}$, are almost
universal for simulation, i.e., any target unitary $U$ can be
approximated as $U\approx\cdots U_{1}^{m_{4}}U_{0}^{m_{3}}U_{1}^{m_{2}}%
U_{0}^{m_{1}}$ for some integers $m_{j}$. Under suitable conditions, it takes
$\mathcal{O}(d^{2}\epsilon^{-d})$ steps to approximate $U$ up to precision
$\epsilon$. The choice between $U_0$ and $U_1$ is done by measuring a 
classical bit. We may introduce a quantum version, where the two different 
unitaries $U_0=e^{i H_0}$ or $U_1=e^{i H_1}$ are chosen depending 
on the state of qubit $R_j$. This results in the conditional gate
\begin{equation}
\hat{U}_{j}=\exp\left(  iH_{0}\otimes|0\rangle_{j}{}_{j}\langle
0|+iH_{1}\otimes|1\rangle_{j}{}_{j}\langle1|\right)  .\label{condAB}%
\end{equation}
Channel simulation is then obtained by replacing the
unitary evolution $U$ in the Stinespring dilation via its simulation.
\begin{figure}[t]
\centering
\includegraphics[width=0.7\linewidth]{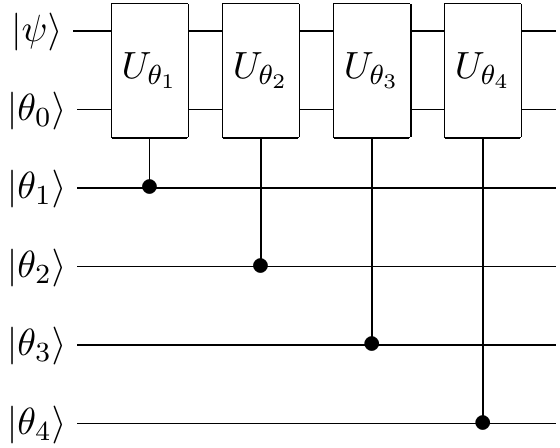} \caption{Simulation of
a quantum channel via Stinespring decomposition together with unitary
simulation as in Fig.~\ref{fig:PQC}. }%
\label{fig:PQC3}%
\end{figure}
The result is illustrated in Fig.~\ref{fig:PQC3},
where the program state $\pi$ is defined over $\mathbf{R}=(R_{0},\dots,R_{N})$
and each $\hat{H}_{j}$ acts on the input system $A$ and two ancillary qubits
$R_{0}$ and $R_{j}$. Following the universality construction of Ref.~\cite{lloyd1995almost}
we show in the Supplementary Note~\ref{s:pqc} 
that the channel shown in Fig.~\ref{fig:PQC3} 
provides a universal processor.  Moreover, 
the channel $\Lambda$ that maps any program $\pi$ to the processor's Choi matrix 
is obtained as
\begin{equation}
\Lambda(\pi)=\mathrm{Tr}_{\mathbf{R}}\left[  \hat{U}_{A\mathbf{R}}\left(
\Phi_{BA}\otimes\pi_{\mathbf{R}}\right)  \hat{U}_{A\mathbf{R}}^{\dagger
}\right]  ,
\label{PQCmap}
\end{equation}
where
$\hat{U}_{A\mathbf{R}}=\openone_{B}\otimes\prod_{j=1}^{N}\hat{U_{j}}%
_{A,R_{0},R_{j}}$,
from which we can identify the optimal
program $|\tilde{\pi}\rangle$ via our methods.

PQCs are not inherently monotonic. A deeper (higher $N$) design may
	simulate a given channel worse than a more shallow design. 
We can design a modified PQC that is monotonic by design,
which we designate a ``monotonic PQC", by replacing the qubits in our program
state with qutrits, and modifying Eq.~\ref{condAB} to read
\begin{equation}
\hat{U}_{j}=\exp\left(  iH_{0}\otimes|0\rangle_{j}{}_{j}\langle
0|+iH_{1}\otimes|1\rangle_{j}{}_{j}\langle1|+\mathbf{0}\otimes|2\rangle_{j}{}_{j}\langle2|\right),\label{condABmod}%
\end{equation}
where $\mathbf{0}$ is a zero operator, 
so that gate $j$ enacts the identity channel if program qutrit $j$ is in the
state $|2\rangle\langle2|$. Then, if it were the case that a PQC with
$N$ program qubits could simulate a given channel better than one with $N+m$, a
monotonic PQC with $N+m$ qutrits in the program state could perform at least as
well as the PQC with $N$ program qubits by setting the first $m$ qutrits to
$|2\rangle\langle2|$. This processor design is both universal and
monotonic. More precisely, let $C(\mathrm{PQC}_N)$ denote the value of a cost function
$C$ for simulating a channel $\mathcal E$ with an $N$-gate PQC, using the optimal
program state, and let $C(\mathrm{mPQC}_N)$ denote the value of $C$ for
simulating $\mathcal E$ with an $N$-gate monotonic PQC, again using the optimal
program state. We are then guaranteed that
\begin{equation}
C(\mathrm{mPQC}_N)\leq \min_{M\leq N}C(\mathrm{PQC}_M).
\end{equation}

\begin{figure}[t]
\vspace{+0.3cm} \centering\includegraphics[width=1.0\linewidth]{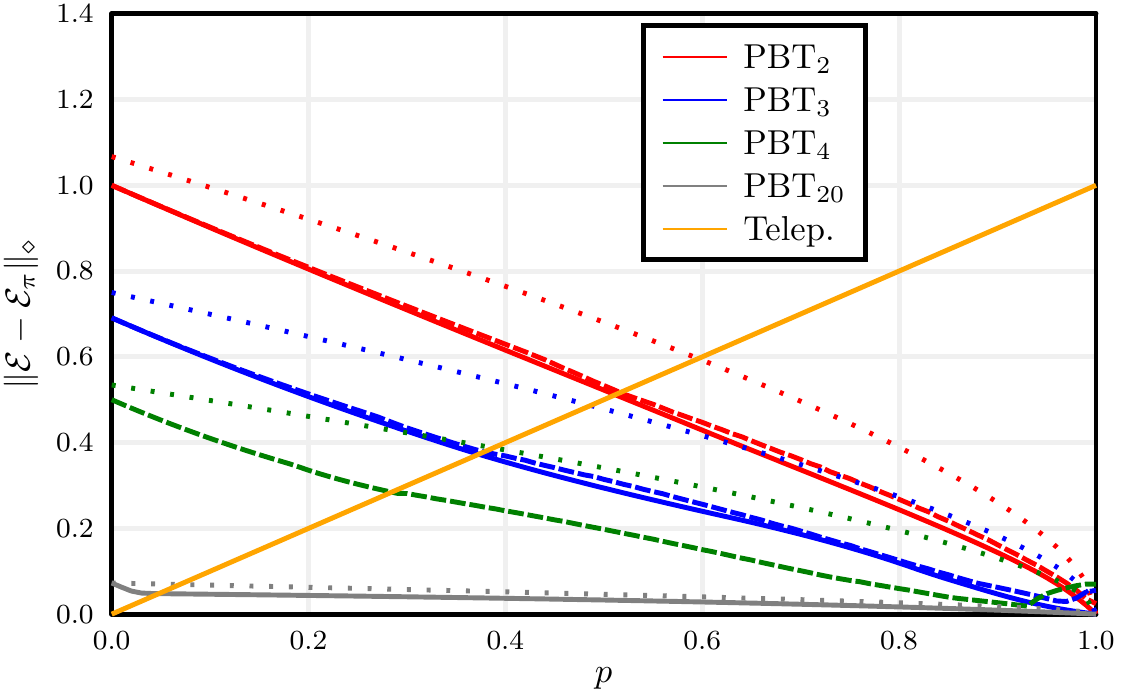}
\vspace{-0.3cm} \caption{
	Diamond-distance error $C_{\diamond}$ in simulating
an amplitude damping channel $\mathcal{E}_{p}$ at various damping rates $p$.
We compare the performance of different designs for the programmable quantum
processor: Standard teleportation and port-based teleportation with $N$ ports
(PBT$_{N}$).
The optimal program $\tilde{\pi}$ is obtained by
either minimizing directly the diamond distance $C_{\diamond}$ (solid lines),
or the trace distance $C_{1}$ (dashed lines) via the projected subgradient
iteration. In both cases, from $\tilde{\pi}$ we then compute $C_{\diamond
}(\tilde{\pi})$. The lowest curves are obtained by optimizing $\pi$ over the
Choi space in Eq.~(\ref{Choispace}). For comparison, we also show the
(non-optimal) performance when the program is the channel's Choi matrix
(dotted lines). 
}%
\label{fig:ad_pbt}%
\end{figure}

\begin{figure}[t]
\vspace{+0.3cm} \centering\includegraphics[width=1.0\linewidth]{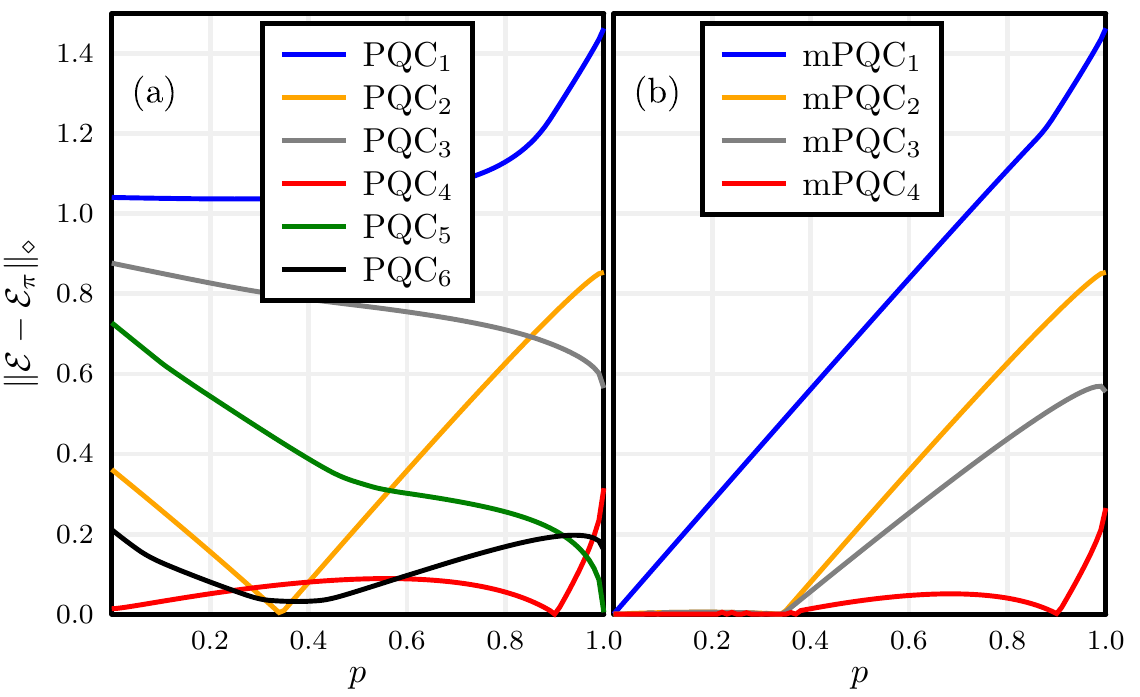}
\vspace{-0.3cm} \caption{Diamond-distance error $C_{\diamond}$ in simulating
an amplitude damping channel $\mathcal{E}_{p}$ at various damping rates $p$.
We compare the performance of two different designs for the programmable quantum
processor: parametric quantum circuits with $N+1$ registers (PQC$_{N}$)
and monotonic parametric quantum circuits with $N+1$ registers (mPQC$_{N}$).
In both cases 
the optimal program $|\tilde{\pi}\rangle$ is obtained
by minimizing the diamond distance $C_{\diamond}$. }%
\label{fig:ad_pqc}%
\end{figure}

\subsection*{Processor benchmarking}
In order to show the performance of the
various architectures, we consider the simulation of an amplitude damping
channel with probability $p$. The reason is because this is the most difficult
channel to simulate, with a perfect simulation only known for infinite
dimension, e.g., using continuous-variable quantum
operations~\cite{pirandola2017fundamental}. In Figs.~\ref{fig:ad_pbt}-\ref{fig:ad_pqc} we compare
teleportation-based, PBT, PQC and ``monotonic PQC'' (mPQC) programmable processors whose program states
have been optimized according to the cost functions $C_{\diamond}$ and $C_{1}%
$. For the PBT processor the trace distance cost $C_{1}$ is remarkably close
to $C_{\diamond}$ and allows us to easily explore high depths. Note that the
optimal program states differ from the naive choice of the Choi matrix of the target
channel. Note too that PQC processors display non-monotonic behaviour when simulating amplitude damping channels, meaning that shallow PQC processors (e.g., for $N=4$) may perform better than deeper processors~\footnote{For the PQC processor we use the universal
Hamiltonians $H_{0}=\sqrt{2}(X\otimes Y-Y\otimes X)$ and $H_{1}=(\sqrt{2}%
Z{+}\sqrt{3}Y{+}\sqrt{5}X)\otimes(Y{+}\sqrt{2}Z)$, where $X$, $Y$, and $Z$ are
Pauli operators.}. Monotonic PQC processors guarantee that deeper designs always perform at least as well as any shallower design.
In Fig.~\ref{fig:ad_pqc} perfect simulation is achievable at specific values of $p$ because of 
our choice of the universal gates $U_0$ and $U_1$. More details are provided in the 
Supplementary Note~\ref{s:appli}. 

Many other numerical simulations are performed in the Supplementary Note~\ref{s:appli}
where we study the convergence rate in learning a unitary operation, 
the exact simulation of Pauli channels, and approximate simulation of both dephasing 
and amplitude damping channels. In particular, we study the performance of the approximate 
solution when optimizing over larger, but easier to compute, cost functions 
such as the trace distance or the infidelity. 

\section*{DISCUSSION}
In this work we have considered a general finite-dimensional model of
a programmable quantum processor, which is a fundamental scheme for quantum
computing and also a primitive tool for other areas of quantum information. By
introducing suitable cost functions, based on the diamond distance, trace
distance and quantum fidelity, we have shown how to characterize the optimal
performance of this processor in the simulation of an arbitrary quantum gate
or channel. In fact, we have shown that the minimization of these cost
functions is a convex optimization problem that can always be solved.

In particular, by minimizing the diamond distance via SDP, we can always
determine the optimal program state for the simulation of an arbitrary
channel. Alternatively, we may minimize the simpler but larger cost functions
in terms of trace distance and quantum fidelity via gradient-based methods 
adapted from ML,
so as to provide a very good approximation of the optimal program state. This
other approach can also provide closed analytical solutions, as is the case
for the simulation of arbitrary unitaries, for which the minimization of the
fidelity cost function corresponds to computing an eigenvector.

We have then applied our results to various designs of programmable quantum
processor, from a shallow teleportation-based scheme to deeper and
asymptotically-universal designs that are based on PBT and PQCs. We have
explicitly benchmarked the performances of these quantum processors by
considering the simulation of unitary gates, depolarizing and amplitude
damping channels, showing that the optimal program states may differ from the
naive choice based on the Choi matrix of the target channel.
Moreover, our results can be applied also for universal quantum measurements \cite{d2005efficient}.


A potential application of our work may be the development
of ``programmable'' model of cloud-based quantum
computation, where a client has an input state to be
processed by an online quantum server which is equipped with a
programmable quantum processor. The client classically
informs the server about what type of computation it
needs (e.g., some specified quantum algorithm) and the
server generates an optimal program state which closely
approximates the overall quantum channel to be applied
to the input. The server then accepts the input from
the client, processes it, and returns the output together
with the value of a cost function quantifying how close
the computation was with respect to the client's request.

Our results may also be useful in areas beyond quantum computing, wherever
channel simulation is a basic problem. For instance, this is the case 
when we investigate the ultimate limits of quantum
communications~\cite{commREVIEW}, design optimal Hamiltonians for one-way
quantum repeaters, and for all those areas of quantum sensing, hypothesis testing
and metrology which are based on quantum channel
simulations~\cite{sensingREVIEW}.
Indeed the study of adaptive protocols of quantum channel
discrimination (or estimation) is notoriously difficult, and their 
optimal performance is not completely understood. Nonetheless, these protocols
can be analyzed by using simulation techniques \cite{pirandola2018fundamental,sensingREVIEW} where 
the channel, encoding the unknown parameter, is replaced by
an approximate simulating channel, and its parameter is mapped into the 
label of a program state (therefore reducing the problem from channel to 
state discrimination/estimation). In this regard, our theory
provides the optimal solution to this basic problem, by determining
the best simulating channel and the corresponding program state.

\section*{METHODS }

\subsection*{Convexity proofs\label{Sec2}}

In this section we provide a proof of Theorem~\ref{t:convexC1CF},
namely we show that the minimization of the main cost functions
$C_{\diamond}$, $C_{1}$ and $C_{F}$ is a convex optimization problem in the
space of the program states $\pi$. This means that we can find the optimal
program state $\tilde{\pi}$ by minimizing $C_{\diamond}$ or, alternatively,
sub-optimal program states can be found by minimizing either $C_{1}$ or
$C_{F}$. For the sake of generality, we prove the result for all of the cost
functions discussed in the previous sections. 
We restate Theorem~\ref{t:convexC1CF} below for completeness:

\begin{theorem*}
The minimization of the generic cost function
$C=C_{\diamond}$, $C_{1}$, $C_{F}$, $C_{R}$ or $C_{p}$ for any $p>1$ is a
convex optimization problem in the space of program states. 
\end{theorem*}

\noindent\textit{Proof.}~~Let us start to show the result for the diamond
distance $C_{\diamond}$. In this case, we can write the following
\begin{align}
&  C_{\diamond}[p\pi+(1-p)\pi^{\prime}]\nonumber\\
&  :=\left\Vert \mathcal{E}-\mathcal{E}_{p\pi+(1-p)\pi^{\prime}}\right\Vert
_{\diamond}\nonumber\\
&  \overset{(1)}{=}\left\Vert (p{+}1{-}p)\mathcal{E}-p\mathcal{E}_{\pi}%
-(1{-}p)\mathcal{E}_{\pi^{\prime}}\right\Vert _{\diamond}\nonumber\\
&  \overset{(2)}{\leq}\left\Vert p\mathcal{E}-p\mathcal{E}_{\pi}\right\Vert
_{\diamond}+\left\Vert (1{-}p)\mathcal{E}-(1{-}p)\mathcal{E}_{\pi^{\prime}%
}\right\Vert _{\diamond}\nonumber\\
&  \overset{(3)}{\leq}p\left\Vert \mathcal{E}-\mathcal{E}_{\pi}\right\Vert
_{\diamond}+(1-p)\left\Vert \mathcal{E}-\mathcal{E}_{\pi^{\prime}}\right\Vert
_{\diamond}\nonumber\\
&  =pC_{\diamond}(\pi)+(1-p)C_{\diamond}(\pi^{\prime}),
\end{align}
where we use $(1)$~the linearity of $\mathcal{E}$, $(2)$~ the triangle
inequality and $(3)$ the property $\Vert xA\Vert_{1}=|x|\Vert A\Vert_{1}$,
valid for any operator $A$ and coefficient $x$.

For any Schatten $p$-norm $C_{p}$ with $p\geq1$, we may prove convexity following
a similar reasoning. Since for any combination
$\bar{\pi}:=p_0\pi_0+p_1\pi_1$, with $p_0+p_1=1$, we have
$\Lambda(\bar{\pi})=p_0\Lambda(\pi_0)+p_1\Lambda(\pi_1)$, then 
by exploiting the triangle inequality, and the property $ \|x A\|_p = |x|\|A\|_p$,
we can show that
\begin{align}
	C_p(p_0\pi_0+p_1\pi_1) 
	&:= \|\chi_{\mathcal E}-\Lambda(p_0\pi_0+p_1\pi_1)\|_p
	\\\nonumber &\le  p_0\|\chi_{\mathcal E}-\Lambda(\pi_0)\|_p+p_1\|\chi_{\mathcal E}-\Lambda(\pi_1)\|_p
	\\\nonumber &= p_0 C_p(\pi_0) +  p_1C_p(\pi_1)~.
\end{align}
To show the convexity of $C_{F}$, defined in Eq.~\eqref{Cf}, we note that the
fidelity function $F(\rho,\sigma)$ satisfies the following concavity relation
\cite{uhlmann1976transition}
\begin{equation}
F\left(  \sum_{k}p_{k}\rho_{k},\sigma\right)  ^{2}\geq\sum_{k}p_{k}F(\rho
_{k},\sigma)^{2}~.
\end{equation}
Due to the linearity of $\chi_{\pi}=\Lambda(\pi)$, the fidelity in
Eq.~\eqref{fidCC} satisfies $F_{\bar{\pi}}^{2}\geq\sum_{k}p_{k}F_{\pi_{k}}%
^{2}$ for $\bar{\pi}:=\sum_{k}p_{k}\pi_{k}$. Accordingly, we get the following
convexity result
\begin{equation}
C_{F}\left(  \sum_{k}p_{k}\pi_{k}\right)  \leq\sum_{k}p_{k}C_{F}(\pi_{k})~.
\end{equation}
For the cost function $C_{R}$, the result comes from the linearity of
$\Lambda(\pi)$ and the joint convexity of the relative entropy. In fact, for
$\bar{\pi}:=p_{0}\pi_{0}+p_{1}\pi_{1}$, we may write
\begin{align}
S[\Lambda(\bar{\pi})||\chi_{\mathcal{E}}] &  =S[p_{0}\Lambda(\pi_{0}%
)+p_{1}\Lambda(\pi_{1})||\chi_{\mathcal{E}}]\nonumber\\
&  =S[p_{0}\Lambda(\pi_{0})+p_{1}\Lambda(\pi_{1})||p_{0}\chi_{\mathcal{E}%
}+p_{1}\chi_{\mathcal{E}}]\nonumber\\
&  \leq p_{0}S[\Lambda(\pi_{0}),\chi_{\mathcal{E}}]+p_{1}S[\Lambda(\pi
_{1}),\chi_{\mathcal{E}}],
\end{align}
with a symmetric proof for $S[\chi_{\mathcal{E}}||\Lambda(\bar{\pi})]$. This
implies the convexity of $C_{R}(\pi)$ in Eq.~(\ref{REcost}).~$\blacksquare$

\subsection*{Convex classical parametrizations}

The result of the theorem~\ref{t:convexC1CF} can certainly be extended to any convex
parametrization of program states. For instance, assume that $\pi
=\pi(\boldsymbol{\lambda})$, where $\boldsymbol{\lambda}=\{\lambda_{i}\}$ is a
probability distribution. This means that, for $0\leq p\leq1$ and any two
parametrizations, $\boldsymbol{\lambda}$ and $\boldsymbol{\lambda}^{\prime}$,
we may write%
\begin{equation}
\pi\lbrack p\boldsymbol{\lambda}+(1-p)\boldsymbol{\lambda}^{\prime}%
]=p\pi(\boldsymbol{\lambda})+(1-p)\pi(\boldsymbol{\lambda}^{\prime}).
\label{eqGIA}%
\end{equation}
Then the problem remains convex in $\boldsymbol{\lambda}$ and we may therefore
find the global minimum in these parameters. It is clear that this global
minimum $\boldsymbol{\tilde{\lambda}}$\ identifies a program state
$\pi(\boldsymbol{\tilde{\lambda}})$ which is not generally the optimal state
$\tilde{\pi}$ in the entire program space $\mathcal{S}$, even though the
solution may be a convenient solution for experimental applications.

Note that a possible classical parametrization consists of using classical
program states, of the form
\begin{equation}
\pi(\boldsymbol{\lambda})=\sum_{i}\lambda_{i}\left\vert \varphi_{i}%
\right\rangle \left\langle \varphi_{i}\right\vert ,
\end{equation}
where $\{\left\vert \varphi_{i}\right\rangle \}$ is an orthonormal basis in
the program space. Convex combinations of probability distributions therefore
define a convex set of classical program states
\begin{equation}
\mathcal{S}_{\text{class}}=\{\pi:\pi=\sum_{i}\lambda_{i}\left\vert \varphi
_{i}\right\rangle \left\langle \varphi_{i}\right\vert ,~\left\langle
\varphi_{i}\right\vert \left.  \varphi_{j}\right\rangle =\delta_{ij}\}.
\end{equation}
Optimizing over this specific subspace corresponds to optimizing the
programmable quantum processor over classical programs. It is clear that
global minima in $\mathcal{S}_{\text{class}}$ and $\mathcal{S}$ are expected
to be very different. For instance, $\mathcal{S}_{\text{class}}$ cannot
certainly include Choi matrices which are usually very good quantum programs.


\subsection*{Gradient-based optimization }\label{s:grad}
As discussed in the main text, 
the SDP formulation allows 
the use of powerful and accurate numerical methods, such as the 
interior point method. However, these algorithms are not suitable 
for high dimensional problems, due to their higher computational 
and memory requirements. Therefore, an 
alternative approach (useful for larger program states) consists of the
optimization of the larger but easier-to-compute cost function $C=C_{1}$
(trace distance) or $C_{F}$ (infidelity), for which we can use first 
order methods. Indeed, according to
Theorem~\ref{t:convexC1CF}, all of the proposed cost functions $C:\mathcal{S}\rightarrow
\mathbb{R}$ are convex over the program space $\mathcal{S}$\ and, therefore, we
can solve the optimization $\min_{\pi\in\mathcal{S}}C(\pi)$ by using
gradient-based algorithms. 

Gradient-based convex optimization is at the heart of many popular ML
techniques such as online learning in a high-dimensional feature space
\cite{duchi2008efficient}, missing value estimation problems
\cite{liu2013tensor}, text classification, image ranking, and optical
character recognition \cite{duchi2011adaptive}, to name a few. In all of the
above applications, \textquotedblleft learning\textquotedblright\ corresponds
to the following minimization problem: $\min_{x\in\mathcal{S}}f(x)$, where
$f(x)$ is a convex function and $\mathcal{S}$ is a convex set. Quantum
learning falls into this category, as the space of program states is convex
due to the linearity of quantum mechanics and the fact that cost functions are typically
convex in this space (see Theorem~\ref{t:convexC1CF}). Gradient-based
approaches are among the most applied methods for convex optimization of
non-linear, possibly non-smooth functions~\cite{nesterov2013introductory}.

When the cost function is not 
differentiable we cannot formally define its gradient. Nonetheless,
we can always define the subgradient $\partial C$ of $C$ as 
in Eq.~\eqref{subgradient}, which in principle contains many points.
When $C$ is not only convex but also differentiable, then
$\partial C(\pi)=\{\nabla C(\pi)\}$, 
i.e. the subgradient contains a single element, the gradient $\nabla C$,
that can be obtained via the Fr\'{e}chet derivative of $C$ (for more details
see Supplementary Note~\ref{a:calculus}). When $C$ is not differentiable,
the gradient still provides an element of the subgradient that can be used in
the minimization algorithm.

In order to compute the gradient $\nabla C$, it is convenient to consider the
Kraus decomposition of the processor map $\Lambda$. Let us write%
\begin{equation}
\Lambda(\pi)=\sum_{k}A_{k}\pi A_{k}^{\dagger},
\end{equation}
with Kraus operators $A_{k}$. We then define the dual map $\Lambda^{\ast}$ of
the processor as the one (generally non-trace-preserving) which is given by
the following decomposition%
\begin{equation}
\Lambda^{\ast}(\rho)=\sum_{k}A_{k}^{\dagger}\rho A_{k}.
\end{equation}
With these definitions in hands, we can now prove Theorem~\ref{t:gradients},
which we rewrite here for convenience. 

\begin{theorem*}
Suppose we use a quantum processor $Q$ with map
$\Lambda(\pi)=\chi_{\pi}$\ in order to approximate the Choi matrix
$\chi_{\mathcal{E}}$ of an arbitrary channel $\mathcal{E}$. Then, the
gradients of the trace distance $C_{1}(\pi)$ and the infidelity $C_{F}(\pi
)$\ are given by the following analytical formulas%
\begin{align}
\nabla C_{1}(\pi) &  =\sum_{k}\mathrm{sign}(\lambda_{k})\Lambda^{\ast}%
(P_{k}),\label{traceDproof}\\
\nabla C_{F}(\pi) &  =-2\sqrt{1-C_{F}(\pi)}\nabla F(\pi),\label{Cfgrad}\\
\nabla F(\pi) &  =\frac{1}{2}\Lambda^{\ast}\left[  \sqrt{\chi_{\mathcal{E}}%
}\left(  \sqrt{\chi_{\mathcal{E}}}\,\Lambda(\pi)\,\sqrt{\chi_{\mathcal{E}}%
}\right)  ^{-\frac{1}{2}}\sqrt{\chi_{\mathcal{E}}}\right]
,\label{fidelitygrad}%
\end{align}
where $\lambda_{k}$ ($P_{k}$) are the eigenvalues (eigenprojectors) of the
Hermitian operator $\chi_{\pi}-\chi_{\mathcal{E}}$. When $C_{1}(\pi)$ or
$C_{F}(\pi)$ are not differentiable at $\pi$, then the above expressions
provide an element of the subgradient $\partial C(\pi)$.
\end{theorem*}

\noindent\textbf{Proof}.~~We prove the above theorem assuming that the
functions are differentiable for program $\pi$. For non-differentiable points,
the only difference is that the above analytical expressions are not unique
and provide only one of the possibly infinite elements of the subgradient.
Further details of this mathematical proof are given in
Supplementary Note~\ref{a:calculus}. Following matrix differentiation,
for any function $f(A)=\Tr[g(A)]$ of a matrix $A$, we may write
\begin{equation}
d\text{\textrm{Tr}}[g(A)]=\text{\textrm{Tr}}[g^{\prime}(A)dA],\label{nablag}%
\end{equation}
and the gradient is $\nabla f(A)=g^{\prime}(A)$. Both the trace-distance and
fidelity cost functions can be written in this form. To find the explicit
gradient of the fidelity function, we first note that, by linearity, we may
write%
\begin{equation}
\Lambda({\pi+\delta\pi})=\Lambda({\pi})+\Lambda(\delta\pi)~,\label{chilinear}%
\end{equation}
and therefore the following expansion
\begin{gather}
\sqrt{\chi_{\mathcal{E}}}\Lambda({\pi+\delta\pi})\sqrt{\chi_{\mathcal{E}}%
}=\nonumber\\
\sqrt{\chi_{\mathcal{E}}}\Lambda({\pi})\sqrt{\chi_{\mathcal{E}}}+\sqrt
{\chi_{\mathcal{E}}}\Lambda({\delta\pi})\sqrt{\chi_{\mathcal{E}}}~.
\end{gather}
From this equation and differential calculations of the fidelity (see
Supplementary Note~\ref{fidelityDIFF} for details), we find
\begin{equation}
dF=\frac{1}{2}\text{\textrm{Tr}}\left[  (\sqrt{\chi_{\mathcal{E}}}\Lambda
({\pi})\sqrt{\chi_{\mathcal{E}}})^{-\frac{1}{2}}\sqrt{\chi_{\mathcal{E}}%
}\Lambda({\delta\pi})\sqrt{\chi_{\mathcal{E}}}\right]  ~,
\end{equation}
where $dF=F(\pi+\delta\pi)-F(\pi)$. Then, using the cyclic property of the
trace, we get
\begin{equation}
dF=\frac{1}{2}\text{\textrm{Tr}}\left[  \Lambda^{\ast}\left[  \sqrt
{\chi_{\mathcal{E}}}(\sqrt{\chi_{\mathcal{E}}}\Lambda(\pi)\sqrt{\chi
_{\mathcal{E}}})^{-\frac{1}{2}}\sqrt{\chi_{\mathcal{E}}}\right]  \delta
\pi\right]  .
\end{equation}
Exploiting this expression in Eq.~\eqref{nablag} we get the gradient $\nabla
F(\pi)$ as in Eq.~(\ref{fidelitygrad}). The other Eq.~\eqref{Cfgrad} simply
follows from applying the definition in Eq.~\eqref{Cf}.

For the trace distance, let us write the eigenvalue decomposition
\begin{equation}
\chi_{\pi}-\chi_{\mathcal{E}}=\sum_{k}\lambda_{k}P_{k}~.\label{tracediag}%
\end{equation}
Then using the linearity of Eq.~\eqref{chilinear}, the definition of a processor map
of Eq.~\eqref{lambdadef} and differential calculations of the trace distance
(see Supplementary Note~\ref{traceDIFF} for details), we can write
\begin{align}
dC_{1}(\pi) &  =\sum_{k}\mathrm{sign}(\lambda_{k})\text{\textrm{Tr}}%
[P_{k}\Lambda(d\pi)]\nonumber\\
&  =\sum_{k}\mathrm{sign}(\lambda_{k})\text{\textrm{Tr}}[\Lambda^{\ast}%
(P_{k})d\pi]\nonumber\\
&  =\text{\textrm{Tr}}\left\{  \Lambda^{\ast}[\mathrm{sign}(\chi_{\pi}%
-\chi_{\mathcal{E}})]d\pi\right\}  ~.
\end{align}
From the definition of the gradient in Eq.~\eqref{nablag}, we finally get%
\begin{equation}
\nabla C_{1}(\pi)=\Lambda^{\ast}[\mathrm{sign}(\chi_{\pi}-\chi_{\mathcal{E}%
})],
\end{equation}
which leads to the result in Eq.~(\ref{traceDproof}).~$\blacksquare$

The above results in Eqs.~(\ref{Cfgrad}) and~(\ref{traceDproof}) can be used
together with the projected subgradient method~\cite{boyd2003subgradient} or
conjugate gradient algorithm~\cite{jaggi2011convex,jaggi2013revisiting} to
iteratively find the optimal program state in the minimization of $\min
_{\pi\in\mathcal{S}}C(\pi)$ for $C=C_{1}$ or $C_{F}$. In the following sections
we present two algorithms, the projected subgradient method and the
conjugate gradient method, and show how they can be adapted to our problem.


Projected subgradient methods have the advantage of simplicity and the ability
to optimize non-smooth functions, but can be slower, with a convergence rate
$\mathcal{O}\left(  \epsilon^{-2}\right)  $ for a desired accuracy $\epsilon$.
Conjugate gradient methods~\cite{jaggi2011convex,jaggi2013revisiting} have a
faster convergence rate $\mathcal{O}\left(  \epsilon^{-1}\right)  $, provided
that the cost function is smooth. This convergence rate can be improved even
further to $\mathcal{O}\left(  \epsilon^{-1/2}\right)  $ for strongly convex
functions~\cite{garber2015faster} or using Nesterov's accelerated gradient
method~\cite{nesterov2005smooth}. The technical difficulty in the adaptation
of these methods for learning program states comes because the latter is a
constrained optimization problem, namely at each iteration step the optimal
program must be a proper quantum state, and the cost functions coming from
quantum information theory are, generally, non-smooth.

\subsection*{Projected subgradient method}\label{s:projsg}

Given the space $\mathcal{S}$ of program states, let us define the projection
$\mathcal{P}_{\mathcal{S}}$ onto $\mathcal{S}$ as%
\begin{equation}
    \mathcal{P}_{\mathcal{S}}(X)=\argmin_{\pi\in S}\Vert X-\pi\Vert
_{2}~,\label{proj}%
\end{equation}
where argmin is the argument of the minimum, namely the closest state $\pi\in\mathcal S$
to the operator $X$.
Then, a first order algorithm to solve $\min_{\pi\in\mathcal{S}}C(\pi)$ is to
apply the projected subgradient
method~\cite{nesterov2013introductory,boyd2003subgradient}, which iteratively
applies the iteration \eqref{projsubgrad}, which we rewrite below for convenience 
\begin{equation}%
\begin{array}
[c]{l}%
1)~\mathrm{Select~an~operator~}g_{i}{~\mathrm{from~}}\partial C(\pi_{i}),\\
2)~\text{\textrm{Update}~}\pi_{i+1}=\mathcal{P}_{\mathcal{S}}\left(  \pi
_{i}-\alpha_{i}g_{i}\right)  ,
\end{array}
\end{equation}
where $i$ is the iteration index and $\alpha_{i}$ a learning rate.

The above algorithm differs from standard gradient methods in two aspects: i)
the update rule is based on the subgradient, which is defined even for
non-smooth functions; ii) the operator $\pi_{i}-\alpha_{i}g_{i}$ is generally
not a quantum state, so the algorithm fixes this issue by projecting that
operator back to the closest quantum state, via Eq.~\eqref{proj}. The
algorithm converges to the optimal solution $\pi_{\ast}$ (approximating the
optimal program $\tilde{\pi}$) as~\cite{boyd2003subgradient}
\begin{equation}
C(\pi_{i})-C(\pi_{\ast})\leq\frac{e_{1}+G\sum_{k=1}^{i}\alpha_{k}^{2}}%
{2\sum_{k=1}^{i}\alpha_{k}}=:\epsilon,
\end{equation}
where $e_{1}=\Vert\pi_{1}-\pi_{\ast}\Vert_{2}^{2}$ is the initial error (in
Frobenius norm) and $G$ is such that $\Vert g\Vert_{2}^{2}\leq G$ for any
$g\in\partial C$. Popular choices for the learning rate that assure
convergence are $\alpha_{k}\propto1/\sqrt{k}$ and $\alpha_{k}=a/(b+k)$ for
some $a,b>0$.

In general, the projection step is the major drawback, which often limits the applicability
of the projected subgradient method to practical
problems. Indeed, projections like Eq.~\eqref{proj} require another full
optimization at each iteration that might be computationally intensive.
Nonetheless, we show in the following theorem that this issue does not occur
in learning quantum states, because the resulting optimization can be solved analytically.

\begin{theorem}
\label{t:proj} Let $X$ be a Hermitian operator in a $d$-dimensional Hilbert
space with spectral decomposition $X=UxU^{\dagger}$, where the eigenvalues
$x_{j}$ are ordered in decreasing order. Then $\mathcal{P}_{\mathcal{S}}(X)$
of Eq.~\eqref{proj} is given by
\begin{equation}
\mathcal{P}_{\mathcal{S}}(X)=U\lambda U^{\dagger},~~\lambda_{i}=\max
\{x_{i}-\theta,0\},
\end{equation}
where $\theta=\frac{1}{s}\sum_{j=1}^{s}\left(  x_{j}-1\right)  $ and
\begin{equation}
s=\max\left\{  k\in\lbrack1,...,d]:x_{k}>\frac{1}{k}\sum_{j=1}^{k}\left(
x_{j}-1\right)  \right\}  .
\end{equation}

\end{theorem}

\noindent\textbf{Proof}.~~Any quantum (program) state can be written in the
diagonal form $\pi=V\lambda V^{\dagger}$ where $V$ is a unitary matrix, and
$\lambda$ is the vector of eigenvalues in decreasing order, with $\lambda
_{j}\geq0$ and $\sum_{j}\lambda_{j}=1$. To find the optimal state, it is
required to find both the optimal unitary $V$ and the optimal eigenvalues
$\lambda$ with the above property, i.e.,
\begin{equation}
\mathcal{P}_{\mathcal{S}}(X)=\argmin_{V,\lambda}\Vert X-V\lambda V^{\dagger
}\Vert_2~.\label{proj2}%
\end{equation}
For any unitarily-invariant norm, the following inequality holds~\cite[Eq.
IV.64]{bhatia2013matrix}
\begin{equation}
\Vert X-\pi\Vert_2\geq\Vert x-\lambda\Vert_2~,
\end{equation}
with equality when $U=V$, where $X=UxU^{\dagger}$ is a spectral decomposition
of $X$ such that the $x_{j}$'s are in decreasing order. This shows that the
optimal unitary in Eq.~\eqref{proj2} is the diagonalization matrix of the
operator $X$. The eigenvalues of any density operator form a probability
simplex. The optimal eigenvalues $\lambda$ are then obtained thanks to
Algorithm 1 from Ref.~\cite{duchi2008efficient}.~$\blacksquare$

In the following section we present an alternative algorithm with faster
convergence rates, but stronger requirements on the function to be optimized.

\subsection*{Conjugate gradient method}

The conjugate gradient method~\cite{jaggi2011convex,nesterov2013introductory},
sometimes called the Frank-Wolfe algorithm, has been developed to provide a better
convergence speed and to avoid the projection step at each iteration. Although
the latter can be explicitly computed for quantum states (thanks to our
Theorem~\ref{t:proj}), having a faster convergence rate is important,
especially with  higher dimensional Hilbert spaces. The downside of this method
is that it necessarily requires a differentiable cost function $C$, with
gradient $\nabla C$.

In its standard form, the conjugate gradient method to approximate the
solution of $\argmin_{\pi\in\mathcal{S}}C(\pi)$ is defined by the following
iterative rule%
\begin{equation}%
\begin{array}
[c]{l}%
1)~\mathrm{Find~}\argmin_{\sigma\in\mathcal{S}}\text{\textrm{Tr}}[\sigma\nabla
C(\pi_{i})],\\
2)~\pi_{i+1}=\pi_{i}+\frac{2}{i+2}(\sigma-\pi_{i})=\frac{i}{i+2}\pi_{i}%
+\frac{2}{i+2}\sigma.
\end{array}
\label{FrankWolfe0}%
\end{equation}
The first step in the above iteration rule is solved by finding the smallest
eigenvector $|\sigma\rangle$ of $\nabla C(\pi_{i})$. Indeed, since $\pi$ is an
operator and $C(\pi)$ a scalar, the gradient $\nabla C$ is an operator with
the same dimension as $\pi$. Therefore, for learning quantum programs
we find the iteration \eqref{FrankWolfe}, that we rewrite below for 
convenience 
\begin{equation}%
\begin{array}
[c]{l}%
1)~\text{\textrm{Find the smallest eigenvalue}}~\left\vert \sigma
_{i}\right\rangle ~\text{\textrm{of}}~\nabla C(\pi_{i}),\\
2)~\pi_{i+1}=\frac{i}{i+2}\pi_{i}+\frac{2}{i+2}\left\vert \sigma
_{i}\right\rangle \left\langle \sigma_{i}\right\vert .
\end{array}
\end{equation}
When the gradient of $C$ is Lipschitz continuous with constant $L$, the
conjugate gradient method converges after $\mathcal{O}(L/\epsilon)$
steps~\cite{jaggi2013revisiting,nesterov2005smooth}. The
following iteration with adaptive learning rate $\alpha_{i}$ has even faster
convergence rates, provided that $C$ is strongly
convex~\cite{garber2015faster}:
\begin{equation}%
\begin{array}
[c]{l}%
1)~\text{\textrm{Find the smallest eigenvalue }}\left\vert \sigma
_{i}\right\rangle \text{\textrm{ of~}}\nabla C(\pi_{i}),\\
2)~\text{\textrm{Find }}\alpha_{i}=\argmin_{\alpha\in\lbrack0,1]}\alpha
\langle\tau_{i},\nabla C(\pi_{i})\rangle\\
~\text{\textrm{~~}}+\alpha^{2}\frac{\beta_C}{2}\Vert\tau_{i}\Vert^{2}_C%
,~\text{\textrm{for }}\tau_{i}=|\sigma_{i}\rangle\langle\sigma_{i}|-\pi_{i},\\
3)~\pi_{i+1}=(1-\alpha_{i})\pi_{i}+\alpha_{i}\left\vert \sigma_{i}%
\right\rangle \left\langle \sigma_{i}\right\vert .
\end{array}
\label{FFrankWolfe}%
\end{equation}
where the constant $\beta_C$ and norm $\|\cdot\|_C$ depend on $C$ \cite{garber2015faster}.

In spite of the faster convergence rate, conjugate gradient methods require
smooth cost functions (so that the gradient $\nabla C$ is well defined at
every point). However, cost functions based on trace distance \eqref{traceD}
are not smooth. For instance, the trace distance in one-dimensional spaces reduces
to the absolute value function $|x|$ that is non-analytic at $x=0$.
When some eigenvalues are close to zero, conjugate gradient methods may display
unexpected behaviors, though we have numerically observed that
convergence is always obtained with a careful choice of the learning rate.
In the next section we show how to formally justify the applicability
of the conjugate gradient method, following
Nesterov's smoothing prescription~\cite{nesterov2005smooth}.

\subsection*{Smoothing: smooth trace distance}

The conjugate gradient method converges to the global optimum after
$\mathcal{O}\left(  \frac{L}{\epsilon}\right)  $ steps, provided that the
gradient of $C$ is $L$-Lipschitz continuous~\cite{nesterov2005smooth}.
However, the constant $L$ can diverge for non-smooth functions like the trace
distance \eqref{traceD} so the convergence of the algorithm cannot be formally
stated, although it may still be observed in numerical simulations. 
 To solidify the convergence proof (see also Supplementary Note~\ref{a:smoothtrace}%
), we introduce a smooth approximation to the trace distance. This is defined
by the following cost function that is differentiable at every point
\begin{equation}
C_{\mu}(\pi)=\Tr\left[  h_{\mu}\left(  \chi_{\pi}-\chi_{\mathcal{E}}\right)
\right]  =\sum_{j}h_{\mu}(\lambda_{j})~,\label{Dmu}%
\end{equation}
where $\lambda_{j}$ are the eigenvalues of $\chi_{\pi}-\chi_{\mathcal{E}}$ and
$h_{\mu}$ is the so-called Huber penalty function
\begin{equation}
h_{\mu}(x):=%
\begin{cases}
\frac{x^{2}}{2\mu} & \mathrm{~if~}|x|<\mu~,\\
|x|-\frac{\mu}{2} & \mathrm{~if~}|x|\leq\mu~.
\end{cases}
\label{HuberPenalty}%
\end{equation}
The previous definition of the trace distance, $C_{1}$ in Eq.~\eqref{traceD},
is recovered for $\mu\rightarrow0$ and, for any non-zero $\mu$, the $C_{\mu}$
bounds $C_{1}$ as follows
\begin{equation}
C_{\mu}(\pi)\leq C_{1}(\pi)\leq C_{\mu}(\pi)+\frac{\mu d}{2},\label{Dmuineq}%
\end{equation}
where $d$ is the dimension of the program state $\pi$. In
Supplementary Note~\ref{a:smoothtrace} we then prove the following result

\begin{theorem}
The smooth cost function $C_{\mu}(\pi)$ is a convex function over program
states and its gradient is given by
\begin{equation}
\nabla C_{\mu}(\pi)=\Lambda^{\ast}[h_{\mu}^{\prime}(\chi_{\pi}-\chi
_{\mathcal{E}})],
\end{equation}
where $h_{\mu}^{\prime}$ is the derivative of $h_{\mu}$. Moreover, the
gradient is $L$-Lipschitz continuous with
\begin{equation}
L=\frac{d}{\mu}~,\label{LipschitzConst}%
\end{equation}
where $d$ is the dimension of the program state.
\end{theorem}

Being Lipschitz continuous, the conjugate gradient algorithm and its
variants~\cite{nesterov2005smooth,garber2015faster} converge up to an accuracy
$\epsilon$ after $\mathcal{O}(L/\epsilon)$ steps. In some applications, it is
desirable to analyze the convergence in trace distance in the limit of large
program states, namely for $d\rightarrow\infty$. The parameter $\mu$ can be
chosen such that the smooth trace distance converges to the trace distance,
namely $C_{\mu}\rightarrow C_{1}$ for $d\rightarrow\infty$. Indeed, given the
inequality \eqref{Dmuineq}, a possibility is to set $\mu=\mathcal{O}%
(d^{-(1+\eta)})$ for some $\eta>0$ so that, from Eq.~\eqref{LipschitzConst},
the convergence to the trace norm is achieved after $\mathcal{O}(d^{2+\eta})$ steps.

\bigskip

\bigskip

\noindent\textbf{Acknowledgements.~} L.B. acknowledges support by
the program ``Rita Levi Montalcini'' for young researchers.
S.P. and J.P. acknowledge support by the EPSRC via the `UK Quantum
Communications Hub' 
(Grants EP/M013472/1 and EP/T001011/1) 
and S.P. acknowledges support by the European Union via the
project `Continuous Variable Quantum Communications' (CiViQ, no
820466). 

%

\clearpage

\begin{widetext}
{\Huge \bf Supplementary Materials}
\vspace{1cm}
\end{widetext}

\setcounter{equation}{0}

\renewcommand\theequation{S\arabic{equation}} 
\renewcommand{\thesection}{\arabic{section}}

\renewcommand{\thesubsection}{\thesection.\arabic{subsection}}

\section{More on programmable simulation\label{Sec1}}

\begin{figure}[b]
\vspace{-0.5cm}
\par
\begin{center}
\includegraphics[width=0.35\textwidth]{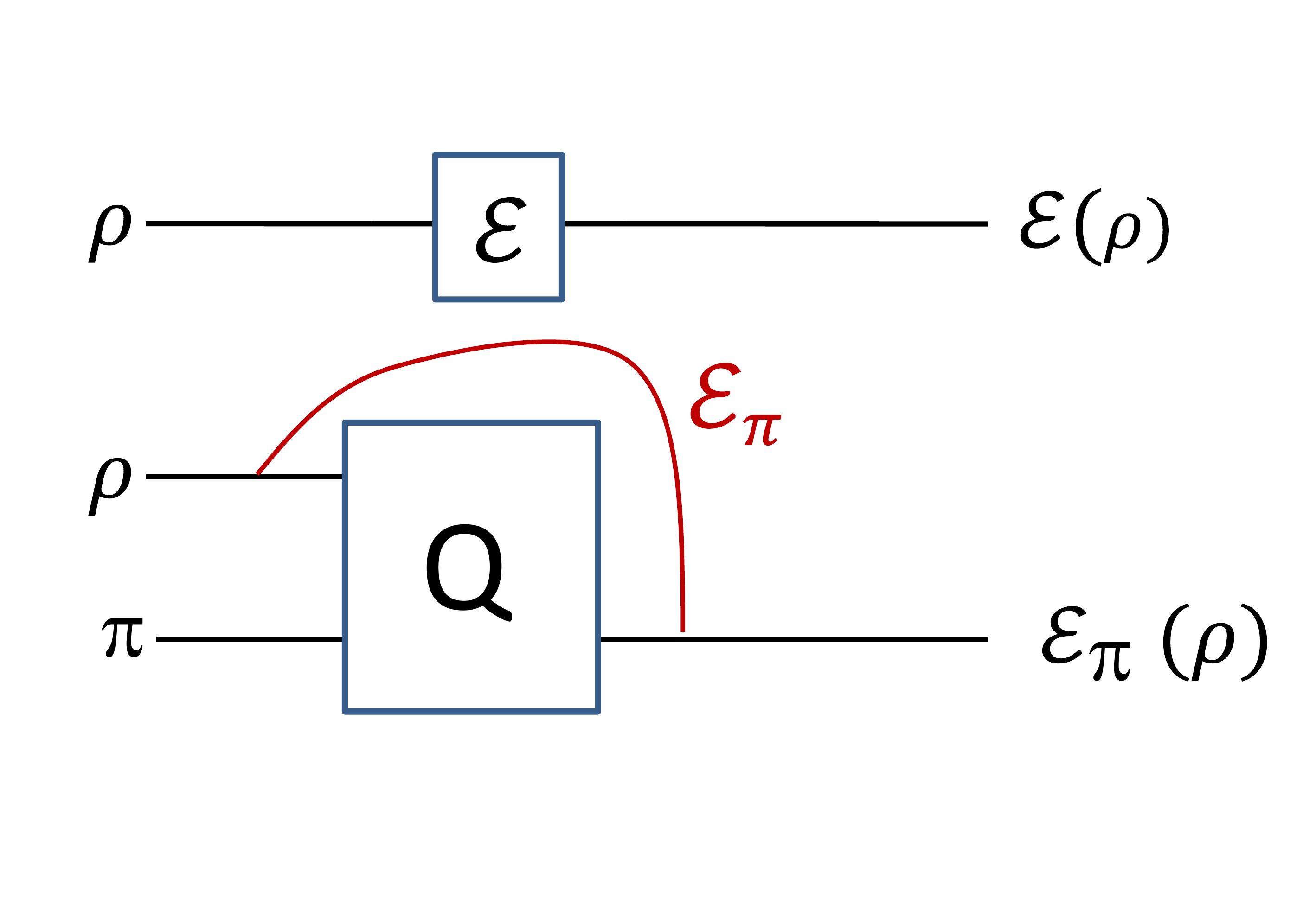}
\end{center}
\par
\vspace{-0.9cm}\caption{Arbitrary quantum channel $\mathcal{E}$ and its
simulation $\mathcal{E}_{\pi}$ via a quantum processor $Q$ applied to a
program state $\pi$.}%
\label{QMLprogram0}%
\end{figure}

As discussed in the main text, 
the task we are interested in is the 
simulation of a channel $\mathcal{E}$ using a programmable quantum
processor~\cite{nielsen1997programmable} that we simply call a \textquotedblleft
quantum processor\textquotedblright\ (see Fig.~\ref{QMLprogram0}). 
This is
represented by a completely positive trace-preserving (CPTP) universal map $Q$
as in Eq.~\eqref{channelDEF}.
Our goal is to find the program state $\pi$ according to \eqref{sol1}, 
namely the state for which the simulation
$\mathcal{E}_{\pi}$ is the closest to $\mathcal{E}$. 
The most appropriate definition of ``closeness'' between two quantum 
channels is via the diamond norm 
$
C_{\diamond}(\pi):=\left\Vert \mathcal{E}-\mathcal{E}_{\pi}\right\Vert
_{\diamond}\leq2
$. 
Recall that the diamond distance is defined by the following maximization%
\begin{equation}
\left\Vert \mathcal{E}-\mathcal{E}_{\pi}\right\Vert _{\diamond}=\max_{\varphi
}\left\Vert \mathcal{I}\otimes\mathcal{E}(\varphi)-\mathcal{I}\otimes
\mathcal{E}_{\pi}(\varphi)\right\Vert _{1}, \label{diamondDEF}%
\end{equation}
where $\left\Vert O\right\Vert _{1}:=\mathrm{Tr}\sqrt{O^{\dagger}O}$ is the
trace norm~\cite{watrous2018theory}. Because the trace norm is convex over
mixed states, one may reduce the maximization in Eq. (\ref{diamondDEF}) to
bipartite pure states $\varphi=\left\vert \varphi\right\rangle \left\langle
\varphi\right\vert $. In general, we therefore need to consider a min-max
optimization, i.e., find $\tilde{\pi}$ and (pure) $\tilde{\varphi}$ such that%
\begin{align}
&  \left\Vert \mathcal{I}\otimes\mathcal{E}(\tilde{\varphi})-\mathcal{I}%
\otimes\mathcal{E}_{\tilde{\pi}}(\tilde{\varphi})\right\Vert _{1}\nonumber\\
&  =\min_{\pi}\max_{\varphi}\left\Vert \mathcal{I}\otimes\mathcal{E}%
(\varphi)-\mathcal{I}\otimes\mathcal{E}_{\pi}(\varphi)\right\Vert _{1}~.
\label{minmax}%
\end{align}


\subsection{Programmable quantum measurements}\label{a:progme}
Programmable quantum measurements \cite{d2005efficient} 
represent a particular instance of the general channel approximation 
problem, where the channel approximates a measurement device. 
Consider a POVM $\{\Pi_j\}$ and a programmable POVM $\{\Pi_j^\pi\}$ with 
program $\pi$ which is obtained by performing a {\it fixed} joint 
POVM $\{Q_j\}$ on both the input state $\rho$ and the 
program $\pi$.  We 
may associate to these measurement devices two measurement channels 
$\mathcal E(\rho) =\sum_j p_j \ket j \bra j $, 
where $p_j=\Tr[\rho\Pi_j]$ is the probability of getting the outcome $j$, 
and similarly 
$\mathcal E_{\pi}(\rho) = \sum_j p_j^\pi \ket j \bra j $, where
\begin{equation}
	p^\pi_j = \Tr[\rho \Pi^{\pi}_j] = \Tr_2[Q_j(\rho\otimes \pi)]~,
\end{equation}
is the probability of getting the outcome $j$
using the programmable measurement $\Pi_j^\pi$.  
The above discussion shows 
that programmable quantum measurements represent a particular instance of the general 
case that we consider for arbitrary channels, so we may optimize the 
program state $\pi$ with the techniques presented in our paper.

Note that we may also consider a different cost function, 
namely the worst-case distance between the two probability 
distributions given by 
\begin{equation}
	C_M(\pi) = \max_\rho \sum_j |p_j-p_j^\pi|~.
\end{equation}
It was shown in \cite{d2005efficient} that 
there exist (fixed) universal POVMs $\{Q_j\}$ such that $\Pi_j^\pi$ approximates 
any arbitrary measurement $\Pi_j$ with an optimal program $\pi$. The 
error in the approximation, as quantified by $C_M$, decreases with the dimension 
of the program. 
Via the measurement channels defined above, it is easy to see that
\begin{equation}
	C_M(\pi) = \max_\rho \|\mathcal E(\rho)-\mathcal E_\pi(\rho)\|_1 \leq C_\diamond(\pi)~,
\end{equation}
so that our theory includes the results of \cite{d2005efficient} as a special case.

\subsection{SDP minimization}\label{s:sdpmin}

We show that some of the convex cost functions that we have introduced 
can be explicitly evaluated via semi-definite programming (SDP). 
This allows us to use standard SDP algorithms for finding the 
optimal program. 


We first fix the program state $\pi$ and show how for fixed $\pi$ it 
is possible to compute $C_\diamond(\pi)$ via semidefinite programming. 
Let us introduce the linear map $\Omega_{\pi}:=\mathcal{E}-\mathcal{E}_{\pi}$
with corresponding Choi matrix
\begin{equation}
\chi_{\Omega_{\pi}}=\chi_{\mathcal{E}}-\chi_{\pi}=\chi_{\mathcal{E}}%
-\Lambda(\pi).
\end{equation}
Thanks to the property of strong duality of the diamond norm, for any program
$\pi$ we can compute the cost function $C_{\diamond}(\pi)=\Vert\Omega_{\pi
}\Vert_{\diamond}$ via the following SDP~\cite{watrous2009semidefinite}
\begin{gather}
\mathrm{Minimize~}\frac{1}{2}\left(  \left\Vert \mathrm{Tr}_{2}M_{0}%
\right\Vert _{\infty}+\left\Vert \mathrm{Tr}_{2}M_{1}\right\Vert _{\infty
}\right)  ,\nonumber\\
\text{\textrm{Subject to}}\mathrm{~}%
\begin{pmatrix}
M_{0} & -d~\chi_{\Omega_{\pi}}\\
-d~\chi_{\Omega_{\pi}}^{\dagger} & M_{1}%
\end{pmatrix}
\geq0,
\end{gather}
where $M_{0}\geq0$ and $M_{1}\geq0$ in $\mathbb{C}^{d\times d^{\prime}}$, and
the spectral norm $\Vert O\Vert_{\infty}$ equals the maximum singular value of
$O$.

Moreover, because $\chi_{\Omega_{\pi}}$ is Hermitian, the above SDP can be
simplified into
\begin{gather}
\mathrm{Minimize~}2\left\Vert \mathrm{Tr}_{2}Z\right\Vert _{\infty
},\nonumber\\
\text{\textrm{Subject to}~}Z\geq0~\text{\textrm{and} }Z\geq d~\chi
_{\Omega_{\pi}}. \label{minZ}%
\end{gather}
Not only does this procedure compute $C_{\diamond}(\pi)$, but it also provides the
upper bound $C_{\diamond}(\pi)\leq d\left\Vert \mathrm{Tr}_{2}\left\vert
\chi_{\mathcal{E}}-\chi_{\pi}\right\vert \right\Vert _{\infty}$~\cite{Karol}.
In fact, it is sufficient to choose $Z=d~\chi_{\Omega_{\pi}}^{+}$, where
$\chi^{+}=(\chi+|\chi|)/2$ is the positive part of $\chi$. Using{
$\mathrm{Tr}_{2}\chi_{\Omega_{\pi}}=0$, we may write $\mathrm{Tr}_{2}Z\leq
d\mathrm{Tr}_{2}\chi_{\Omega_{\pi}}^{+}=\frac{d}{2}\mathrm{Tr}_{2}%
|\chi_{\Omega_{\pi}}|$.

The SDP\ form in Eq.~(\ref{minZ}) is particularly convenient for finding the
optimal program. In fact, suppose now that $\pi$ is not fixed but we want to
optimize on this state too, so as to compute the optimal program state
$\tilde{\pi}$ such that 
$\tilde{\pi}={\rm argmin}_{\pi\in\mathcal{S}%
}C_{\diamond}(\pi)$. The problem is therefore mapped into the following unique
minimization%
\begin{gather}
\mathrm{Minimize~}2\left\Vert \mathrm{Tr}_{2}Z\right\Vert _{\infty
},\nonumber\\
\text{\textrm{Subject to}~}Z\geq0,~\pi\geq0,~\mathrm{Tr}(\pi)=1,~{Z}\geq
d~\chi_{\Omega_{\pi}}.
\end{gather}
Unlike the min-max optimization of Eq.~\eqref{minmax}, the above SDP is much
simpler as it contains a unique minimization. Therefore,
this algorithm can be used to optimize the performance of any programmable
quantum processor.

Using a similar argument, and exploiting known convex programming 
formulations for the trace norm and the fidelity cost 
\cite{watrous2013simpler,recht2010guaranteed}, we can also compute 
the optimal program states $\tilde \pi$ as 
$\tilde\pi_1 = {\rm argmin}_{\pi\in S} C_1(\pi)$ and 
$\tilde\pi_F = {\rm argmax}_{\pi\in S} F(\pi)$.
Note indeed that, clearly, the fidelity has to be maximized, rather 
than minimized. 
The optimal program $\tilde\pi_1$ and its associated cost $C_1(\tilde \pi_1)$ 
can be computed with the following minimization
\begin{gather}
	{\rm minimize~} \Tr[P+Q]~,
	\\ \nonumber 
	{\rm subject~to~} \chi_{\Omega_\pi} = P-Q , P\geq 0, Q\geq 0, \pi\geq 0, \Tr[\pi]=1~.
\end{gather}
The optimal program $\tilde\pi_F$ and its associated cost $F(\tilde \pi_F)$ 
can be computed with the following maximization
\begin{gather}
	{\rm maximize~} \frac{\Tr[X+X^\dagger]}2,
	\nonumber \\
	{\rm subject~to~} 
	\begin{pmatrix}
		\chi_{\mathcal E} & X \\ X^\dagger & \chi_{\Omega_\pi}
	\end{pmatrix} \geq 0,  \pi\geq 0, \Tr[\pi]=1~.
\end{gather}

\subsection{Cost functions and their relative dependence}\label{a:cost}
For completeness, we report here some inequalities between 
the different cost functions introduced in the paper. 
For example, 
we may connect the minimization of the diamond distance $C_{\diamond}(\pi)$ to
the minimization of the trace distance $C_1(\pi)$
via the sandwich relation~\cite{watrous2018theory}%
\begin{equation}
C_{1}(\pi)\leq C_{\diamond}(\pi)\leq d~C_{1}(\pi).\label{mainEQ}%
\end{equation}
While the lower bound is immediate from the definition of
Eq.~(\ref{diamondDEF}), the upper bound can be proven using the following
equivalent form of the diamond distance
\begin{equation}
\Vert\mathcal{E}-\mathcal{E}_{\pi}\Vert_{\diamond}=\sup_{\rho_{0},\rho_{1}%
}d\Vert(\sqrt{\rho_{0}}\otimes\openone)(\chi_{\mathcal{E}}-\chi_{\pi}%
)(\sqrt{\rho_{1}}\otimes\openone)\Vert_{1},\label{diamondSDPstate}%
\end{equation}
where the optimization is carried out over the density matrices $\rho_{0}$ and
$\rho_{1}$~\cite[Theorem 3.1]{Watrous}. In fact, consider the Frobenius norm
$\Vert A\Vert_{2}:=\sqrt{\Tr[A^\dagger A]}$ and the spectral norm
\begin{equation}
\left\Vert A\right\Vert _{\infty}:=\max\{\left\Vert Au\right\Vert
:u\in\mathbb{C}^{d},\left\Vert u\right\Vert \leq1\},
\end{equation}
which satisfy the following properties~\cite{watrous2018theory}
\begin{align}
\Vert ABC\Vert_{1} &  \leq\Vert A\Vert_{\infty}\Vert B\Vert_{1}\Vert
C\Vert_{\infty}~,\label{prop1}\\
\Vert A\otimes\openone\Vert_{\infty} &  =\Vert A\Vert_{\infty}\leq\Vert
A\Vert_{2}.\label{prop2}%
\end{align}
Then, from Eqs.~(\ref{diamondSDPstate}), (\ref{prop1}) and~(\ref{prop2}), one
gets
\begin{align}
\Vert\mathcal{E}-\mathcal{E}_{\pi}\Vert_{\diamond} &  \leq\sup_{\rho_{0}%
,\rho_{1}}d\sqrt{\mathrm{Tr}\rho_{0}\mathrm{Tr}\rho_{1}}\Vert\chi
_{\mathcal{E}}-\chi_{\pi}\Vert_{1}\nonumber\\
&  =d\Vert\chi_{\mathcal{E}}-\chi_{\pi}\Vert_{1}.
\end{align}

Thanks to Eq.~(\ref{mainEQ}), we may avoid the maximization step in the
definition of the diamond distance and simplify the original problem to
approximating the Choi matrix $\chi_{\mathcal{E}}$ of the channel by varying
the program state $\pi$. This is a process of learning Choi matrices as
depicted in Fig.~\ref{QMLprogram2}. Because the simpler cost function
$C_{1}(\pi)$ is an upper bound, its minimization generally provides a
sub-optimal solution for the program state. \begin{figure}[t]
\vspace{-0.1cm}
\par
\begin{center}
\includegraphics[width=0.40\textwidth] {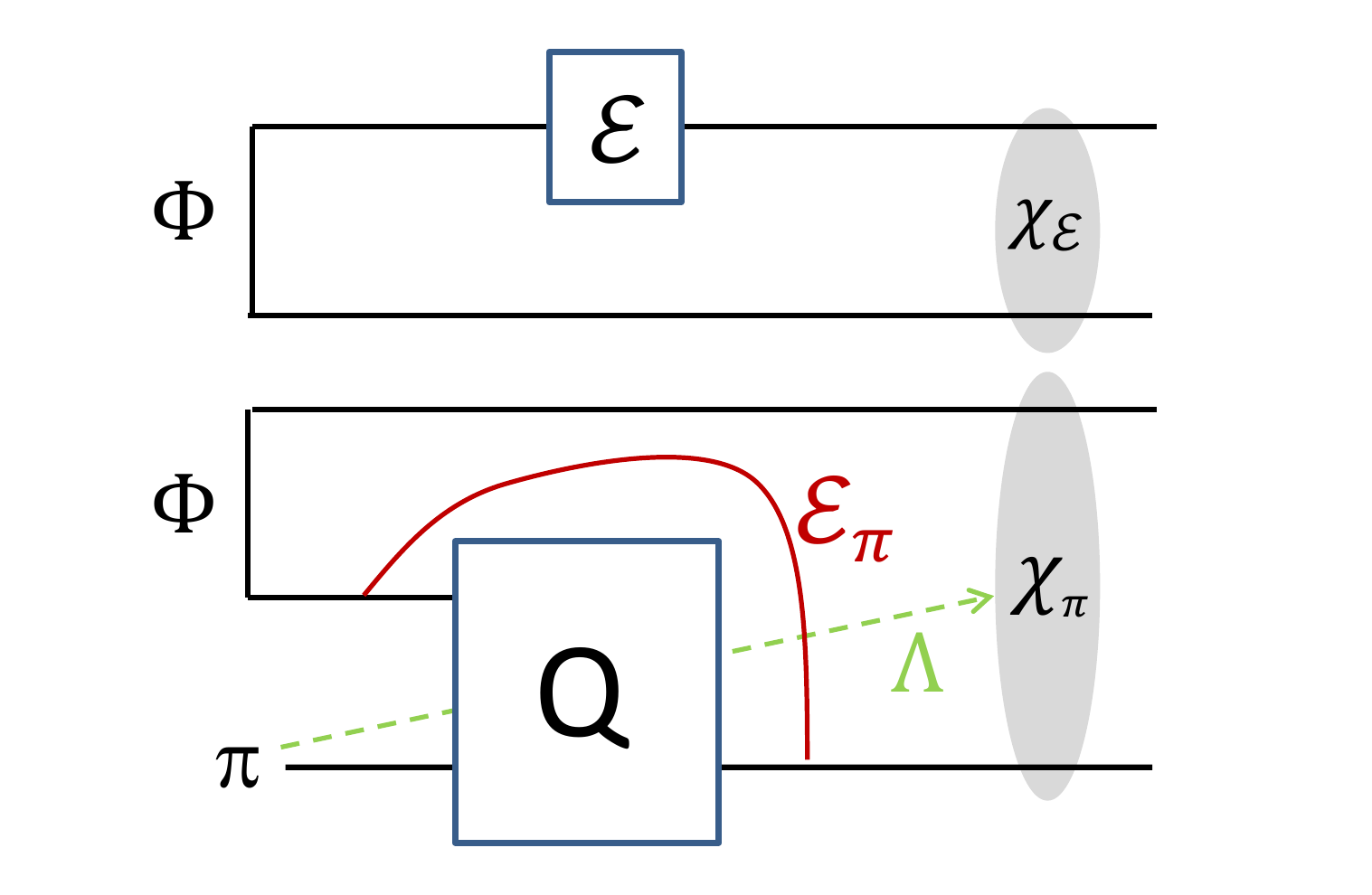}
\end{center}
\par
\vspace{-0.5cm}\caption{Map of the processor and learning of Choi matrices.
Consider an arbitrary (but known) quantum channel $\mathcal{E}$ and its
associated Choi matrix $\chi_{\mathcal{E}}$, generated by propagating part of
a maximally-entangled state $\Phi$. Then, consider a quantum processor $Q$
with program state $\pi$\ which generates the simulated channel $\mathcal{E}%
_{\pi}$\ and, therefore, the corresponding Choi matrix $\chi_{\pi}%
:=\chi_{\mathcal{E}_{\pi}}$ upon propagating part of $\Phi$ as the input state.
The map of the processor is the CPTP map $\Lambda$ from the program state
$\pi$ to the output Choi matrix $\chi_{\pi}$. In a simplified version of our
problem, we may optimize the program $\pi$ in such a way as to minimize the trace
distance $C_{1}(\pi):=\left\Vert \chi_{\mathcal{E}}-\chi_{\pi}\right\Vert
_{1}$.}%
\label{QMLprogram2}%
\end{figure}

Finally we may consider other cost functions in terms of any Shatten p-norm
$C_{p}(\pi):=\Vert\chi_{\mathcal{E}}-\chi_{\pi}\Vert_{p}$, even though this
option provides lower bounds instead of upper bounds for the trace distance.
Recall that, given an operator $O$ and a real number $p\geq1$, we may define
its Schatten p-norm as~\cite{watrous2018theory}%
\begin{equation}
\left\Vert O\right\Vert _{p}=(\mathrm{Tr}|O|^{p})^{1/p},
\end{equation}
where $|O|=\sqrt{O^{\dagger}O}$. For any $1\leq p\leq q\leq\infty$, one has
the monotony $\left\Vert O\right\Vert _{p}\geq\left\Vert O\right\Vert _{q}$,
so that $\left\Vert O\right\Vert _{\infty}\leq\ldots\leq\left\Vert
O\right\Vert _{1}$. An important property is duality. For each pair of
operators $A$ and $B$, and each pair of parameters $p,q\in\lbrack1,\infty]$
such that $p^{-1}+q^{-1}=1$, we may write~\cite{watrous2018theory}
\begin{equation}
\left\Vert A\right\Vert _{p}=\sup_{\left\Vert B\right\Vert _{q}\leq
1}\left\vert \left\langle B,A\right\rangle \right\vert
\equiv \sup_{\left\Vert B\right\Vert _{q}\leq
1}\left\langle B,A\right\rangle
,
\label{dualityNORM}%
\end{equation}
where $\left\langle B,A\right\rangle =\mathrm{Tr}(B^{\dagger}A)$ is the
Hilbert-Schmidt product, and the second inequality follows since we can arbitrarily
change the sign of $B$.

\section{Convergence in learning arbitrary unitaries\label{Sec5}}
The simulation of quantum gates or, more generally, unitary transformations
is crucial for quantum computing applications \cite{lloyd1996universal} so
ML techniques have been developed for this purpose
\cite{banchi2016quantum,innocenti2018supervised,mitarai2018quantum}.
In the main text we have shown that,
for learning arbitrary unitaries, the fidelity cost function provides a
convenient choice for which the optimal program can be found analytically. 
Indeed, the optimal program is always a pure state and is given by 
the eigenstate of $\Lambda^{\ast}\left[  |\chi
_{U}\rangle\langle\chi_{U}|\right]  $ with the maximum eigenvalue.
Here we consider the convergence of the Frank-Wolfe iteration 
Eq.~\eqref{FrankWolfe} towards that state. 

Let $\pi_{1}$ be the initial guess
for the program state. After $k$ iterations of Eq.~\eqref{FrankWolfe}, we find
the following approximation to the optimal program state
\begin{equation}
\pi_{k}=\frac{2}{k+k^{2}}\pi_{1}+\left(  1-\frac{2}{k+k^{2}}\right)
\tilde{\pi}_{F}~,
\end{equation}
where $\frac{2}{k+k^{2}}=\prod_{j=1}^{k-1}\frac{j}{j+2}$. The above equation
shows that $\pi_{k}\rightarrow\tilde{\pi}_{F}$ for $k\rightarrow\infty$, with
error in trace distance
\begin{equation}
\Vert\pi_{k}-\tilde{\pi}_{F}\Vert_{1}=\frac{2}{k+k^{2}}\left\Vert \pi
_{1}-\tilde{\pi}_{F}\right\Vert _{1}=\mathcal{O}(k^{-2})~.\label{itera}%
\end{equation}

This example shows that the convergence rate $\mathcal{O}%
(\epsilon^{-1})$ of the conjugate method provides a worst case instance
that can be beaten in some applications with some suitable cost functions.
From Eq.~\eqref{itera} we see that $\epsilon=k^{-2}$ for learning arbitrary unitaries
via the minimization of $C_F$,
meaning that convergence is obtained with the faster rate $\mathcal{O}%
(\epsilon^{-1/2})$. On the other hand, there are
no obvious simplifications for the optimization of the trace distance, since
the latter still requires the diagonalization of Eq.~\eqref{tracediag}. For
the trace distance, or its smooth version, only numerical approaches are feasible.

\section{Applications}\label{s:appli}
\subsection{Learning a unitary with the teleportation processor}\label{a:tele}

Here we consider the following example.
Assume that the target channel is a unitary $U$, so that its
Choi matrix is $\chi_{U}:=|\chi_{U}\rangle\langle\chi_{U}|$ with $|\chi
_{U}\rangle=\openone\otimes U|\Phi\rangle$ and where $|\Phi\rangle$ is maximally
entangled. By using Eq.~(\ref{eqAB}), the fact that $\Lambda_{\rm tele}=\Lambda_{\rm tele}^*$
and $U^{\ast}\otimes\openone|\Phi
\rangle=\openone\otimes U^{\dagger}|\Phi\rangle$, we may write the dual
processor map
\begin{align}
&  \Lambda_{\text{tele}}^{\ast}[|\chi_{U}\rangle\langle\chi_{U}|]\nonumber\\
&  =\frac{1}{d^{2}}\sum_{i}\left(  \openone\otimes V_{i}^{U}\right)
|\Phi\rangle\langle\Phi|\left(  \openone\otimes V_{i}^{U}\right)  ^{\dagger
},\label{telelambdadual}%
\end{align}
where $V_{i}^{U}=U_{i}UU_{i}^{\dagger}$. The maximum eigenvector of
$\Lambda_{\text{tele}}^{\ast}[|\chi_{U}\rangle\langle\chi_{U}|]$ represents
the optimal program state $\tilde{\pi}_{F}$ for simulating the unitary
$U$\ via the teleportation processor (according to the fidelity cost
function).
In some cases, the solution is immediate. For instance, this happens when
$V_{i}^{U}\propto U$ is independent of $i$. This is the case when $U$ is a
teleportation unitary, because it satisfies the Weyl-Heisenberg
algebra~\cite{pirandola2017fundamental}. For a teleportation unitary $U$, we
simply have%
\begin{equation}
\Lambda^{\ast}[|\chi_{U}\rangle\langle\chi_{U}|]=|\chi_{U}\rangle\langle
\chi_{U}|~,
\label{fidelitygradpure}
\end{equation}
so that the unique optimal program is $\tilde{\pi}_{F}=|\chi_{U}\rangle
\langle\chi_{U}|$.

\begin{figure}[t]
\centering
\includegraphics[width=0.99\linewidth]{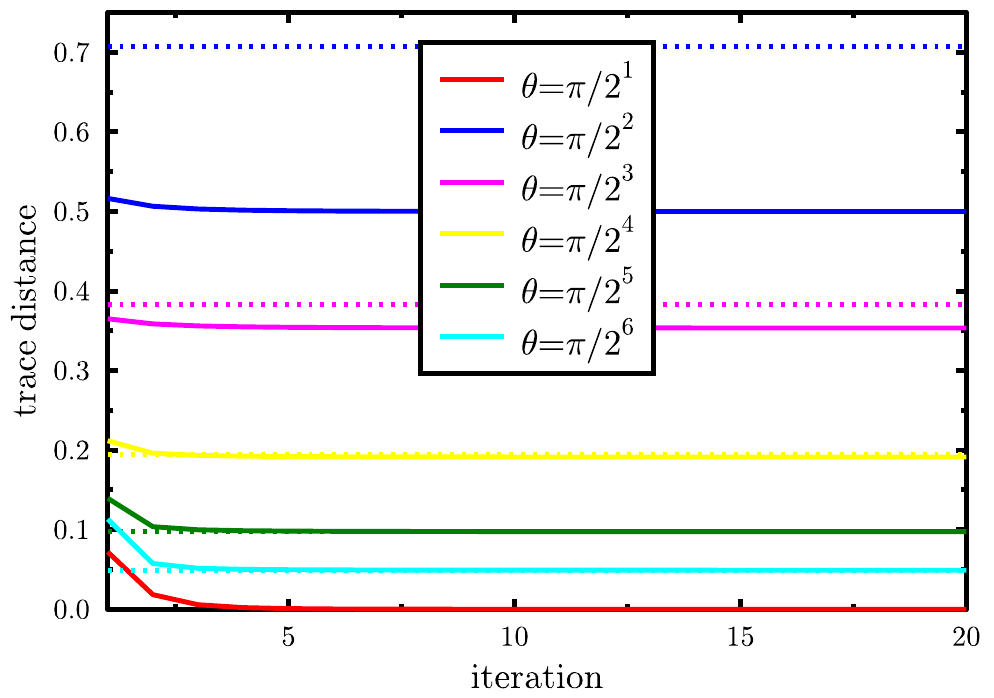}
\caption{Optimization of program states for simulating the rotation
$R(\theta)=e^{i\theta X}$ with a teleportation processor. The optimization is
via the minimization of trace distance $C_{1}$ of Eq.~\eqref{traceD} with the
projected subgradient method in Eq.~\eqref{projsubgrad}. The dashed lines
correspond to the upper bound $\sqrt{1-F[\Lambda(\tilde{\pi}_{F}%
),\chi_{\mathcal{E}}]^{2}}$ of the trace distance, where $\tilde{\pi}_{F}$ is
the optimal program that maximizes the fidelity, namely the eigenvector of
$\Lambda^{\ast}[|\chi_{U}\rangle\langle\chi_{U}|]$
with the maximum eigenvalue. }%
\label{fig:teleunitary}%
\end{figure}

In Fig.~\ref{fig:teleunitary} we show the convergence of the projected
subgradient algorithm using the teleportation processor and target unitaries
$R(\theta)=e^{i\theta X}$, for different values of $\theta$. When $\theta$ is
a multiple of $\pi/2$, then the above unitary is teleportation covariant and
the Frank-Wolfe algorithm converges to zero trace distance. For other values
of $\theta$ perfect simulation is impossible, and we notice that the algorithm
converges to a non zero value of the trace distance \eqref{traceD}. For
comparison, in Fig.~\ref{fig:teleunitary} we also plot the value of the
fidelity upper bound $\sqrt{1-F[\Lambda(\tilde{\pi}_{F}),\chi_{\mathcal{E}%
}]^{2}}$, where $\tilde{\pi}_{F}$ is the optimal program that maximizes the
fidelity of Eq.~\eqref{fidCC}, namely the eigenvector of
Eq.~\eqref{telelambdadual} with the maximum eigenvalue.
We note that for $\theta=\pi/2^{\ell}$, the trace
distance decreases for larger $\theta$. The limit case $\ell\rightarrow\infty$
is perfectly simulable as $R(0)$ is teleportation covariant.

\subsection{Pauli channel simulation}

Pauli channels are defined as~\cite{nielsen2000quantum}
\begin{equation}
\mathcal{P}(\rho)=\sum_{i}p_{i}U_{i}\rho U_{i}^{\dagger}~,
\end{equation}
where $U_{i}$ are generalized Pauli operators and $p_{i}$ are some probabilities.
For $d=2$ the Pauli operators are the four Pauli matrices $I,X,Y,Z$ and in any
dimension they form the Weyl-Heisenberg group~\cite{nielsen2000quantum}. These
operators are exactly the teleportation unitaries $U_{j}$ defined in the
previous section. The Choi matrix $\chi_{\mathcal{P}}$ of a Pauli channel
$\mathcal{P}$ is diagonal in the Bell basis, i.e., we have
\begin{equation}
\chi_{\mathcal{P}}=\sum_{i}p_{i}|\Phi_{i}\rangle\langle\Phi_{i}|~,
\end{equation}
where $\Phi_{i}=\openone\otimes U_{i}|\Phi\rangle$ and $|\Phi\rangle
=\sum_{j=1}^{d}|jj\rangle/\sqrt{d}$.

We now consider the simulation of a Pauli channel with the teleportation
quantum processor introduced in the previous section. Let
\begin{equation}
\pi=\sum_{ij}\pi_{ij}|\Phi_{i}\rangle\langle\Phi_{j}|~,
\end{equation}
be an arbitrary program state expanded in the Bell basis. For any program
state, the Choi matrix of the teleportation-simulated channel is given by
Eq.~\eqref{eqAB}. Using standard properties of the Pauli matrices, we find
\begin{equation}
\chi_{\pi}\equiv\Lambda(\pi)=\sum_{i}\pi_{ii}|\Phi_{i}\rangle\langle\Phi
_{i}|~,
\end{equation}
namely a generic state is transformed into a Bell diagonal state. Therefore,
the cost function
\begin{equation}
C_{1}^{\mathrm{Pauli}}=\Vert\chi_{P}-\chi_{\pi}\Vert_{1}~,
\end{equation}
can be minimized analytically for any Pauli channel by choosing $\pi
_{ij}=p_{i}\delta_{ij}$. With this choice we find $C_{1}^{\mathrm{Pauli}}=0$,
meaning that the simulation is perfect.

From theory~\cite{bowen2001teleportation,cope2017simulation,BDSW} we know that
only Pauli channels can be perfectly simulated in this way. No matter how much more
general we make the states $\pi$, it is
proven~\cite{bowen2001teleportation,cope2017simulation} that these are the
only channels we can perfectly simulate. This is true even if we apply the
Pauli corrections in a probabilistic way, i.e., we assume a classical channel
from the Bell outcomes to the corresponding label of the Pauli correction
operator~\cite{cope2017simulation}.

\subsection{PBT: Numerical examples}

\begin{figure}[t]
\centering
\includegraphics[width=0.95\linewidth]{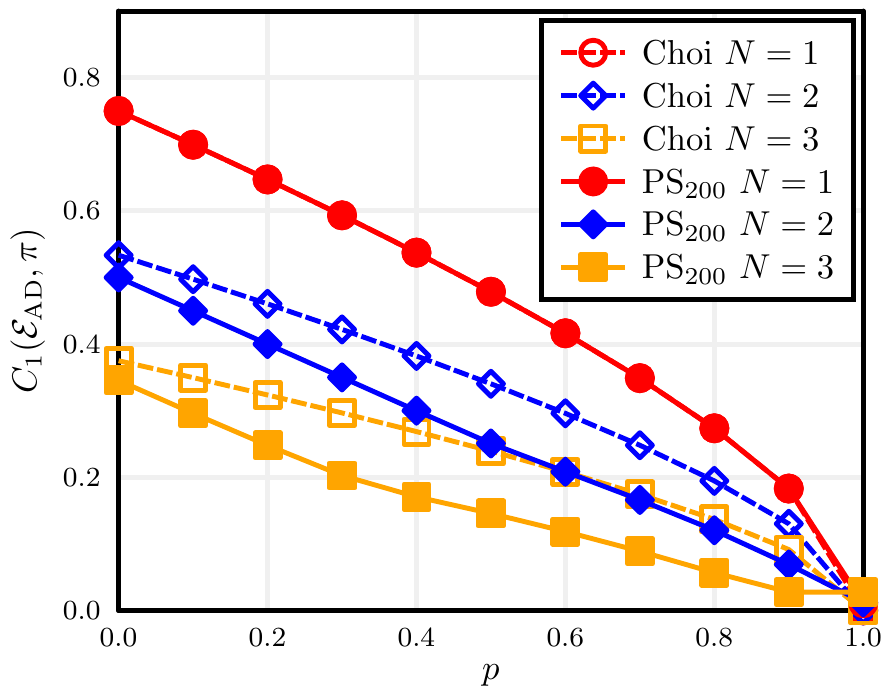} \caption{PBT
Simulation of the amplitude damping channel $\mathcal{E}_{\mathrm{AD}}$ for
various damping rates $p$. Minimization of the trace distance $C_{1}%
(\mathcal{E}_{\mathrm{AD}},\pi)=\Vert\chi_{\mathcal{E}_{\mathrm{AD}}}%
-\chi_{\pi}\Vert_{1}$ between the target channel's Choi matrix and its PBT
simulation with program state $\pi$, for different number of ports $N$. We
consider $N=1,2,3$ and two kinds of programs: copies of the channel's Choi
matrix $\chi_{\mathcal{E}_{\mathrm{AD}}}^{\otimes N}$ and the state
$\tilde{\pi}_{1}$ obtained from the minimization of $C_{1}$ via the projected
subgradient (PS) method after 200 iterations. Note that the simulation error
$C_{1}$ is maximized for the identity channel ($p=0$) and goes to zero for
$p\rightarrow1$.}%
\label{fig:AD_vs_p}%
\end{figure}
We first consider the simulation of an amplitude damping channel
$\mathcal{E}_{\mathrm{AD}}(\rho)=\sum_{i}K_{i}^{\mathrm{AD}}\rho
K_{i}^{\mathrm{AD}\dagger}$, which is defined by the Kraus operators
\begin{equation}
K_{0}^{\mathrm{AD}}=%
\begin{pmatrix}
1 & 0\\
0 & \sqrt{1-p}%
\end{pmatrix}
,~~K_{1}^{\mathrm{AD}}=%
\begin{pmatrix}
0 & \sqrt{p}\\
0 & 0
\end{pmatrix}
.\label{KrausAD}%
\end{equation}
In Fig.~\ref{fig:AD_vs_p} we study the performance of the PBT simulation of
the amplitude damping channel $\mathcal{E}_{\mathrm{AD}}$ for different
choices of $p$. For $p=0$ the amplitude damping channel is equal to the identity
channel, while for $p=1$ it is a \textquotedblleft reset\textquotedblright%
\ channel sending all states to $|0\rangle$. We compare the simulation error
with program states $\pi$ either made by products of the channel's Choi matrix
$\chi_{\mathcal{E}_{\mathrm{AD}}}^{\otimes N}$ as in
Eq.~\eqref{e:pbtchoiprogram} or obtained from the minimization of the trace
distance cost function of Eq.~\eqref{traceD} with the projected subgradient
iteration in Eq.~\eqref{projsubgrad}. Alternative methods, such as the conjugated
gradient algorithm, perform similarly for this channel. We observe that,
surprisingly, the optimal program $\tilde{\pi}_{1}$ obtained by minimizing the
trace distance $C_{1}$ is always better than the natural choice $\chi
_{\mathcal{E}_{\mathrm{AD}}}^{\otimes N}$.

\begin{figure}[t]
\centering
\includegraphics[width=0.95\linewidth]{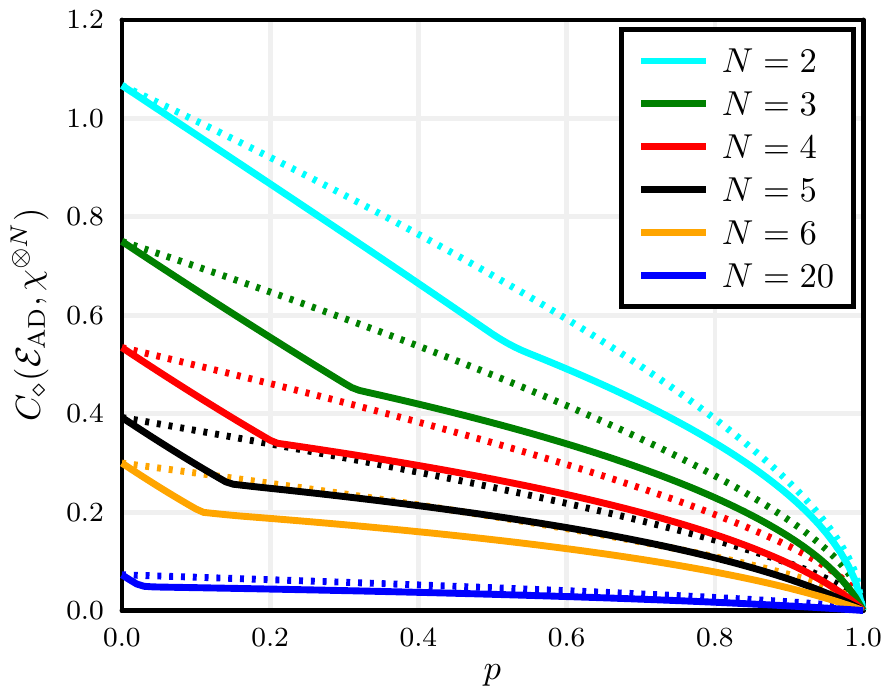} \caption{ PBT
Simulation of the amplitude damping channel $\mathcal{E}_{\mathrm{AD}}$ for
various damping rates $p$. We plot the diamond distance cost function
$C_{\diamond}(\mathcal{E}_{\mathrm{AD}},\pi)=\Vert\mathcal{E}_{\mathrm{AD}%
}-\mathcal{E}_{\mathrm{AD}\text{,}\pi}\Vert_{\diamond}$ between the target
channel $\mathcal{E}_{\mathrm{AD}}$ and its PBT\ simulation $\mathcal{E}%
_{\mathrm{AD}\text{,}\pi}$ with program state $\pi$. In particular, for the
program state we compare the naive choice of the channel's Choi matrix
$\pi=\chi_{\mathcal{E}_{\mathrm{AD}}}^{\otimes N}$ (dotted lines) with the SDP
minimization over the set of generic Choi matrices $\pi=\chi^{\otimes N}$
(solid lines). Different values of $N=2,\dots,6$ and $N=20$ are shown.
}%
\label{fig:ADchoi}%
\end{figure}

In Fig.~\ref{fig:ADchoi} we study the PBT\ simulation of the amplitude damping
channel by considering the subset of program states $\pi=\chi^{\otimes N}$
which is made of tensor products of the $4\times4$ generic Choi matrices
$\chi$ (satisfying $\mathrm{Tr}_{2}\chi=\openone/2$). As discussed
in Sec.~\ref{compressSEC}, this is equivalent to optimizing over the
Choi set $\mathcal{C}_{N}$\ and it practically reduces to the convex
optimization of the channel $\tilde{\Lambda}$ over the generic single-copy
Choi matrix $\chi$. Moreover, $\tilde{\Lambda}$ itself can be simplified, as
shown in Appendix \ref{compressSEC}, so that all of the operations depend polynomially on
the number $N$ of ports. This allows us to numerically explore much larger
values of $N$, even for the minimization of $C_{\diamond}$. In
Fig.~\ref{fig:ADchoi} the dotted lines correspond to the value of
$C_{\diamond}$ when the program $\pi=\chi_{\mathcal{E}_{\mathrm{AD}}}^{\otimes
N}$ is employed, where $\chi_{\mathcal{E}_{\mathrm{AD}}}$ is the channel's
Choi matrix. As Fig.~\ref{fig:ADchoi} shows, the cost $C_{\diamond}$ may be
significantly smaller with an optimal $\chi$, thus showing that the optimal
program may be different from the channel's Choi matrix, especially when $p$
is far from the two boundaries $p=0$ and $p=1$.\begin{figure}[t]
\centering
\includegraphics[width=0.95\linewidth]{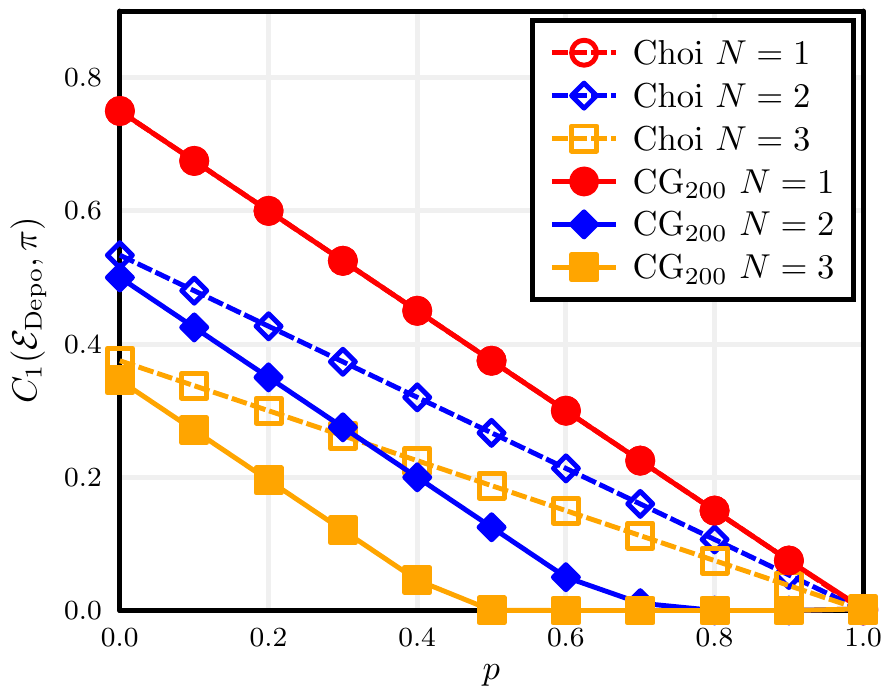} \caption{PBT
Simulation of the qubit depolarizing channel versus probability of
depolarizing $p$. Trace distance $C_{1}(\mathcal{E}_{\mathrm{dep}},\pi
)=\Vert\chi_{\mathcal{E}_{\mathrm{dep}}}-\chi_{\pi}\Vert_{1}$ between the
target channel's Choi matrix and its PBT simulation with program state $\pi$,
for different number of ports $N$. We consider $N=1,2,3$ and two kinds of
programs: copies of the channel's Choi matrix $\pi=\chi_{\mathcal{E}%
_{\mathrm{dep}}}^{\otimes N}$ and the optimal program state $\tilde{\pi}_{1}$
obtained from the minimization of $C_{1}$ via the conjugate gradient (CG)
method after 200 iterations. Note that the simulation error $C_{1}$ is maximized
for the identity channel ($p=0$) and eventually goes to zero for a finite
value of $p$ that decreases for increasing $N$. }%
\label{fig:Depo_vs_p}%
\end{figure}\begin{figure}[t]
\centering
\includegraphics[width=0.95\linewidth]{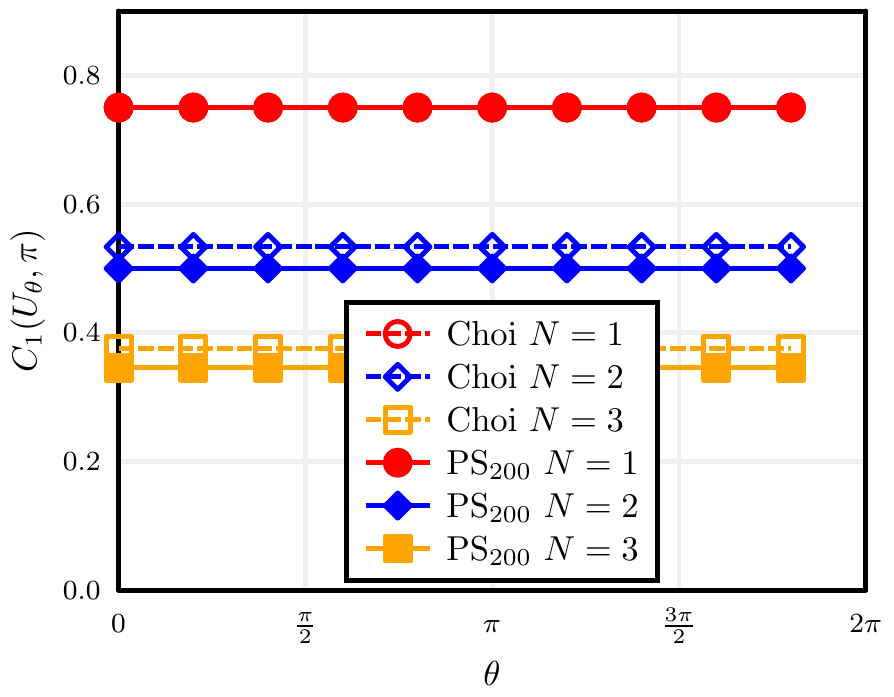} \caption{PBT Simulation
of the unitary gate $U_{\theta}=e^{i\theta X}$ for different angles $\theta$,
where $X$ is the bit-flip Pauli matrix. Trace distance $C_{1}(U_{\theta}%
,\pi)=\Vert\chi_{U_{\theta}}-\chi_{\pi}\Vert_{1}$ between the target Choi
matrix of the unitary and its PBT simulation with program state $\pi$, for
different number of ports $N$. We consider $N=1,2,3$ and two kinds of
programs: copies of the Choi matrix of the unitary $\chi_{U_{\theta}}^{\otimes
N}$ and the program state $\tilde{\pi}_{1}$ obtained from the minimization of
$C_{1}$ via the projected subgradient (PS) method after 200 iterations. }%
\label{fig:UX2}%
\end{figure}

As an other example, we consider the simulation of the depolarizing channel
defined by
\begin{equation}
\mathcal{E}_{\text{\textrm{dep}}}(\rho)=(1-p)\rho+\frac{p}{d}\openone.
\end{equation}
It was shown in \cite{ishizaka2009quantum,pirandola2018fundamental} that PBT generates 
a depolarizing channel, whose depolarizing probability $p_{\rm th}$ depends on $N$. This implies
that a quantum processor based on PBT can perfectly simulate a depolarizing channel
when $p \geq p_{\rm th}$. 
In Fig.~\ref{fig:Depo_vs_p} we study the performance of PBT simulation of the
depolarizing channel in terms of $p$. Perfect simulation is possible 
for $p\geq p_{\rm th} \simeq 0.71$ for $N=2$ and  
$p\geq p_{\rm th} = 0.5$ for $N=3$, where the explicit values of $p_{\rm th}$ 
are taken from Ref.~\cite{pirandola2018fundamental}.
For $p=0$ the depolarizing channel is
equal to the identity channel, while for $p=1$ it sends all states to the
maximally mixed state. Again we compare the simulation error with program
states either composed of copies of the channel's Choi matrix $\chi_{\mathcal{E}%
_{\mathrm{dep}}}^{\otimes N}$ or obtained from the minimization of $C_{1}$
with the conjugate gradient method of Eq.~\eqref{FrankWolfe}, which performs
significantly better than the projected subgradient for this channel. Also for
the depolarizing channel we observe that, for any finite $N$, we obtain a
lower error by optimizing over the program states instead of the naive choice
$\chi_{\mathcal{E}_{\mathrm{dep}}}^{\otimes N}$.

Finally, in Fig.~\ref{fig:UX2} we study the PBT simulation of a unitary gate
$U_{\theta}=e^{i\theta X}$ for different values of $\theta$. Unlike the
previous non-unitary channels, in Fig.~\ref{fig:UX2} we observe a flat error
where different unitaries have the same simulation error of the identity
channel $\theta=0$. This is expected because both the trace distance and the
diamond distance are invariant under unitary transformations. In general, we
have the following.

\begin{proposition}
Given a unitary $\mathcal{U}(\rho)=U\rho U^{\dagger}$ and its PBT\ simulation
$\mathcal{U}_{\pi}$ with program $\pi$\ we may write
\begin{equation}
\min_{\pi}||\mathcal{U}-\mathcal{U}_{\pi}||_{\diamond}=\min_{\pi}%
||\mathcal{I}-\mathcal{I}_{\pi}||_{\diamond},
\end{equation}
where $\mathcal{I}_{\pi}$ is the PBT\ simulation of the identity channel.
\end{proposition}

\noindent\textit{Proof}.~~In fact, we simultaneously prove
\begin{equation}
\min_{\pi}||\mathcal{I}-\mathcal{I}_{\pi}||_{\diamond}\overset{(1)}{\leq}%
\min_{\pi}||\mathcal{U}-\mathcal{U}_{\pi}||_{\diamond}\overset{(2)}{\leq}%
\min_{\pi}||\mathcal{I}-\mathcal{I}_{\pi}||_{\diamond},
\end{equation}
where (1) comes from the fact that $||\mathcal{U}-\mathcal{U}_{\pi
}||_{\diamond}=||\mathcal{U}^{-1}\mathcal{U}-\mathcal{U}^{-1}\mathcal{U}_{\pi
}||_{\diamond}=||\mathcal{I}-\mathcal{U}^{-1}\mathcal{U}_{\pi}||_{\diamond}$
and $\mathcal{U}^{-1}\mathcal{U}_{\pi}$ is a possible PBT\ simulation of the
identity $\mathcal{I}$ with program state $\mathcal{I}\otimes(\mathcal{U}%
^{-1})^{\otimes N}(\pi)$ once $\mathcal{U}^{-1}$ is swapped with the filtering
of the ports; then (2) comes from the fact that the composition $\mathcal{U}%
\circ\mathcal{I}_{\pi}$ is a possible simulation of the unitary $\mathcal{U}$
with program state $\mathcal{I}\otimes\mathcal{U}^{\otimes N}(\pi)$ and we
have the inequality$\ ||\mathcal{U}\circ\mathcal{I}-\mathcal{U}\circ
\mathcal{I}_{\pi}||_{\diamond}\leq||\mathcal{I}-\mathcal{I}_{\pi}||_{\diamond
}$.~$\blacksquare$



\begin{figure}[t]
    \centering
    \includegraphics[width=0.95\linewidth]{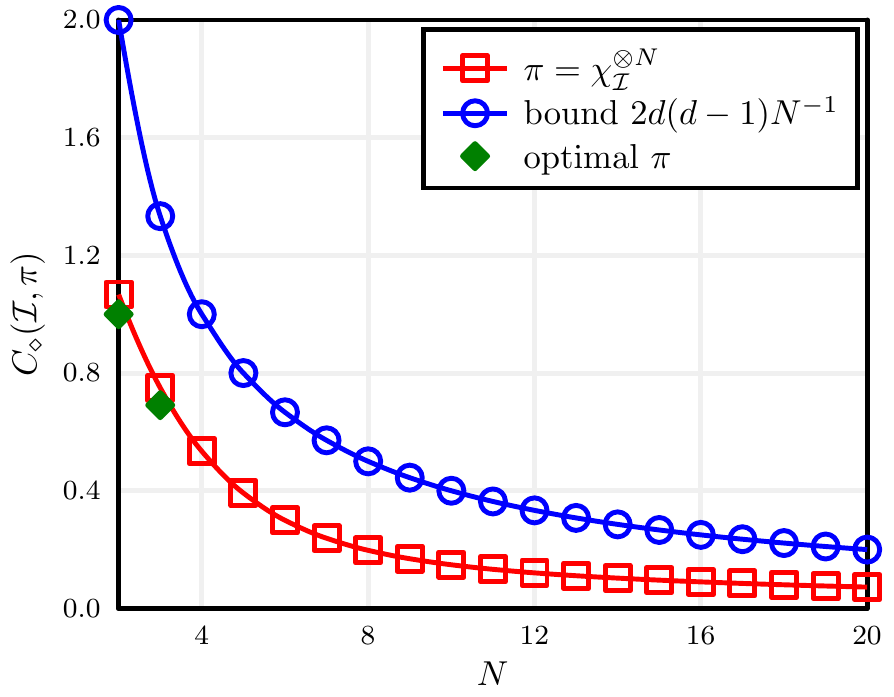}
    \caption{ PBT Simulation of the identity channel for different number of ports $N$.
        For the identity channel the optimal Choi matrix coincides with the channel's Choi
        matrix $\chi_{\mathcal I}\equiv \Phi$. The optimal $\pi$ has been obtained by minimising
        $C_\diamond$ via SDP. The upper bound corresponds to Eq.~\eqref{simerr}.
    }
    \label{fig:id}
\end{figure}

The scaling of $||\mathcal{I}-\mathcal{I}_{\pi}||_{\diamond}$ for different values of
$N$ is plotted in Fig.~\ref{fig:id} where numerical values are obtained from SDP, while
the upper bound is given by Eq.~\eqref{simerr}.

\subsection{Parametric quantum circuits\label{Sec8}}\label{s:pqc}

We now study another design of universal quantum processor that can simulate
any target quantum channel in the asymptotic limit of an arbitrarily large
program state. This is based on a suitable reformulation of the PQCs, which
are known to simulate any quantum computation with a limited set of quantum
gates~\cite{lloyd1995almost,lloyd1996universal}.

A PQC is composed of a sequence of unitary matrices $U_{j}(\theta_{j})$, each
depending on a classical parameter $\theta$. The resulting unitary operation
is then
\begin{equation}
U(\theta)=U_{N}(\theta_{N})\dots U_{2}(\theta_{2})U_{1}(\theta_{1}%
).\label{Utheta}%
\end{equation}
A convenient choice is $U_{j}(\theta_{j})=\exp(i\theta_{j}H_{j})$, where
each elementary gate corresponds to a Schr\"{o}dinger evolution with
Hamiltonian $H_{j}$ for a certain time interval $\theta_{j}$. For certain
choices of $H_{j}$ and suitably large $N$ the above circuit is
universal~\cite{lloyd1996universal}, namely any unitary can be obtained with
$U(\theta)$ and a suitable choice of $\theta_{j}$. The optimal parameters can
be found via numerical algorithms~\cite{khaneja2005optimal}, e.g. by minimizing
the cost function $C(\theta)=|\Tr[U_{\rm target}^\dagger U(\theta)]|$.
However, the above cost function is not convex, so the numerical algorithms
are not guaranteed to converge to the global optimum.

As a first step, we show that the task of learning the optimal parameters in a
PQC can be transformed into a convex optimization problem by using a quantum
program. This allows us to use SDP and gradient-based ML methods to find
the global optimum solution.

\subsubsection{Convex reformulation}

Consider a program state $|\pi\rangle=|\theta_{1},\dots,\theta_{N}\rangle$
composed of $N$ registers $R_{j}$, each in a separable state $|\theta
_{j}\rangle$. We can transform the classical parameters in Eq.~\eqref{Utheta}
into quantum parameters via the conditional gates
\begin{equation}
\hat{U}_{j}=\exp\left(  iH_{j}\otimes\sum_{\theta_{j}}\theta_{j}|\theta
_{j}\rangle\langle\theta_{j}|\right)  ,\label{condUni}%
\end{equation}
that act non-trivially on systems and registers $R_{j}$. If the parameters
$\theta_{j}$ are continuous, then we can replace the sum with an integral.
\begin{figure}[t]
\centering
\includegraphics[width=0.9\linewidth]{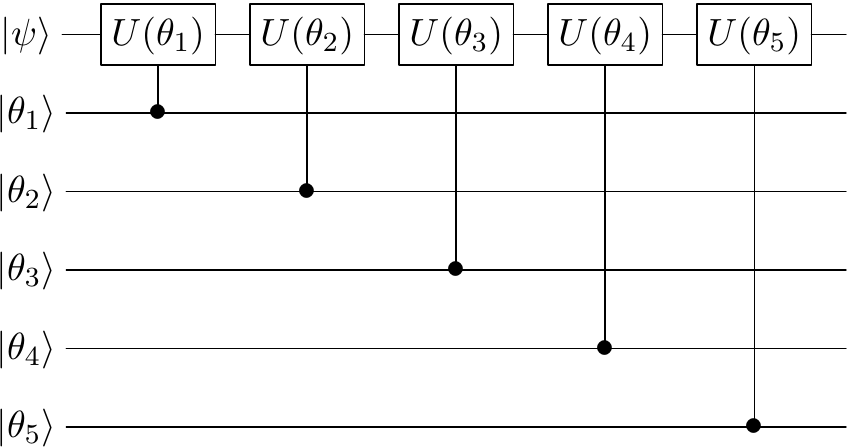} \caption{Convex
reformulation of a PQC as a coherent programmable quantum processor that
applies a sequence of conditional gates, as in Eq.~\eqref{condUni}, depending on
the program state $|\pi\rangle=|\theta_{1},\dots,\theta_{N}\rangle$. The
program state is not destroyed and can be reused. }%
\label{fig:PQC}%
\end{figure}With the above gates, we define the parametric quantum channel
\begin{equation}
Q_{\pi}(\rho)=\mathrm{Tr}_{R}\left[  \prod_{j=1}^{N}\hat{U_{j}}\left(
\rho\otimes\pi\right)  \prod_{j=1}^{N}\hat{U_{j}}^{\dagger}\right]
,\label{PQCchan}%
\end{equation}
whose action on a generic state $|\psi\rangle$ is shown in Fig.~\ref{fig:PQC}.
For a pure separable program $|\pi\rangle=|\theta_{1},\dots,\theta_{N}\rangle
$, we obtain the standard result, i.e.,
\begin{equation}
Q_{|\theta_{1},\dots,\theta_{N}\rangle}(\rho)=U(\theta)\rho U(\theta
)^{\dagger},\label{Qchan0}%
\end{equation}
where $U(\theta)$ is defined in Eq.~\eqref{Utheta}. The parametric quantum
processor $Q_{\pi}$ in Eq.~(\ref{PQCchan}) is capable of simulating any
parametric quantum channel, but it is more general, as it allows entangled
quantum parameters and also parameters in quantum superposition.

An equivalent measurement-based protocol is obtained by performing the trace
in Eq.~\eqref{Qchan0} over the basis $|\theta_{1},\dots,\theta_{N}\rangle$, so
that
\begin{equation}
Q_{\pi}(\theta)=\sum_{\{\theta_{j}\}}U(\theta)\rho U(\theta)^{\dagger}%
\langle\theta_{1},\dots,\theta_{N}|\pi|\theta_{1},\dots,\theta_{N}%
\rangle,\label{Mchan}%
\end{equation}
where $U(\theta)$ is defined in Eq.~\eqref{Utheta}. In this alternative yet
equivalent formulation, at a certain iteration $j$, the processor measures the
qubit register $R_{j}$. Depending on the measurement outcome $\theta_{j}$, the
processor then applies a different unitary $U(\theta_{j})$ on the system.
However, in this formulation the program state $|\pi\rangle$ is destroyed
after each channel use. From Eq.~\eqref{Mchan} we note that $Q_{\pi}$ depends
on $\pi$ only via the probability distribution $\langle\theta_{1},\dots
,\theta_{N}|\pi|\theta_{1},\dots,\theta_{N}\rangle$. As such, any advantage in
using quantum states can only come from the capability of quantum systems to
model computationally hard probability distributions
\cite{boixo2018characterizing}.

\subsubsection{Universal channel simulation via PQCs}

The universality of PQCs can be employed for universal channel simulation, 
thanks to Stinespring's dilation theorem: 
\begin{equation}
\mathcal{E}(\rho_{A})=\mathrm{Tr}_{R_{0}}[U(\rho_{A}\otimes\theta
_{0})U^{\dagger}],\label{stinespring}%
\end{equation}
where $\theta_{0}$ belongs to $R_{0}$, and $U$ acts on system $A$ and register
$R_{0}$. In Ref.~\cite{lloyd1995almost} it was shown that two quantum gates
are universal for quantum computation. Specifically, given $U_{0}%
=e^{it_{0}H_{0}}$ and $U_{B}=e^{it_{1}H_{1}}$ for fixed times $t_{i}$ and
Hamiltonians $H_{j}$, it is possible to write any unitary as
\begin{equation}
U\approx\cdots U_{1}^{m_{4}}U_{0}^{m_{3}}U_{1}^{m_{2}}U_{0}^{m_{1}%
},\label{qcdecomposition}%
\end{equation}
for some integers $m_{j}$. Under suitable conditions, it was shown that with
$M=\sum_{j}m_{j}=\mathcal{O}(d^{2}\epsilon^{-d})$ it is possible to
approximate any unitary $U$ with a precision $\epsilon$. More precisely, the
conditions are the following

\begin{enumerate}
\item[i)] The Hamiltonians $H_{0}$ and $H_{1}$ are generators of the full Lie
algebra, namely $H_{0}$, $H_{1}$ and their repeated commutators generate all
the elements of su(d).

\item[ii)] The eigenvalues of $U_{0}$ and $U_{1}$ have phases that are
irrationally related to $\pi$.
\end{enumerate}

The decomposition in Eq.~\eqref{qcdecomposition} is a particular case of
Eq.~\eqref{Utheta} where $\theta_{j}$ can only take binary values $\theta
_{j}=0,1$. As such we can write the conditional gates of Eq.~\eqref{condUni}
as in Eq.~\ref{condAB}, which is rewritten below
\begin{equation}
\hat{U}_{j}=\exp\left(  it_{0}H_{0}\otimes|0\rangle_{j}{}_{j}\langle
0|+it_{1}H_{1}\otimes|1\rangle_{j}{}_{j}\langle1|\right)  ,\label{condAB2}%
\end{equation}
for some times $t_{j}$. Channel simulation is then obtained by replacing the
unitary evolution $U$ of Eq.~\eqref{stinespring} with the approximate form in
Eq.~\eqref{qcdecomposition} and its simulation in Eq.~\eqref{Qchan}.
The result is illustrated in Fig.~\ref{fig:PQC3} and described by
the following channel
\begin{equation}
Q_{\pi}(\rho)=\mathrm{Tr}_{\mathbf{R}}\left[  \prod_{j=1}^{N}\hat{U_{j}%
}_{A,R_{0},R_{j}}\left(  \rho_{A}\otimes\pi\right)  \prod_{j=1}^{N}\hat{U_{j}%
}_{A,R_{0},R_{j}}^{\dagger}\right]  ,
\label{Qchan}
\end{equation}
where the program state $\pi$ is defined over $\mathbf{R}=(R_{0},\dots,R_{N})$
and each $\hat{H}_{j}$ acts on the input system $A$ and two ancillary qubits
$R_{0}$ and $R_{j}$. The decomposition of Eq.~\eqref{qcdecomposition} assures
that, with the program
\begin{equation}
|\pi\rangle=|\theta_{0}\rangle\otimes\cdots\otimes|1\rangle^{\otimes m_{2}%
}\otimes|0\rangle^{\otimes m_{1}},
\end{equation}
the product of unitaries approximates $U$ in Eq.~\eqref{stinespring} with
precision $\epsilon$. This is possible in general, provided that the program
state has dimension $\mathcal{O}(d^{2}\epsilon^{-d})$.
However, the channel \eqref{Qchan} is more general, as it allows both quantum
superposition and entanglement.

The processor map $\Lambda$, written in Eq.~\eqref{PQCmap} easily follows 
from this construction~,
while the (non-trace-preserving) dual channel may be written as
\begin{equation}
\Lambda^{\ast}(X)=\langle\Phi_{BA}|\hat{U}_{A\mathbf{R}}^{\dagger}\left(
X_{BA}\otimes\openone_{\mathbf{R}}\right)  \hat{U}_{A\mathbf{R}}|\Phi
_{BA}\rangle.
\end{equation}
This channel requires $2N$ quantum gates at each iteration and can be employed
for the calculation of gradients, following Theorem~\ref{t:gradients}. When we
are interested in simulating a unitary channel $U$ via the quantum fidelity,
then following the results of Section~\ref{Sec5}, the corresponding optimal
program $\tilde{\pi}_{F}$ is simply the eigenvector $\Lambda^{\ast}[|\chi
_{U}\rangle\langle\chi_{U}|]$ with the maximum eigenvalue, where $|\chi_{U}%
\rangle=\openone\otimes U|\Phi\rangle$. Note also that $\Lambda^{\ast}%
[|\chi_{U}\rangle\langle\chi_{U}|]=Z^{\dagger}Z$ where
\begin{equation}
Z=\left(  \langle\chi_{U}|_{BA}\otimes\openone_{\mathbf{R}}\right)  \hat
{U}_{A\mathbf{R}}\left(  |\Phi_{BA}\rangle\otimes\openone_{\mathbf{R}}\right)
,
\end{equation}
so the optimal program $\tilde{\pi}_{F}$ is the principal component of $Z$.
Since there are quantum algorithms for principal component analysis
\cite{lloyd2014quantum}, the optimization may be efficiently performed on a
quantum computer.

\subsubsection{Monotonicity by design}

Two unitaries are sufficient for universality, however such a design may not
be monotonic as a function of $N$ when simulating a given channel. By adding
the identity as a third possible unitary, as in Eq.~\eqref{condABmod},
we get a processor that is monotonic
by design. Note that the identity cannot be one of the initial choices of
unitaries, due to the condition that the eigenvalues of $U_0$ and $U_1$ have
phases that are irrationally related to $\pi$. Because we want a single program
qudit to control the application of one of three unitaries, we use qutrits to
control the gates.

Such a design is more powerful than the original design, because it is
guaranteed to both be monotonic as a function of $N$, where $N$ is the number
of controlled gates, but also to be able to simulate any channel at least as
well as the original $M$-gate processor can, for any $M \leq N$. This is
because a valid program state for the monotonic processor is $\pi_M\otimes
|2\rangle\langle2|^{\otimes (N-M)}$, where $\pi_M$ is any program state for the
original $M$-gate processor. Such a program state would result in the same cost
function that the original processor obtains using $\pi_M$.

Due to having a larger program state space, and more unitaries to choose
from, the $N$-gate monotonic design can also potentially perform not just as
well as but better than any $M$-gate processor using the original design, when
simulating many channels. This comes at the cost of higher dimensionality. The
number of parameters to optimise over scales with order $\mathcal{O}(3^{2N})$
for the monotonic processor, whilst it scales with order $\mathcal{O}(2^{2N})$
for the original design.

\subsection{PQC: Numerical examples}

As an example we study the simulation of an amplitude damping channel, with
Kraus operators in Eq.~\eqref{KrausAD}. A possible Stinespring dilation for
this channel is obtained with $|\theta_{0}\rangle=|0\rangle$ and
\begin{equation}
U=%
\begin{pmatrix}
1 & 0 & 0 & 0\\
0 & \sqrt{1-p} & \sqrt{p} & 0\\
0 & -\sqrt{p} & \sqrt{1-p} & 0\\
0 & 0 & 0 & 1
\end{pmatrix}
=e^{iH_{\mathrm{AD}}},
\end{equation}
where the Hamiltonian is given by
\begin{equation}
H_{\mathrm{AD}}=\frac{\arcsin(\sqrt{p})}{2}(Y\otimes X-X\otimes Y),
\end{equation}
with $X$ and $Y$ being Pauli operators. We may construct a PQC simulation by
taking
\begin{equation}
U_{0}=e^{i\alpha(Y\otimes X-X\otimes Y)},
\end{equation}
for some $\alpha$ and taking $U_{1}$ to be a different unitary that makes the
pair $U_{0}$, $U_{1}$ universal. Here we may choose $\alpha=\sqrt{2}$ and
$U_{1}=e^{iH_{1}}$ with
\begin{equation}
H_{1}=(\sqrt{2}Z+\sqrt{3}Y+\sqrt{5}X)\otimes(Y+\sqrt{2}Z).
\end{equation}
\begin{figure}[t]
\centering
\includegraphics[width=0.9\linewidth]{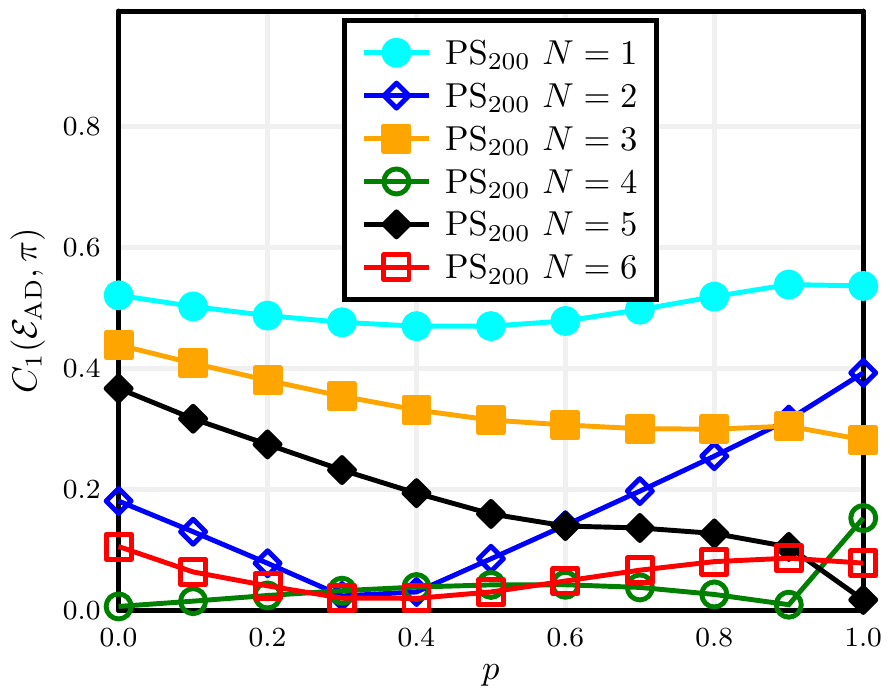} \caption{ PQC
simulation of the amplitude damping channel. Trace distance $C_{1}%
(\mathcal{E}_{\mathrm{AD}},\pi)=\Vert\chi_{\mathcal{E}_{\mathrm{AD}}}%
-\chi_{\pi}\Vert_{1}$ between the target channel's Choi matrix and its PQC
simulation with program state $\pi$, for different numbers of register qubits
$N$. The optimal program is obtained from the minimization of $C_{1}$ via the
projected subgradient (PS) method after 200 iterations. }%
\label{fig:UQC_AD_vs_p}%
\end{figure}

Results are shown in Fig.~\ref{fig:UQC_AD_vs_p}. Compared with the similar PBT
simulation of Fig.~\ref{fig:AD_vs_p}, we observe that the PQC simulation (using the non-monotonic design) displays
a non-monotonic behavior as a function of $N$. PBT with $N$ pairs requires a
register of $2N$ qubits, while PQC requires $N+1$ qubits, namely $N$ qubits
for the conditional gates and an extra auxiliary qubit coming from the Stinespring
decomposition (see Fig.~\ref{fig:PQC3}). We observe that, with a comparable
yet finite register size, PQC can outperform PBT in simulating the amplitude
damping channel. 
\begin{figure}[t]
\centering
\includegraphics[width=0.9\linewidth]{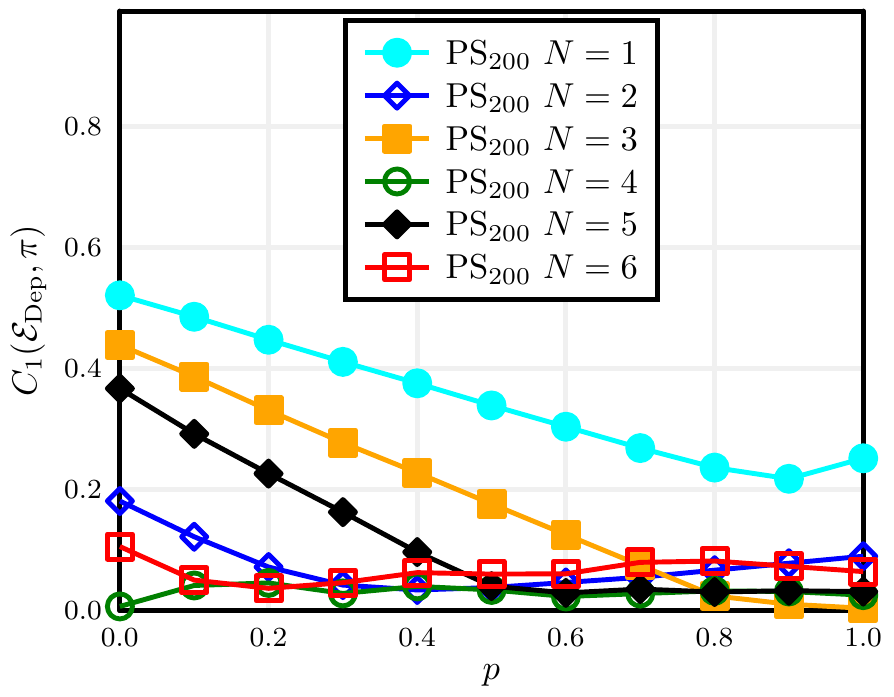} \caption{ PQC
simulation of the depolarizing channel. Trace distance $C_{1}(\mathcal{E}%
_{\mathrm{Dep}},\pi)=\Vert\chi_{\mathcal{E}_{\mathrm{Dep}}}-\chi_{\pi}%
\Vert_{1}$ between the target channel's Choi matrix and its PQC simulation
with program state $\pi$, for different numbers of register qubits $N$. The
optimal program is obtained from the minimization of $C_{1}$ via the projected
subgradient (PS) method after 200 iterations. }%
\label{fig:UQC_Dep_vs_p}%
\end{figure}In Fig.~\ref{fig:UQC_Dep_vs_p} we also study the PQC simulation of
the depolarizing channel for different values of $p$. Although the gates
$U_{0}$ and $U_{1}$ were chosen with inspiration from the Stinespring
decomposition of the amplitude damping channel, those gates are universal and
capable of simulating other channels. Indeed, we observe in
Fig.~\ref{fig:UQC_Dep_vs_p} that a depolarizing channel is already well
simulated with $N=4$ for all values of $p$.

\section{Matrix calculus\label{a:calculus}}

\subsection{Matrix differentiation\label{matrixDIFF}}

For a general overview of these techniques, the reader may consult
Ref.~\cite{stickel1987frechet}. Thanks to Cauchy's theorem, a matrix function
can be written as
\begin{equation}
f(A)=\frac{1}{2\pi i}\int_{\Gamma}d\lambda\,f(\lambda)(\lambda\openone-A)^{-1}%
~.\label{e.f}%
\end{equation}
For the same reason
\begin{equation}
f^{\prime}(A)=\frac{1}{2\pi i}\int_{\Gamma}d\lambda\,f(\lambda)(\lambda
\openone-A)^{-2}~.\label{e.fp}%
\end{equation}
Applying a basic rule of matrix differentiation, $d(A^{-1})=-A^{-1}(dA)A^{-1}$
we obtain
\begin{equation}
df(A)=\frac{1}{2\pi i}\int_{\Gamma}d\lambda\,f(\lambda)(\lambda
\openone-A)^{-1}dA(\lambda\openone-A)^{-1}~.\label{e.df}%
\end{equation}
Clearly, $df(A)=f^{\prime}(A)dA$ only when $[A,dA]=0$. In general $df(A)$ is a
superoperator that depends on $A$ and is applied to $dA$. The explicit form is
easily computed using the eigenvalue decomposition or other techniques
\cite{stickel1987frechet}. Note that in some cases the expressions are simple.
Indeed, using the cyclic invariance of the trace, we have
\begin{equation}
d\text{\textrm{Tr}}[f(A)]=\text{\textrm{Tr}}[f^{\prime}(A)dA],\label{e:nablag}%
\end{equation}
while in general $d$\textrm{Tr}$[Bf(A)]\neq$\textrm{Tr}$[Bf^{\prime}(A)dA]$.

\subsection{Differential of the quantum fidelity\label{fidelityDIFF}}

The quantum fidelity can be expanded as
\begin{align}
F(X,Y) &  =\text{\textrm{Tr}}\sqrt{\sqrt{X}Y\sqrt{X}}\\
&  =\frac{1}{2\pi i}\int_{\Gamma}d\lambda\,\sqrt{\lambda}\text{\textrm{Tr}%
}[(\lambda\openone-\sqrt{X}Y\sqrt{X})^{-1}]~,\nonumber
\end{align}
where in the second line we have applied Eq.~\eqref{e.f}. Taking the
differential with respect to $Y$ and using the cyclic property of the trace we
get%
\begin{align}
d_{Y}F &  :=F(X,Y+dY)-F(X,Y)\nonumber\\
&  \overset{(1)}{=}\frac{1}{2\pi i}\int_{\Gamma}d\lambda\,\sqrt{\lambda
}\text{\textrm{Tr}}[(\lambda\openone-\sqrt{X}Y\sqrt{X})^{-2}\sqrt{X}dY\sqrt
{X}]\nonumber\\
&  \overset{(2)}{=}\frac{1}{2}\text{\textrm{Tr}}[(\sqrt{X}Y\sqrt{X}%
)^{-\frac{1}{2}}\sqrt{X}dY\sqrt{X}]\nonumber\\
&  \overset{(3)}{=}\frac{1}{2}\text{\textrm{Tr}}[\sqrt{X}(\sqrt{X}Y\sqrt
{X})^{-\frac{1}{2}}\sqrt{X}\;dY]~,\label{dFdY}%
\end{align}
where in (1) we use Eq.~\eqref{e.df} and the cyclic property of the trace; in
(2) we use Eq.~\eqref{e.fp} with $f(\lambda)=\sqrt{\lambda}$, so $f^{\prime
}(\lambda)=\frac{1}{2}\lambda^{-1/2}$; and in (3) we use the cyclic property
of the trace. See also Lemma 11 in \cite{coutts2018certifying}.

\subsection{Differential of the trace distance\label{traceDIFF}}

The trace norm for a Hermitian operator $X$ is defined as%
\begin{align}
t(X)  &  =\Vert X\Vert_{1}:=\text{\textrm{Tr}}\sqrt{X^{\dagger}X}%
=\text{\textrm{Tr}}[\sqrt{X^{2}}]\nonumber\\
&  =\frac{1}{2\pi i}\int_{\Gamma}d\lambda\,\sqrt{\lambda}\text{\textrm{Tr}%
}[(\lambda\openone-XX)^{-1}]~,
\end{align}
where in the second line we applied Eq.~\eqref{e.f}. From the spectral
decomposition $X=U\lambda U^{\dagger}$, we find $t(X)=\sum_{j} |\lambda_{j}|$,
so the trace distance reduces to the absolute value function for
one-dimensional Hilbert spaces. The absolute value function $|\lambda|$ is
differentiable at every point, except $\lambda=0$. Therefore, for any
$\lambda\neq0$, the subgradient of the absolute value function is composed of only its
gradient, i.e.
\begin{equation}
\partial|\lambda| = \{ \mathrm{sign}(\lambda) \} ~~~ \mathrm{for} ~~
\lambda\neq0~. \label{dabsvaln}%
\end{equation}
For $\lambda=0$ we can use the definition \eqref{subgradient} to write
\begin{equation}
\partial|\lambda|_{\lambda=0} = \{z: |\sigma|\ge z \sigma\mathrm{~~ for~
all~}\sigma\}~,
\end{equation}
which is true iff $-1\le z\le1$. Therefore,
\begin{equation}
\partial|\lambda|_{\lambda=0} = [-1,1]~.
\end{equation}
The sign function in \eqref{dabsvaln} can be extended to $\lambda=0$ in
multiple ways (common choices are $\mathrm{sign}(0)=-1,0,1$). From the above
equation, it appears that for any extension of the sign function, provided
that $\mathrm{sign}(0)\in[-1,1]$ we may write the general form
\begin{equation}
\mathrm{sign}(\lambda) \in\partial|\lambda|~, \label{signinabs}%
\end{equation}
which is true for any value of $\lambda$.

With the same spirit we extend the above argument to any matrix dimension,
starting from the case where $X$ is an invertible operator (no zero
eigenvalues). Taking the differential with respect to $X$ we find%
\begin{align}
dt(X) &  :=t(X+dX)-t(X)=\nonumber\\
&  \overset{(1)}{=}\frac{1}{2\pi i}\int_{\Gamma}d\lambda\,\sqrt{\lambda
}\text{\textrm{Tr}}[(\lambda\openone-X^{2})^{-2}(X(dX)+(dX)X)]\nonumber\\
&  \overset{(2)}{=}\frac{1}{2}\text{\textrm{Tr}}[(X^{2})^{-\frac{1}{2}%
}(X(dX)+(dX)X)]\nonumber\\
&  \overset{(3)}{=}\text{\textrm{Tr}}[(X^{2})^{-\frac{1}{2}}X\;(dX)]
\end{align}
where in (1) we use Eq.~\eqref{e.df}, the cyclic property of the trace and the
identity $dX^{2}=X(dX)+(dX)X$; in (2) we use Eq.~\eqref{e.fp} with
$f(\lambda)=\sqrt{\lambda}$, so $f^{\prime}(\lambda)=\frac{1}{2}\lambda
^{-1/2}$; and in (3) we use the cyclic property of the trace and the
commutation of $X$ and $\sqrt{X^{2}}$. Let
\begin{equation}
X=\sum_{k}\lambda_{k}P_{k}~,
\end{equation}
be the eigenvalue decomposition of $X$ with eigenvalues $\lambda_{k}$ and
eigenprojectors $P_{k}$. For non-zero eigenvalues we may write
\begin{equation}
(X^{2})^{-\frac{1}{2}}X=\sum_{k}\mathrm{sign}(\lambda_{k})P_{k}=:\mathrm{sign}%
(X)~,
\end{equation}
and accordingly
\begin{align}
dt(X) &  :=\Vert X+dX\Vert_{1}-\Vert X\Vert_{1}\nonumber\label{dtdX}\\
&  =\sum_{k}\mathrm{sign}(\lambda_{k})\text{\textrm{Tr}}[P_{k}\;dX]~.
\end{align}
Therefore, for invertible operators we may write
\[
\partial t(X)=\{\nabla t(X)\}~,\nabla t(X)=\mathrm{sign}(X)~.
\]
We now consider the general case where some eigenvalues of $X$ may be zero. We
do this by generalizing Eq.~\eqref{signinabs}, namely we show that even if
$\partial t(X)$ may contain multiple elements, it is always true that $\nabla
t\in\partial t$, provided that $-\openone\leq\mathrm{sign}(X)\leq\openone$.
Following \eqref{subgradient} we may write, for fixed $X$ and arbitrary $Y$,%
\begin{gather}
t(Y)-t(X)-\text{\textrm{Tr}}[\nabla t(X)(Y-X)]\nonumber\\
\overset{(1)}{=}t(Y)-t(X)-\text{\textrm{Tr}}[\nabla t(X)Y]+t(X)\nonumber\\
\overset{(2)}{\geq}t(Y)-\text{\textrm{Tr}}[Y]=\sum_{j}(|\lambda_{j}%
|-\lambda_{j})\geq0~,\label{tsubdiffineq}%
\end{gather}
where in (1) we use the property $\Vert X\Vert_{1}=\Tr[{\rm sign}(X)X]$ and in
(2) we use the assumption $-\openone\leq\mathrm{sign}(X)\leq\openone$.
From the definition of the subgradient \eqref{subgradient}, the above equation
shows that $\mathrm{sign}(X)\in\partial t(X)$, so we may always use $\nabla
t(X)=\mathrm{sign}(X)$ in the projected subgradient algorithm
\eqref{projsubgrad}.

\section{Smoothing techniques\label{SmoothAPP}}

\subsection{Stochastic smoothing\label{a:stocsmooth} }

The conjugate gradient algorithm converges after $\mathcal{O}(c/\epsilon)$
steps \cite{jaggi2011convex,jaggi2013revisiting}, where $\epsilon$ is the
desired precision and $c$ is a curvature constant that depends on the
function. However, it is known that $c$ could diverge for non-smooth
functions. This is the case for the trace norm, as shown in Example 0.1 in
\cite{ravi2017deterministic}.

A general solution, valid for arbitrary functions, is stochastic smoothing
\cite{yousefian2012stochastic}. In this approach the non-smooth function
$C(\pi)$ is replaced by the average
\begin{equation}
C_{\eta}(\pi)=\mathbb{E}_{\sigma}[C(\pi+\eta\sigma)]~.
\end{equation}
where $\sigma$ is such that $\Vert\sigma\Vert_\infty\leq1$. If $|C(x)-C(y)|\leq
M\Vert x-y\Vert_{\infty}$, then
\begin{equation}
C(\pi)\leq C_{\eta}(\pi)\leq C(\pi)+M\eta~,
\end{equation}
so that $C_{\eta}(\pi)$ provides a good approximation for $C(\pi)$. Moreover,
$C_{\eta}$ is differentiable at any point, so we may apply the conjugate
gradient algorithm. A modified conjugate gradient algorithm with adaptive
stochastic approximation was presented in Ref.~\cite{lan2013complexity}. At
each iteration $k$ the algorithm reads
\[%
\begin{array}
[c]{l}%
1)~\text{\textrm{Sample some operators }}\sigma_{1},\dots,\sigma_{k},\\
2)~\text{\textrm{Evaluate }}\bar{g}_{k}=\frac{1}{k}\sum_{j=1}^{k}g(\pi
_{k}+\eta_{k}\sigma_{j})\text{\textrm{ for }}\eta_{k}\propto k^{-1/2},\\
3)~\text{\textrm{Find the smallest eigenvalue }}\left\vert \sigma
_{k}\right\rangle \text{\textrm{ of~}}\bar{g}_{k},\\
4)~\pi_{k+1}=\frac{k}{k+2}\pi_{k}+\frac{2}{k+2}\left\vert \sigma
_{k}\right\rangle \left\langle \sigma_{k}\right\vert .
\end{array}
\]
where $g$ denotes any element of the subgradient $\partial C$. The above
algorithm converges after $\mathcal{O}(\epsilon^{2})$ iterations. Since
Eqs.~\eqref{fidelitygrad} and \eqref{traceDproof} provide an element of the
subgradient, the above algorithm can be applied to both fidelity and trace
distance. However, this algorithm requires $k$ evaluations of the subgradient
to perform the averages, so it may be impractical when the number of
iterations get larger. In the following we study an alternative that does not
require any average.

\subsection{Nesterov's smoothing\label{a:smoothtrace}}

An alternative smoothing scheme is based on Nesterov's dual
formulation~\cite{nesterov2005smooth}. Suppose that the non-smooth objective
function $f$ admits a dual representation as follows
\begin{equation}
f(x)=\sup_{y}[\langle x,y\rangle-g(y)],
\end{equation}
for some inner product $\langle\cdot,\cdot\rangle$. Nesterov's approximation
consists of adding a strongly convex function $d$ to the dual
\begin{equation}
f_{\mu}(x)=\sum_{y}[\langle x,y\rangle-g(y)-\mu d(y)].
\end{equation}
The resulting $\mu$-approximation is smooth and satisfies
\begin{equation}
f_{\mu}(x)\leq f(x)\leq f_{\mu}(x)+\mu\sup_{y}d(y).\label{fmuineq}%
\end{equation}

The trace norm admits the dual representation~\cite{watrous2018theory}
\begin{equation}
t(X)=\Vert X\Vert_{1}=\sup_{\Vert Y\Vert_\infty\leq1}\langle Y,X\rangle,
\end{equation}
where $\langle Y,X\rangle$ is the Hilbert Schmidt product. This can be
regularized with any strongly convex function $d$. A convenient choice
\cite{liu2013tensor} that enables an analytic solution is $d(X)=\frac
{1}{2}\Vert X\Vert_{2}^{2}:=\frac{1}{2}\langle X,X\rangle$, so
\begin{equation}
t_{\mu}(X)=\max_{\Vert Y\Vert_\infty\leq1}\left[  \langle Y,X\rangle-\frac{\mu}%
{2}\Vert Y\Vert_{2}^{2}\right]  .\label{tmu}%
\end{equation}
This function is smooth and its gradient is given by \cite{liu2013tensor}%
\begin{align*}
\nabla t_{\mu}(X) &  =\argmax_{\Vert Y\Vert_\infty\leq1}\left[  \langle
Y,X\rangle-\frac{\mu}{2}\Vert Y\Vert_{2}^{2}\right]  \\
&  =\argmin_{\Vert Y\Vert_\infty\leq1}\Vert\mu Y-X\Vert_{2}^{2}=U\Sigma_{\mu
}V^{\dagger},
\end{align*}
where $X=U\Sigma V^{\dagger}$ is the singular value decomposition of $X$ and
$\Sigma_{\mu}$ is a diagonal matrix with diagonal entries $(\Sigma_{\mu}%
)_{i}=\min\{\Sigma_{i}/\mu,1\}$. Plugging this into Eq.~\eqref{tmu} we get
\begin{equation}
t_{\mu}(X)=\Tr\left[  \Sigma_{\mu}\left(  \Sigma-\frac{\mu}{2}\Sigma_{\mu
}\right)  \right]  .\label{e.tmu}%
\end{equation}
For a diagonalizable matrix $X$ with spectral decomposition $X=U\lambda
U^{\dagger}$, the singular value decomposition is obtained with $\Sigma
=|\lambda|$ and $V=U\mathrm{sign}(\lambda)$. Inserting these expressions in
\eqref{e.tmu} we find
\begin{equation}
t_{\mu}(X)=\sum_{j}h_{\mu}(\lambda_{j})=\Tr[h_\mu(X)],
\end{equation}
where $h_{\mu}$ is the so called Huber penalty function
\begin{equation}
h_{\mu}(x)=%
\begin{cases}
\frac{x^{2}}{2\mu} & \mathrm{~if~}|x|<\mu,\\
|x|-\frac{\mu}{2} & \mathrm{~if~}|x|\geq\mu.
\end{cases}
\end{equation}
The gradient $\nabla t_{\mu}$ is then $h_{\mu}^{\prime}(X)\equiv Uh^{\prime
}(\lambda)U^{\dagger}$, where
\begin{equation}
h_{\mu}^{\prime}(x)=%
\begin{cases}
\frac{x}{\mu} & \mathrm{~if~}|x|<\mu,\\
\mathrm{sign}(x) & \mathrm{~if~}|x|\geq\mu.
\end{cases}
\end{equation}

We then find that,
via the smooth trace norm $t_{\mu}$, we can define the smooth trace distance of
Eq.~\eqref{Dmu} that is differentiable at every point
\begin{equation}
C_{\mu}(\pi)=\Tr\left[  h_{\mu}\left(  \chi_{\pi}-\chi_{\mathcal{E}}\right)
\right]  .
\end{equation}
Thanks to the inequalities in \eqref{fmuineq}, the smooth trace distance
bounds the cost $C_{1}$ as
\begin{equation}
C_{\mu}(\pi)\leq C_{1}(\pi)\leq C_{\mu}(\pi)+\frac{\mu d}{2},
\end{equation}
where we employed the identity $\sup_{\Vert Y\Vert_\infty\leq1}\Vert Y\Vert^2_{2}\le d$ to
get the upper bound. Moreover, we find the following

\begin{lemma}
The smooth trace distance, defined in Eq.~\eqref{Dmu}, is a convex function of
$\pi$.
\end{lemma}

\noindent\textit{Proof.}~~From the definition and Eq.~\eqref{tmu} we find%
\begin{align}
C_{\mu}(\pi)  & =t_{\mu}\left[  \Lambda(\pi)-\chi_{\mathcal{E}}\right]
\nonumber\\
& =\max_{\Vert Y\Vert_\infty\leq1}\left[  \langle Y,\Lambda(\pi)-\chi_{\mathcal{E}%
}\rangle-\frac{\mu}{2}\Vert Y\Vert_{2}^{2}\right]  .
\end{align}
Now for $\bar{\pi}=p\pi_{1}+(1-p)\pi_{2}$ linearity implies $f(\bar{\pi
}):=\langle Y,\Lambda(\bar{\pi})-\chi_{\mathcal{E}}\rangle=pf(\pi
_{1})+(1-p)f(\pi_{2})$. Therefore%
\begin{align}
C_{\mu}(\bar{\pi})  & =\max_{\Vert Y\Vert_\infty\leq1}\left[  pf(\pi_{1}%
)+(1-p)f(\pi_{2})-\frac{\mu}{2}\Vert Y\Vert_{2}^{2}\right]  \nonumber\\
& \leq p\max_{\Vert Y\Vert_\infty\leq1}\left[  \langle Y,\Lambda(\pi_{1}%
)-\chi_{\mathcal{E}}\rangle-\frac{\mu}{2}\Vert Y\Vert_{2}^{2}\right]
\nonumber\\
& +(1-p)\max_{\Vert Z\Vert_\infty\leq1}\left[  \langle Z,\Lambda(\pi_{2}%
)-\chi_{\mathcal{E}}\rangle-\frac{\mu}{2}\Vert Z\Vert_{2}^{2}\right]
\nonumber\\
& =pC_{\mu}(\pi_{1})+(1-p)C_{\mu}(\pi_{2}),
\end{align}
showing the convexity.~$\blacksquare$

Then, using the definitions from \cite{nesterov2005smooth}, the following
theorem bounds the growth of the gradient

\begin{theorem}
The gradient of the smooth trace norm is Lipschitz continuous
with Lipschitz constant
\begin{equation}
L=\frac{d}{\mu}.
\end{equation}
\end{theorem}

In particular, if the gradient is Lipschitz continuous, the smooth trace norm
satisfies the following inequality for any state $\pi,\sigma$
\begin{equation}
C_{\mu}(\sigma)\leq C_{\mu}(\pi)+\langle\nabla C_{\mu}(\pi),\sigma-\pi
\rangle+\frac{L}{2}\Vert\sigma-\pi\Vert_2^{2}.
\end{equation}
\textit{Proof.}~~Given the linearity of the quantum channel $\Lambda$, we can
apply theorem 1 from \cite{nesterov2005smooth} to find
\begin{equation}
L=\frac{1}{\mu}\sup_{\Vert x\Vert_2=1,\Vert y\Vert_2=1}\langle y,\Lambda
(x)\rangle.
\end{equation}
Since all eigenvalues of $y$ are less than or equal to 1, we can write $y\leq1$
and as such
\begin{equation}
L\leq\frac{1}{\mu}\sup_{\Vert x\Vert_2=1}\Tr[\Lambda(x)]=\frac{1}{\mu}%
\sup_{\Vert x\Vert_2=1}\Tr[x]\leq\frac{d}{\mu}.~
\end{equation}
\hfill$\blacksquare$

\section{PBT: program state compression\label{compressSEC}}

The dimension of the program state grows exponentially with the number of ports $N$ as $d^{2N}$
where $d$ is the dimension of the Hilbert space. However, as also discussed in
the original proposal~\cite{ishizaka2008asymptotic,ishizaka2009quantum} and more recently in
Ref.~\cite{christandl2018asymptotic}, the resource state of PBT can be chosen
with extra symmetries, so as to reduce the number of free parameters. In
particular, we may consider the set of program states that are symmetric under
the exchange of ports, i.e., such that rearranging any $A$ modes and the
corresponding $B$ modes leaves the program state unchanged.

Let $P_{s}$ be the permutation operator swapping labels 1 to $N$ for the
labels in the sequence $s$, which contains all the numbers 1 to $N$ once each
in some permuted order. Namely $P_{s}$ exchanges all ports according to the
rule $i\mapsto s_{i}$. Since PBT is symmetric under exchange of ports, we may
write%
\begin{equation}
\mathcal{P}_{P_{s}\pi P_{s}^{\dag}}(\rho)=\mathcal{P}_{\pi}\left(
\rho\right)  ~\text{for any }s\text{.}%
\end{equation}
Consider then an arbitrary permutation-symmetric resource state $\pi
_{\mathrm{sym}}$ as
\[
\pi_{\mathrm{sym}}=\frac{1}{N!}\sum_{s}P_{s}\pi P_{s}^{\dag},
\]
where the sum is over all possible sequences $s$ that define independent
permutations and $N!$ is the total number of possible permutations. Clearly
$\mathcal{P}_{\pi_{\mathrm{sym}}}=\mathcal{P}_{\pi}$, so any program state
gives the same PBT channel as some symmetric program state. It therefore
suffices to consider the set of symmetric program states. This is a convex
set: any linear combination of symmetric states is a symmetric state.

To construct a basis of the symmetric space, we note that each element of a
density matrix is the coefficient of a dyadic (of the form $\left\vert
x\right\rangle \left\langle y\right\vert $). If permutation of labels maps one
dyadic to another, the coefficients must be the same. This allows us to
constrain our density matrix using fewer global parameters. For instance, for
$d=2$ we can define the 16 parameters $n_{00,00}$, $n_{00,01}$, $n_{00,10}$,
etc., corresponding to the number of ports in the dyadic of the form
$\left\vert 0_{A}0_{B}\right\rangle \left\langle 0_{A}0_{B}\right\vert $,
$\left\vert 0_{A}0_{B}\right\rangle \left\langle 0_{A}1_{B}\right\vert $,
$\left\vert 0_{A}0_{B}\right\rangle \left\langle 1_{A}0_{B}\right\vert $, etc.
Each element of a symmetric density matrix can then be defined solely in terms
of these parameters, i.e., all elements corresponding to dyadics with the same
values of these parameters have the same value.

For the general qudit case, in which our program state consists of $N$ ports,
each composed of two $d$-dimensional qudits, we can find the number of
independent parameters from the number of independent dyadics. Each port in a
dyadic can be written as $|a_{A},b_{B}\rangle\langle c_{A},d_{B}|$ where the
extra indices $A$ and $B$ describe whether those states are modeling either
qudit $A$ or $B$. There are $d^{4}$ different combinations of $\{a,b,c,d\}$,
so we can place each qudit into one of $d^{4}$ categories based on these
values. If two elements in the density matrix correspond to dyadics with the
same number of ports in each category, they must take the same value. Hence,
the number of independent coefficients is given by the number of ways of
placing $N$ (identical) ports into $d^{4}$ (distinguishable) categories.
This is exactly the binomial coefficient
\begin{equation}
\binom{N+d^{4}-1}{d^{4}-1}=\mathcal{O}(N^{d^{4}-1})~.
\end{equation}
Consequently, exploiting permutation symmetry of the PBT\ protocol, we can
\textit{exponentially} reduce the number of parameters for the optimization
over program states.

The number of parameters can be reduced even further by considering products
of Choi matrices. We may focus indeed on the Choi set, first defined 
in Eq.~\eqref{Choispace}, 
\begin{equation}
\mathcal{C}_{N}=\left\{  \pi:\pi=\sum_{k}p_{k}\chi_{k}^{\otimes N}\right\}
~,\label{Cset}%
\end{equation}
where each $\chi_{k}=\chi_{AB}^{k}$ is a generic Choi matrix, therefore
satisfying $\mathrm{Tr}_{B}\chi_{k}=d^{-1}\openone$, and $p_{k}$ form a
probability distribution. Clearly $\mathcal{C}$ is a convex set. We now show
that this set can be further reduced to just considering $N=1$.

%
When the program state $\pi=\chi^{\otimes N}$ is directly used in the PBT 
protocol we find 
\begin{align}
\Lambda(\pi) &  =\sum_{i=1}^{N}\mathrm{Tr}_{\mathbf{A\bar{B}}_{i}C}\left[
\Pi_{i}\left(  \chi_{AB}^{\otimes N}\otimes\Phi_{DC}\right)  \right]
_{B_{i}\rightarrow B_{\mathrm{out}}}\\
&  =\frac{1}{d^{N-1}}\sum_{i=1}^{N}\mathrm{Tr}_{A_{i}C}\left[  \Pi_{i}\left(
\chi_{A_{i}B_{\mathrm{out}}}\otimes\Phi_{DC}\right)  \right]  \\
&  :=\tilde{\Lambda}(\chi)~,\label{e:LambdaChoi}%
\end{align}
namely that the optimization can be reduced to the $\mathcal{O}(d^{4}$)
dimensional space of Choi matrices $\chi$. Note that, in the above equation,
we used the identity
\begin{equation}
\mathrm{Tr}_{\mathbf{\bar{B}}_{i}}\chi_{AB}^{\otimes N}=\chi_{A_{i}B_{i}%
}\otimes\frac{\openone_{\mathbf{\bar{A}}_{i}}}{d^{N-1}},
\end{equation}
where $\mathbf{\bar{A}}_{i}=\mathbf{A}\backslash A_{i}$.

Now let $\pi$ be a linear combination of tensor products of Choi matrix
states, $\chi_{k}^{\otimes N}$, each with probability $p_{k}$ as in
Eq.~(\ref{Cset}). Then we can write
\begin{align}
\mathrm{Tr}_{\mathbf{\bar{B}}_{i}}\pi_{AB} &  =\mathrm{Tr}_{\mathbf{\bar{B}%
}_{i}}\sum_{k}p_{k}\chi_{k}^{\otimes N}\\
&  =\sum_{k}p_{k}\left(  \chi_{A_{i}B_{i}}^{k}\otimes\frac
{\openone_{\mathbf{\bar{A}}_{i}}}{d^{N-1}}\right)  .
\end{align}
However, this is precisely the partial trace over the tensor product
$\chi^{\prime\otimes N}$ of some other Choi matrix $\chi^{\prime}=\sum
_{k}p_{k}\chi_{k}$. Hence, the program state $\pi=\sum_{k}p_{k}\chi
_{k}^{\otimes N}$ simulates the same channel as the resource state
$\pi^{\prime}=\left(  \sum_{k}p_{k}\chi_{k}\right)  ^{\otimes N}$.

Therefore, the optimization over the convex set $\mathcal{C}_{N}$ can be
reduced to the optimization over products of Choi matrices $\chi^{\otimes N}$.
From Eq.~\eqref{e:LambdaChoi} this can be further reduced to the optimization
of the quantum channel $\tilde{\Lambda}$ over the convex set of single-copy
Choi matrices $\chi$
\begin{equation}
\mathcal{C}_{1}=\left\{  \pi:\pi=\chi_{AB},~\mathrm{Tr}_{B}\chi_{AB}%
=\openone/2\right\}  ,
\end{equation}
which is $\mathcal{O}(d^{4})$. Using $\mathcal{C}_{1}$ drastically reduces the
difficulty of numerical simulations, thus allowing the exploration of
significantly larger values of $N$. 



Finally, we provide an explicit expression for the reduced map $\tilde{\Lambda}%
$\ of Eq.~(\ref{e:LambdaChoi}) in the case of qubits. For $d=2$ we can rewrite
PBT in a language that can be more easily formulated from representations of
SU(2). For simplicity of notation, here we do not use bold letters for
vectorial quantities.
Let us modify the POVM in Eq.~\eqref{PBTchan} as
\begin{align}
\tilde{\Pi}_{i} &  =\sigma_{AC}^{-1/2}\Psi_{A_{i}C}^{-}{\sigma}_{AC}^{-1/2},\\
\sigma_{AC} &  =\sum_{i=1}^{N}\Psi_{A_{i}C}^{-},\\
\Pi_{i} &  =\tilde{\Pi}_{i}+\Delta,\\
\Delta &  =\frac{1}{N}\left(  \openone-\sum_{j}\tilde{\Pi}_{j}\right)
,\label{DeltaOp}%
\end{align}
where $|\Psi^{-}\rangle=(|01\rangle-|10\rangle)/\sqrt{2}$ is a singlet state.
For $\pi=\chi^{\otimes n}$ the quantum channel is simplified. In fact, since
$\Tr_{B}\chi=\openone/2$, we may write
\begin{align}
\mathcal{P}_{\pi} &  =\sum_{i=1}^{N}\frac{1}{2^{N-1}}\mathrm{Tr}_{AC}\left[
\sqrt{\Pi_{i}}\left(  \rho_{C}\otimes\chi_{A_{i}B}\otimes\openone_{\bar{A}%
_{i}}\right)  \sqrt{\Pi_{i}}\right]  \nonumber\\
&  =\sum_{\ell}K_{\ell}^{0}(\rho_{C}\otimes\chi)K_{\ell}^{0}{}^{\dagger}%
+\sum_{\ell^{\prime}}K_{\ell}^{1}(\rho_{C}\otimes\chi)K_{\ell}^{1}{}^{\dagger
},\label{PBTasd}%
\end{align}
where $\ell$ and $\ell^{\prime}$ are multi-indices and, in defining the Kraus
operators, we have separated the contributions from $\tilde{\Pi}_{i}$ and
$\Delta$ (see below).

In order to express these operators, we write
\begin{equation}
|\psi_{CA_{i}}^{-}\rangle\!\langle\psi_{CA_{i}}^{-}|=\frac{\openone-\vec
{\sigma}_{C}\cdot\vec{\sigma}_{A_{i}}}{4},
\end{equation}
so that%
\begin{align}
\sigma_{AC}  & =\sum_{i=1}^{N}|\psi_{CA_{i}}^{-}\rangle\!\langle\psi_{CA_{i}%
}^{-}|=\frac{N}{4}-\vec{S}_{C}\cdot\vec{S}_{A}\nonumber\\
& =\frac{N}{4}-\frac{\vec{S}_{\text{\textrm{tot}}}^{2}-\vec{S}_{C}^{2}-\vec
{S}_{A}^{2}}{2},
\end{align}
where $\vec{S}=\vec{\sigma}/2$ is a vector of spin operators, $\vec{S}%
_{A}=\sum_{j}\vec{S}_{A_{j}}$ and $\vec{S}_{\mathrm{tot}}=\vec{S}_{C}+\vec
{S}_{A}$. The eigenvalues of $\sigma_{AC}$ are then obtained from the
eigenvalues of the three commuting Casimir operators
\begin{equation}
\lambda(s_{A})=\frac{N}{4}-\frac{S_{\mathrm{tot}}(S_{\mathrm{tot}}%
+1)-s_{A}(s_{A}+1)-3/4}{2}~,
\end{equation}
where $S_{\mathrm{tot}}=s_{A}\pm1/2$.

Substituting the definition of $S_{\mathrm{tot}}$, we find two classes of
eigenvalues
\begin{equation}
\lambda^{+}(s_{A})=\frac{N-2s_{A}}{4}~,\lambda^{-}(s_{A})=\frac{N+2s_{A}+2}%
{4}~,\label{eigv2}%
\end{equation}
with corresponding eigenvectors
\begin{equation}
|\pm,s_{A},M,\alpha\rangle=\sum_{k,m}\Gamma_{s_{A}\pm\frac{1}{2},s_{A}%
}^{M,m,k}|k\rangle_{C}|s_{A},m,\alpha\rangle_{A}~,\label{esigma}%
\end{equation}
where $-\frac{N+1}{2}\leq M\leq\frac{N+1}{2}$, $\alpha=1,\dots,g^{[N]}(s)$
describes the degeneracy, $g^{[N]}(s)$ is the size of the degenerate subspace,
and
\begin{equation}
\Gamma_{S,s}^{M,m,k}=\langle S,M;s,1/2|1/2,1/2-k;s,m\rangle\label{CG}%
\end{equation}
are Clebsch-Gordan coefficients.

Note that the Clebsch-Gordan coefficients define a unitary transformation
between the two bases $|s_{1},m_{1};s_{2};m_{2}\rangle$ and $|S,M;s_{1}%
,s_{2}\rangle$. From the orthogonality relations, of these coefficients we find
the equalities%
\begin{align}
\sum_{S,M}\Gamma_{S,s}^{M,m,i}\Gamma_{S,s}^{M,m^{\prime},i^{\prime}}  &
=\delta_{i,i^{\prime}}\delta_{m,m^{\prime}},\\
\sum_{m,i}\Gamma_{S,s}^{M,m,i}\Gamma_{S^{\prime},s}^{M^{\prime},m,i}  &
=\delta_{M,M^{\prime}}\delta(S,S^{\prime},s),
\end{align}
where $\delta(S,S^{\prime},s)=1$ iff $S=S^{\prime}$ and $|s-1/2|\leq S\leq
s+1/2$. The eigenvalues in Eq.~\eqref{eigv2} are zero iff $S_{\mathrm{tot}%
}=S_{A}+1/2$ and $S_{A}=N/2$. These eigenvalues have degeneracy
$2S_{\mathrm{tot}}+1=N+2$ and the corresponding eigenvectors are
\begin{equation}
|\perp,M,\alpha\rangle=|+,N/2,M,\alpha\rangle~.
\end{equation}
Thus, the operator $\Delta$ from Eq.~\eqref{DeltaOp} may be written as
\begin{equation}
\Delta=\frac{1}{N}\sum_{M=-\frac{N+1}{2}}^{\frac{N+1}{2}}\sum_{\alpha}%
|\perp,M,\alpha\rangle\!\langle\perp,M,\alpha|~.
\end{equation}

To finish the calculation we need to perform the partial trace over all spins
except those in port $i$.
We use $s_{\bar{A}_{i}}$, $m_{\bar{A}_{i}}$ and $\alpha_{i}$ to model the
state of the total spin in ports ${A_{j}}$ with $j\neq i$. These refer to the
value of total spin and the projection along the $z$ axis, as well as the
degeneracy. Moreover, since $S_{\bar{A}_{i}}$ commutes with both $S_{A}^{2}$
and $S_{A}^{z}$, we may select a basis for the degeneracy that explicitly
contains $s_{\bar{A}_{i}}$. We may write then $\alpha=(s_{\bar{A}_{i}}%
,\tilde{\alpha}_{i})$ where $\tilde{\alpha}_{i}$ represents some other degrees
of freedom.

With the above definitions, when we insert several resolutions of the identity
in Eq.~\eqref{PBTasd}, we may write the Kraus operators as \begin{widetext}
\begin{align}
    K^0_{i,s_{\bar{A}_i},m_{\bar{A}_i},\alpha_i,s'_{\bar{A}_i},m'_{\bar{A}_i},\alpha'_i} &=
    2^{-\frac{N-1}2}
    \bra{s_{\bar{A}_i},m_{\bar{A}_i},\alpha_i}\otimes \bra{\psi^-_{A_iC}}\sigma_{AC}^{-1/2}
    \ket{s'_{\bar{A}_i},m'_{\bar{A}_i},\alpha'_i}
    \cr \nonumber &=
    2^{-\frac{N-1}2} \sum_{\pm,s_A,M,\alpha} \lambda_\pm(s_A)^{-1/2}
\bra{\psi^-_{A_iC}}
    \bra{s_{\bar{A}_i},m_{\bar{A}_i},\alpha_i}  {\pm,s_A,M,\alpha}\rangle \!
    \bra{\pm,s_A,M,\alpha} {s'_{\bar{A}_i},m'_{\bar{A}_i},\alpha'_i}\rangle,
\\
    K^1_{i,M,\alpha,s'_{\bar{A}_i},m'_{\bar{A}_i},\alpha'_i} &=
    2^{-\frac{N-1}2} N^{-1/2}
    \bra{+,N/2,M,\alpha} {s'_{\bar{A}_i},m'_{\bar{A}_i},\alpha'_i}\rangle,
\end{align}
\end{widetext}where each set of states $|s_{\bar{A}_{i}},m_{\bar{A}_{i}%
},\alpha_{i}\rangle$ represents a basis of the space corresponding to all ports
$j$ with $j\neq i$. To simplify the Kraus operators we study the overlap%
\begin{align}
& \langle s_{\bar{\imath}},m_{\bar{\imath}},\alpha_{i}|{\pm,S,M,\alpha}%
\rangle\nonumber\\
& =\sum_{k,m}|k\rangle_{C}\langle s_{\bar{\imath}},m_{\bar{\imath}},\alpha
_{i}|\Gamma_{S\pm\frac{1}{2},S}^{M,m,k}|S,m,\alpha\rangle_{A}\nonumber\\
& =\sum_{k,m}|k\rangle_{C}\langle s_{\bar{\imath}},m_{\bar{\imath}},\alpha
_{i}|\Gamma_{S\pm\frac{1}{2},S}^{M,m,k}\sum_{\ell}|\ell\rangle_{i}%
|s_{\bar{\imath}}^{\prime},m_{\bar{\imath}}^{\prime},\alpha_{i}^{\prime
}\rangle_{\bar{\imath}}\Gamma_{S,s_{\bar{\imath}}^{\prime}}^{m,m_{\bar{\imath
}}^{\prime},k}\nonumber\\
& =\sum_{k,\ell,m}|k\rangle_{C}|\ell\rangle_{A_{i}}\Gamma_{S\pm\frac{1}{2}%
,S}^{M,m,k}\Gamma_{S,s_{\bar{\imath}}}^{m,m_{\bar{\imath}},\ell}\equiv\hat
{Q}_{\pm,s,M}^{s_{\bar{\imath}},m_{\bar{\imath}}}.
\end{align}
In the last line we find that the overlap is independent of $\alpha$ and
$\alpha_{i}$, though with constraints $\alpha=(s_{\bar{\imath}},\alpha_{i})$,
which requires $\alpha_{i}=\alpha_{i}^{\prime}$. Therefore, different Kraus
operators provide exactly the same operation and, accordingly, we can sum over
these equivalent Kraus operators to reduce the number of indices. After this
process, we get%
\begin{align}
K_{\ell}^{0}  & \equiv K_{s_{\bar{\imath}},m_{\bar{\imath}},m_{\bar{\imath}%
}^{\prime}}^{0}\nonumber\\
& =2^{-\frac{N-1}{2}}\sqrt{N}\sum_{\pm,s_{A},M}\lambda_{\pm}(s_{A}%
)^{-1/2}\sqrt{g^{[N-1]}(s_{\bar{\imath}})}\times\nonumber\\
& \times\left(  \langle\psi_{AC}^{-}|\hat{Q}_{\pm,s_{A},M}^{s_{\bar{\imath}%
},m_{\bar{\imath}}}\hat{Q}_{\pm,s_{A},M}^{s_{\bar{\imath}},m_{\bar{\imath}%
}^{\prime}}{}^{\dagger}\right)  \otimes\openone_{B},\\
K_{\ell}^{1}  & \equiv K_{M,s_{\bar{\imath}},m_{\bar{\imath}}}^{1}\nonumber\\
& =\sqrt{\frac{g^{[N-1]}(s_{\bar{\imath}})}{2^{N-1}}}\;\hat{Q}_{+,N/2,M}%
^{s_{\bar{\imath}},m_{\bar{\imath}}^{\prime}}{}^{\dagger}\otimes\openone_{B}.
\end{align}
The Kraus operators of the reduced channel $\tilde{\Lambda}$ are obtained as
$(K_{\ell}^{u}\otimes\openone_{D})(|\Psi_{CD}^{-}\rangle\otimes\openone_{AB}%
)$. It is simple to check that the above operators define a CPTP-map.

\end{document}